\documentclass[preprint,showpacs,amsmath,amssymb,floatfix,superscriptaddress,longbibliography,aps,prb,nofootinbib]{revtex4-1}

\usepackage{graphicx,xcolor}
\usepackage{bm}
\usepackage{paralist}

\usepackage[utf8]{inputenc} 
\newlength{\alphabet}
\usepackage{hyperref} 
\usepackage{verbatim} 
\usepackage{amsmath} 
\usepackage{amsfonts}
\usepackage{txfonts}
\usepackage{amssymb}
\usepackage{tabularx}  
\usepackage{units} 
\usepackage{physics} 
\usepackage{bbold} 
\usepackage{slashed}
\usepackage{float}
\usepackage[english]{babel}

\usepackage{caption}

\begin{document}

\renewcommand{\floatpagefraction}{0.5}
\bibliographystyle{nature}


\title{Symmetry-enforced topological nodal planes at the Fermi surface of a chiral magnet}


\author{Marc A. Wilde\footnote{marc.wilde@ph.tum.de}}
\affiliation{Physik Department, Technische Universit\"at M\"unchen, D-85748 Garching, Germany}
\affiliation{Centre for QuantumEngineering (ZQE), Technische Universit\"at M\"unchen, D-85748 Garching, Germany}

\author{Matthias Dodenh\"oft}
\affiliation{Physik Department, Technische Universit\"at M\"unchen, D-85748 Garching, Germany}

\author{Arthur Niedermayr}
\affiliation{Physik Department, Technische Universit\"at M\"unchen, D-85748 Garching, Germany}

\author{Andreas Bauer}
\affiliation{Physik Department, Technische Universit\"at M\"unchen, D-85748 Garching, Germany}
\affiliation{Centre for QuantumEngineering (ZQE), Technische Universit\"at M\"unchen, D-85748 Garching, Germany}

\author{Moritz~M. Hirschmann}
\affiliation{Max-Planck-Institute for Solid State Research, Heisenbergstrasse 1, D-70569 Stuttgart, Germany}

\author{Kirill Alpin}
\affiliation{Max-Planck-Institute for Solid State Research, Heisenbergstrasse 1, D-70569 Stuttgart, Germany}

\author{Andreas P. Schnyder\footnote{a.schnyder@fkf.mpg.de}}
\affiliation{Max-Planck-Institute for Solid State Research, Heisenbergstrasse 1, D-70569 Stuttgart, Germany}

\author{Christian Pfleiderer\footnote{christian.pfleiderer@tum.de}}
\affiliation{Physik Department, Technische Universit\"at M\"unchen, D-85748 Garching, Germany}
\affiliation{Centre for QuantumEngineering (ZQE), Technische Universit\"at M\"unchen, D-85748 Garching, Germany}
\affiliation{MCQST, Technische Universit\"at M\"unchen, D-85748 Garching, Germany}

\date{\today}

\maketitle

\textbf{
Following over a decade of intense efforts to enable major progress in spintronics devices and quantum information technology by means of materials in which the electronic structure exhibits non-trivial topological properties, three key challenges are still unresolved\cite{neumann_wigner_z_physik, herring_PR_37,armitage_mele_vishwanath_review, chiu_RMP_16, burkov_review_Weyl, wang_hechang_nat_comm_intrinsic_AHE, huang_chiral_anomaly_PRX_15, ong_chiral_anomaly_PRX_18, huang_hasan_TaAs_Fermi_Arc_nat_comm_15}. First, the identification of topological band degeneracies that are generically rather than accidentally located at the Fermi level. Second, the ability to easily control such topological degeneracies. And third, to identify generic topological degeneracies in large, multi-sheeted Fermi surfaces. Combining de Haas -- van Alphen spectroscopy with density functional theory and band-topology calculations, we report here that the non-symmorphic symmetries\cite{MICHEL2001377, young_kane_rappe_PRL_12, FURUSAKI2017788, zhao_schnyder_PRB_16,zhangPRMat2018, 2019_PRB_Yuxin, PRB_Yuxin, turker_moroz_PRB_18} in ferromagnetic MnSi generate nodal planes (NPs)~\cite{young_kane_rappe_PRL_12, FURUSAKI2017788}, which enforce topological protectorates (TPs) with substantial Berry curvatures at the intersection of the NPs with the Fermi surface (FS) regardless of the complexity of the FS. We predict that these TPs will be accompanied by sizeable Fermi arcs subject to the direction of the magnetization. Deriving the symmetry conditions underlying topological NPs, we show that the 1651 magnetic space groups comprise 7 grey groups and 26 black-and-white groups with topological NPs, including the space group of ferromagnetic MnSi. Thus, the identification of symmetry-enforced TPs on the FS of MnSi that may be controlled with a magnetic field suggests the existence of similar properties, amenable for technological exploitation, in a large number of materials.}


\newpage
Nearly a century ago Wigner, von Neumann, and Herring \cite{neumann_wigner_z_physik, herring_PR_37} were first to address the conditions under which Bloch states form degenerate band crossings, but their topological character and technological relevance was only recognized recently\cite{chiu_RMP_16, armitage_mele_vishwanath_review, burkov_review_Weyl}.  To be useful\cite{armitage_mele_vishwanath_review, burkov_review_Weyl, wang_hechang_nat_comm_intrinsic_AHE, huang_chiral_anomaly_PRX_15, ong_chiral_anomaly_PRX_18, huang_hasan_TaAs_Fermi_Arc_nat_comm_15} tiny changes of a control parameter must generate a large response underscoring the lack of control over the band-filling as the unresolved key challenge in materials with band crossings known to date. This raises the question, if topological band crossings exist that are, (i) generically at the Fermi level, (ii) separated sufficiently in the BZ, and, (iii) easy to control. 

Natural candidates are systems with non-symmorphic symmetries, e.g., screw rotations, generating positions in reciprocal space at which band-crossings are symmetry-enforced. The associated key characteristics include: \cite{MICHEL2001377, young_kane_rappe_PRL_12, FURUSAKI2017788, zhao_schnyder_PRB_16,zhangPRMat2018, 2019_PRB_Yuxin, PRB_Yuxin, turker_moroz_PRB_18} 
(i) the crossings are due to symmetry alone, i.e., they occur on all bands independent of details such as chemical composition,
(ii) pairs of band-crossings with opposite chirality are separated in $k$-space by about half a reciprocal lattice vector, 
(iii)~the band crossings may be enforced on entire planes~\cite{young_kane_rappe_PRL_12, FURUSAKI2017788},  forming so-called nodal planes (NPs) with non-zero topological charge, and 
(iv) their existence may be controlled by means of symmetry breaking.
Thus, if in a material the Fermi surfaces (FSs) cross such topological NPs, they enforce pairwise FS degeneracies with large Berry curvatures. The topology of these FS degeneracies, which we refer to as "topological protectorates" (TPs), will be independent of material-specific details and, moreover, may be controlled by symmetry breaking.  
The putative existence of topological NPs was studied in phononic metamaterials \cite{2017_Xiao_arXiv,2019_Yang_NatComm, 2020_Xiao_SciAdv}, and mentioned in a study of non-magnetic chiral systems focussing on Kramers-Weyl fermions \cite{2018_Chang_NatMat}. 

To demonstrate the formation of symmetry-enforced TPs at the intersection of NPs with the FS, we decided to study the ferromagnetic state of MnSi, which attracts great interest for its itinerant-electron magnetism \cite{1988_Lonzarich_JMMM}, helimagnetism, skyrmion lattice\cite{2009_Muehlbauer_Science}, and quantum phase transition\cite{1997_Pfleiderer_PRB}. Crystallizing in space group (SG) 198, MnSi is a magnetic sibling of nonmagnetic RhSi\cite{RhSi_Nature_hasan_19}, CoSi \cite{CoSi_Nature_hong_19}, and PdGa \cite{2020_Schroeter_Science}, in which sizeable Fermi arcs and multi-fold fermions were recently inferred from ARPES. 
MnSi is ideally suited for our study as magnetic fields exceeding $\sim {0.7}$~T stabilize ferromagnetism with magnetic screw rotation symmetries enforcing NPs. 

\section*{Initial Assessment}
A first theoretical assessment establishes that a ferromagnetic spin-polarization along a high-symmetry direction, e.g., $[010]$, reduces the symmetries from space group (SG) 198 (P2$_1$3) of paramagnetic MnSi to the magnetic SG 19.27 (P$2_1 2'_1 2'_1$) (Supplementary Note~S1, Extended Data Fig.~\ref{EDI:Theory:Fig1}). This SG contains two magnetic screw rotations $\theta  \tilde{C}^x_2$ and $\theta  \tilde{C}^z_2$ (Fig.~\ref{fig:1}a), i.e., 180$^\circ$ screw rotations around the $x$ and $z$ axes combined with time-reversal symmetry $\theta$. These rotations act like mirror symmetries, since they relate Bloch wave functions at $(k_x, k_y, k_z)$ to those at $(-k_x, k_y, k_z)$ and $(k_x, k_y, -k_z)$, respectively, leaving the planes $k_x=0$ and $k_z=0$ and the BZ boundaries $k_x=\pm\pi$ and $k_z=\pm\pi$ invariant. 
Squaring $\theta  \tilde{C}^x_2$ and $\theta  \tilde{C}^z_2$ and letting them operate on the Bloch state $| \psi ( {\bf k}) \rangle$ one finds that $(\theta  \tilde{C}^x_2 )^2  | \psi ( {\bf k}) \rangle  =   e^{i k_x} | \psi ( {\bf k}) \rangle$  and $(\theta  \tilde{C}^z_2 )^2  | \psi ( {\bf k}) \rangle  =   e^{i k_z} | \psi ( {\bf k}) \rangle$. Hence, by Kramers theorem~\cite{kramers_theorem_paper}, all Bloch states on planes with $k_x=\pm\pi$ or $k_z=\pm\pi$ are two-fold degenerate. Moving away from these BZ boundaries, the symmetries are lowered such that the Bloch states become non-degenerate.  Therefore, all bands in ferromagnetic MnSi are forced to cross at $k_x=\pm\pi$ and $k_z=\pm\pi$, representing a duo of NPs. 
	
The topological charge $\nu$ of this duo of NPs (Fig.~\ref{fig:1}b) may be determined with the fermion doubling theorem (FDT) \cite{nielsen_no_go}, which states that $\nu$ summed over all band crossings must be zero. We note that besides the NPs there is an odd number of symmetry-enforced band crossings on the Y$_1-\Gamma-$Y and R$_1-$U$-$R lines forming Weyl points (WPs, $\nu = \pm 1$) and fourfold points (FPs, $\nu = \pm 2$), respectively (Figs.~\ref{fig:1}c and \ref{fig:1}d, Extended Data Fig.~\ref{EDI:Theory:Fig2}, Supplementary Note~S1). Moreover, due to the effective mirror symmetries accidental Weyl points away from these high-symmetry lines must form pairs or quadruplets with the same $\nu$. Because the sum over $\nu$ of all of these Weyl and fourfold points is odd, the duo of NPs must carry a nonzero topological charge to satisfy the FDT. Hence, the duo of NPs at the BZ boundary is the topological partner of a single Weyl point on the Y$_1-\Gamma-$Y line (Fig.~\ref{fig:1}b). This is a counter-example to Weyl semimetals, in which Weyl points occur always in pairs. 
	
Shown in Fig.~\ref{fig:1}d is the band structure of a generic tight-binding model satisfying SG 19.27 (Supplementary Note~S2), where pairs of bands form NPs on the BZ boundaries $k_x =\pm \pi$ and $k_z = \pm \pi$, while on the Y$_1-\Gamma-$Y and R$_1-$U$-$R lines there are Weyl and fourfold points, respectively. Explicit calculation of the Chern numbers shows that all of these band-crossings, including those at the nodal planes, exhibit nonzero topological charges as predicted above. In turn, all of the FSs carry substantial Berry curvatures. The numerical analysis shows that these Berry curvatures become extremal at the NPs and close to the fourfold and Weyl points (Extended Data Fig.~\ref{EDI:Theory:Fig3}). By the bulk boundary correspondence~\cite{chiu_RMP_16,armitage_mele_vishwanath_review}, the nontrivial topology of these band crossings generates large Fermi arcs on the surface, which extend over half of the BZ of the surface (Extended Data Fig.~\ref{EDI:Theory:Fig4}).
These arguments may be extended to 254 of the 1651 magnetic SGs of which 33 have nodal planes whose topological charges are enforced to be non-zero by symmetry alone (Supplementary Note~S3).  

\section*{Calculated Electronic Structure}
Shown in Fig.~\ref{fig:1}e is the Density Functional Theory (DFT) band structure of MnSi taking into account spin-orbit coupling (SOC), for the experimental moment of $0.41\mu_B/$Mn along $[010]$ (Methods, Extended Data Fig.~\ref{EDI:Exp:Fig6}). Ten bands are found to cross the Fermi level (Fig.~1e). In agreement with our symmetry analysis and the tight-binding model (Fig.\,\ref{fig:1}d), we find the same generic band crossings, namely: (i) NPs on the BZ boundaries $k_x=\pm\pi$ and $k_z=\pm\pi$, (ii) an odd number of Weyl points along Y$_1-\Gamma-$Y, and, (iii) an odd number of fourfold points along R$_1-$U$ -$R. 

The calculated FSs as matched to experiment are shown in Fig.~\ref{fig:1}f highlighting the NPs at the BZ boundaries at $k_x=\pm\pi$ and $k_z=\pm\pi$ (see Extended Data Table~\ref{EDI:Exp:tab1} for key parameters and Extended Data Fig.~\ref{EDI:Exp:Fig6}). Eight FS sheets centred at $\Gamma$ comprise two small isolated hole pockets (sheets $1$ and $2$), two intersecting hole pockets with avoided crossings and magnetic breakdown due to SOC (sheets $3$ and $4$), and two pairs of jungle-gym type sheets, ($5$,$6$) and ($7$,$8$). Sheets $9$ and $10$ are centered at $R$, comprising eight three-fingered electron pockets around the $[111]$ axes and a tiny electron pocket, respectively. 
The sheet pairs ($5$,$6$), ($7$,$8$) and ($9$,$10$) extend beyond the BZ boundaries with pairwise sticking at the NPs. They represent TPs, marked in red, with extremal Berry curvatures protected by the magnetic screw rotations $\theta  \tilde{C}^x_2$ and $\theta  \tilde{C}^z_2$. In contrast, sheets $5$ through $10$ do not form TPs at the BZ boundary $k_y=\pm\pi$, because the moment pointing along $[010]$ breaks $\theta  \tilde{C}^y_2$.

Rotating the direction of the magnetization away from $[010]$ distorts the FS sheets, where TPs exist only on those BZ boundaries parallel to the magnetization (cf. Supplementary Video). For instance, rotating the moments within the $x-y$ plane away from $[010]$ breaks the magnetic screw rotation $\theta  \tilde{C}^x_2$, but keeps $\theta  \tilde{C}^z_2$ intact. In turn the TPs gap out on the  $k_y=\pm\pi$ and $k_x=\pm\pi$ planes, while they remain degenerate at $k_z=\pm\pi$ planes (cf. Extended Data Fig.~\ref{EDI:Theory:Fig1}, Suppl. Note~S1).  

\section*{Experimental Results}
To experimentally prove the mechanism causing generic TPs at the intersection of the FS with symmetry-enforced NPs and their dependence on the direction of the magnetization, we mapped out the FS by means of the dHvA effect using capacitive cantilever magnetometry (Methods, Extended Data Fig.~\ref{EDI:Exp:Fig6} and Suppl. Note~S4). 
In the following we focus on magnetic field rotations in the $(001)$ plane, where $\varphi$ denotes the angle of the field with respect to $[100]$. This plane proves to be sufficient to infer the main FS features. Complementary data for the $(001)$ and $(\bar 1 \bar 1 0)$ planes are presented in the Extended Data Fig.\,\ref{EDI:Exp:Fig6}. 
Typical torque data at different temperatures for $\varphi=82.5^{\circ}$ (Fig.~\ref{fig:2}a,b) display pronounced dHvA oscillations for fields exceeding $B \sim 0.7$\,T. The hysteretic behavior below $ \sim 0.7$\,T (inset of Fig.~\ref{fig:2}a) originates from the well-understood helimagnetic and conical phases \cite{2017_Bauer_PRB}.  Figure~\ref{fig:2}b shows the oscillatory high-field part of the torque $\tau (1/B)$ at $T=35$\,mK with the low-frequency components removed for clarity. To extract the dHvA frequencies an FFT analysis of $\tau (1/B)$ was carried out, where the effects of demagnetizing fields and the unsaturated magnetization were taken into account (Methods). The FFT frequencies correspond to extremal FS cross-sections in low effective fields of $\sim 0.7-1.9\,{\rm T}$ (Methods).

Typical dHvA frequencies and FFT amplitudes, shown for $\varphi=82.5^{\circ}$ in Fig.~\ref{fig:2}c, display five different regimes of dHvA frequencies labelled I to V. They comprise over $40$ dHvA frequencies corresponding to different extremal FS orbits as denoted by greek letters (Fig.~\ref{fig:2}c and Extended Data Table~\ref{EDI:Exp:tab2}). In our data analysis we delineated artefacts due to the finite FFT window, such as the side lobes between $\kappa_2$ and $2\kappa_1$, or $3\kappa_2$ and $\xi_1$ (cf. Methods).
Fitting the temperature dependence of the FFT amplitudes within Lifshitz-Kosevich theory \cite{Shoenberg1984}, the effective masses for each of the orbits were deduced ranging from $m^*=0.4\,m_e$ to $17\,m_e$, where $m_e$ is the bare electron mass (Fig.~\ref{fig:2}d). 

To relate the dHvA frequencies to the calculated FS orbits, the torque amplitude was inferred from the DFT band structure by means of the Lifshitz-Kosevich formalism, using small rigid band shifts of the order of $10$~meV to improve the matching following convention (cf. Methods and Extended Data Table~\ref{EDI:Exp:tab1}). The assignment to experiment was based on the consistency between dHvA frequency, angular dispersion, strength of torque signal, field-dependence of the dHvA frequencies, effective masses, and presence of magnetic breakdown as explained in Methods, Extended Data Table~\ref{EDI:Exp:tab2}, Extended Data Fig.\,\ref{EDI:Exp:Fig9} and \ref{EDI:Exp:Fig10}, and Supplementary Note~S5. 

Shown in Fig.~\ref{fig:3}a1 and \ref{fig:3}a2 is an intensity maps of the experimental data of the $(001)$ plane as a function of $\varphi$, where the theoretical dHvA branches are depicted by colored lines (colors correspond to the FS sheets in Fig.~\ref{fig:1}). For comparison,  Fig.~\ref{fig:3}b1 and \ref{fig:3}b2 displays an intensity map of the calculated dHvA spectra, where the experimental frequencies are marked by grey crosses. 

For regimes I through IV, featuring contributions of the large FS sheets (5,6) and (7,8), all frequencies may be assigned unambiguously (Extended Data Fig.\,\ref{EDI:Exp:Fig9} and \ref{EDI:Exp:Fig10}, Suppl. Note~S5). Namely, regime I contains the loop orbits around U associated with pair (5,6) (blue and orange) and the neck orbit of sheet 8 (yellow). Regime II exhibits the dHvA branches originating from neck orbits around  $\Gamma$-$Y$-$\Gamma$ on sheet 7 (purple). The neck orbits of sheet 8 (yellow), which evades detection because of the large slope of the dispersion, its high mass, and the suppression of the magnetic torque near $[010]$ high symmetry direction is consistent with an anomalous frequency splitting at the expected crossing with the loop orbits of pair (5,6) (blue and orange) around 6.5\,kT. (Supplementary Note~S5). Regime III arises from both, pairs (5,6) and (7,8), i.e., neck orbits around $\Gamma$-$Y$-$\Gamma$ of (5,6) and loop orbits around U of (7,8). The remaining cascade of frequencies in regime III reflects breakdown orbits (translucent yellow) arising from avoided crossings between sheets 3 and 4 (red and green). Regime IV is, finally, dominated by sheet 2 of the isolated hole pocket and the $1^{\rm{st}}$ harmonic of sheet 2.

As the magnetic torque generically vanishes at high-symmetry directions, which corresponds to the $\langle100\rangle$ axes in regimes I through IV, the associated FS sheets are centered at the $\Gamma$-point. Likewise, the lowest frequency in regime V corresponds to a $\Gamma$-centered FS sheet, which can be assigned to the small hole pocket of sheet 1. In stark contrast, for regime V above $\sim 0.05\,{\rm kT}$ the high-symmetry directions correspond to the $\langle111\rangle$ axes, while the torque for the $\langle100\rangle$ axes is finite (see also Fig.~\ref{fig:3}a2 and \ref{fig:3}b2 and Extended Data Fig.\,\ref{EDI:Exp:Fig6}g).  Hence, regime V is related to FS pockets in the vicinity of the R-point that may be assigned to FS sheets (9,10) without the need for a detailed account of their shape, completing the assignment. The calculations demonstrate the presence of symmetry-enforced crossings of sheets (9,10) if they intersect the NPs (cf. Fig.~\ref{fig:1}). 

To confirm that we observed the entire FS, we calculated the Sommerfeld coefficient of the specific heat from the density of states at the Fermi level as rescaled by the measured mass enhancements (Extended Data Table~\ref{EDI:Exp:tab1}). Excellent agreement is observed within a few percent of experiment\cite{Bauer2010}, $\gamma\approx28$~mJmol$^{-1}$K$^{-2}$ at $B=12$~T. This analysis reveals, that sheets (5,6), (7,8), and (9,10), which form TPs, contribute 86\% to the total density of states at the Fermi level. 

\section*{Topological Nodal Planes}
Spectroscopic evidence of the symmetry-enforced topological band degeneracies at the BZ boundaries may be inferred from FS sheets (5,6). Identical characteristics are observed for FS sheets (7,8) (cf Extended Data Fig.\,\ref{EDI:Exp:Fig10} and Supplementary Note~S5). We note that the dHvA cyclotron orbits are perpendicular to the NPs for fundamental reasons, piercing through them at specific points of the TPs.
As shown in Fig.~\ref{fig:4}a, a magnetic field parallel to $[010]$ leads to extremal cross-sections for FS sheets (5,6) supporting cyclotron orbits in the vicinity of the U and the Y$_1$ point on planes depicted by blue and green shading, respectively. Centred with respect to the U point are possible cyclotron orbits comprising different segments of FS sheets 5 and 6, which interact at TP1 to TP4 with the BZ boundaries at $k_x=\pm\pi$ and $k_z=\pm\pi$. In the absence of the non-symmorphic symmetries, these intersections would exhibit anticrossing and magnetic breakdown leading to several orbits with different cross-sections and hence several dHvA frequencies. Instead the behaviour is distinctly different to magnetic breakdown or Klein tunneling \cite{2017_Alexandradinata_PRL,2018_vanDelft_PRL}.

As the BZ boundaries at $k_x=\pm\pi$ and $k_z=\pm\pi$ represent symmetry-enforced NPs, the crossing-points of sheets 5 and 6 at TP1 to TP4 are, hence, protected band-degeneracies at which the wave-functions are orthogonal, i.e., TP1 to TP4 are part of the topological protectorates that suppress transitions between orbits (we call orbits containing at least one TP "topological orbits"). In turn, two independent topological orbits (topological orbits 1 and 2) with identical areas and hence the same dHvA frequencies are expected (Figs.~\ref{fig:4}b1 and \ref{fig:4}b2). This is in excellent agreement with experiment, which shows a single dHvA frequency for field parallel $[010]$ ($\varphi = 90^{\circ}$ in Fig.~4c).
Rotating the direction of the magnetic field within the $xy$ plane away from $[010]$, the NP at $k_x = \pm \pi$ gaps out, while the NP at $k_z = \pm \pi$ remains protected. Thus, the associated loop orbits around U (Figs.\,\ref{fig:4}b3 and \ref{fig:4}b4) continue to include two points on the FS at $k_z = \pm \pi$ (TP3 and TP4), leading to two additional topological orbits (topological orbits 3 and 4) of identical cross-section with the same dHvA frequency, in perfect agreement with the observed spectra (Fig.\,\ref{fig:4}c). 

Comparing the extremal cross sections of the neck orbits around $\Gamma$-Y$_1$-$\Gamma$ with those around $\Gamma$-{X}-$\Gamma$, the latter crosses a NP while the former does not. With respect to $\Gamma$-{X}-$\Gamma$  there would be two extremal cross sections with identical areas, positioned symmetrically with respect to X (Fig.~\ref{fig:4}d1), whereas for the cross sections with respect to $\Gamma$-Y$_1$-$\Gamma$ there are two extremal orbits with different areas positioned asymmetrically with respect to Y$_1$ (Fig.~\ref{fig:4}d2). Thus within our symmetry analysis and our DFT calculations we expect a single dHvA branch for neck orbits parallel to a NP as compared to two dHvA branches for neck orbits that are not parallel to a NP (Fig.~\ref{fig:4}d3). Keeping in mind that only neck orbits around Y$_1$ are accessible experimentally, we clearly observe two branches, giving strong evidence that there are no NPs on the $k_y = \pm \pi$ BZ boundary (Fig.~\ref{fig:4}e). 

\section*{Concluding Remarks}
The symmetry-enforced NPs and TPs that are generically located at the Fermi level, which support large Berry curvatures, may account for various properties, such as anomalous Hall currents ~\cite{niu_review_RMP_10} or the nonlinear optical responses~\cite{morimoto_nagaosa_science_advances_16}. Indeed, large anomalous contributions to the Hall response are in excellent \textit{quantitative} agreement with ab initio calculations, where the calculated FS and Berry curvatures were essentially identical to the FS we report here \cite{2014_Franz_PRL}.
Our calculations imply also sizeable Fermi arcs at the surface of MnSi and related magnetic compounds such as FeGe and Fe$_{1-x}$Co$_{x}$Si, connecting the topological charge of the NP directly with a Weyl point (cf. Extended Data Fig.\,\ref{EDI:Theory:Fig4}). These Fermi arcs reflect the presence of duos of NPs. Analogous Fermi arcs will not exist in non-magnetic materials with SG 198\cite{RhSi_Nature_hasan_19, CoSi_Nature_hong_19, 2020_Schroeter_Science} which support trios of NPs (Supplementary Note\,S1). 

In systems with symmetry-enforced NPs and TPs tiny changes of the direction of the magnetization will control the topological band crossing in the bulk and the Fermi arcs, causing massive changes of Berry curvature that may be exploited technologically. The formation of TPs irrespective of the complexity of the FS, raises the question of whether they affect the transport properties \cite{2006_Smith_PRB} and enable exotic states of matter \cite{2014_Grover_JSM}. Extending the analysis presented here to all 1651 magnetic SGs, we find that there is a large number of candidate materials, such as CoNb$_3$S$_6$ \cite{2020_Tenasini_PRM} or Nd$_5$Si$_3$ \cite{2001_Boulet_JAC} with similar topological protectorates (Extended Data Table~\ref{EDI:Theory:tab1}, Supplementary Note~S3), which await to be explored from a fundamental point of view and harnessed for future technologies. 



\newpage
\section*{References}


\newpage
\section*{Acknowledgements}

We wish to thank D. Grundler, F. Rucker, A. Leonhardt, A. Rosch, T. Rapp, S.G. Albert and S.M. Sauther for support and discussions. Preliminary band structure calculations for a limited number of field orientations using FLEUR and JuDFT KKR-GGA were carried out in collaboration with F. Freimuth, B. Zimmermann, and Y. Mokrousov in the very early stages of this study. M.A.W., A.B. and C.P. were supported through DFG TRR80 (project-id 107745057, project E1 and E3), DFG SPP 2137 (Skyrmionics) under grant number PF393/19 (project-id 403191981), DFG GACR Projekt WI 3320/3-1, ERC Advanced Grants No. 291079 (TOPFIT) and 788031 (ExQuiSid), and Germany's excellence strategy EXC-2111 390814868.


\section*{Author Contributions}
M.A.W. and C.P. conceived the experiment and devised its interpretation together with A.P.S..
A.B. prepared and characterized the samples. 
M.D. and M.A.W. conducted the measurements and analyzed the data. 
M.A.W., A.N., and K.A. performed comprehensive band structure calculations. 
M.A.W. connected the experimental data with the calculated band structure.  
M.M.H., K.A., and A.P.S. performed the symmetry analysis and identified the topological properties of the band structure. 
M.M.H. and K.A.  calculated the surface states and the Berry curvatures.
M.A.W., A.P.S., and C.P. wrote the manuscript with contributions from M.M.H. and K.A..
All authors discussed the data and commented on the manuscript.


\section*{Supplementary information}

The supplementary information comprises a text file and a video. The text file reports comprehensive information on the theoretical framework, theoretical analysis, and magnetic space groups featuring topological nodal planes, as well as the analysis of the experimental data. The video highlights the evolution of a cut-away view of the Fermi surface akin Fig.\,\ref{fig:1}\,f  as a function of the direction of the magnetization tracking an applied magnetic field. 


\section*{Data availability} 
Materials and additional data related to this paper are available from the corresponding authors upon reasonable request.


\section*{Competing interests} 
The authors declare no competing interests.


\section*{Correspondence and requests for materials} 
Correspondence and requests for materials should be addressed to M.A.W., A.S. or C.P.

\clearpage \thispagestyle{empty}

\captionsetup[figure]{labelfont={bf},name={Fig.},labelsep=space}

\begin{figure}[h]
\centerline{\includegraphics[width=10.6cm,clip=]{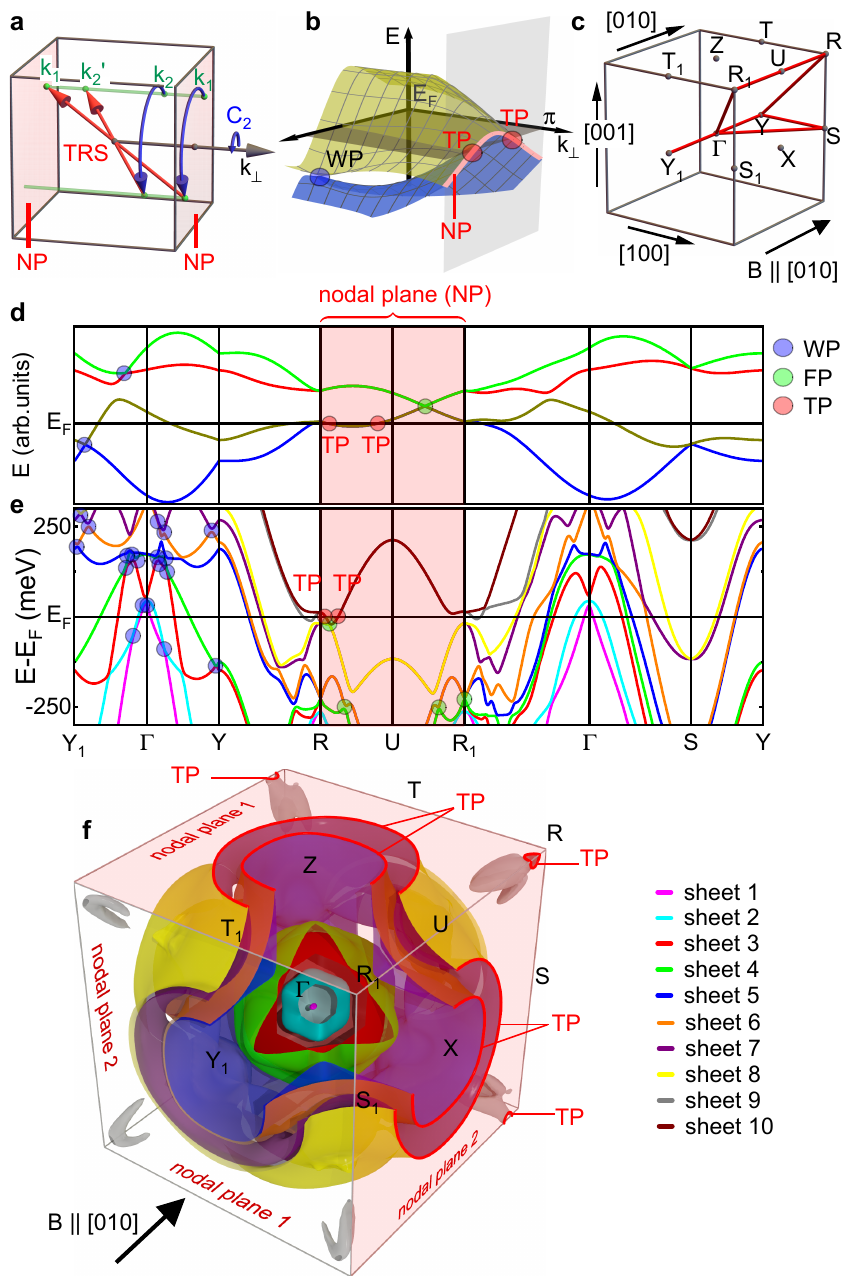}}
\linespread{1.0}\selectfont{}
\caption{\raggedright
{\bf $|$ Symmetries, band topology, Fermi surface protectorates, and band structure of ferromagnetic MnSi.}
{\bf a},~Action of the magnetic screw rotations and time-reversal symmetry (TRS) on the k-points in the Brillouin zone (BZ). 
{\bf b}, Pairs of energy bands $E(k)$ close to the Fermi energy $E_{\mathrm{F}}$ forming a topological nodal plane (NP, red line) on the BZ boundary that is perpendicular to the screw rotation axis. This NP is the topological partner of a single Weyl point in the bulk (blue dot) of opposite topological charge.
{\bf c}, High-symmetry paths in the cubic primitive BZ. Special k-points are denoted by the orthorhombic primitive notation with subscripts for easier identification.
\textbf{d}, Generic tight-binding band structure illustrating generic band degeneracies of ferromagnetic MnSi with its magnetic space group, SG 19.27, namely Weyl points (WP), fourfold degenerate points (FP),  nodal planes (NP), and topological protectorates (TP). 
\textbf{e}, Band structure of ferromagnetic MnSi for magnetization along $[010]$ as calculated in DFT.
Ten bands cross the Fermi level, distinguished by different colors corresponding to the FS sheets numbered in \textbf{f}. 
\textbf{f}, Calculated FS sheets adapted to match the experimental data under magnetic field along $[ 010]$, as discussed in Methods. 
Note the presence of NPs on the Brillouin zone boundaries, $k_x=\pm\pi$ and $k_z=\pm\pi$, as well as TPs marked in red.
}
\label{fig:1}
\end{figure}

\clearpage \thispagestyle{empty}

\begin{figure}[h]
\centerline{\includegraphics[width=1.0\textwidth,clip=]{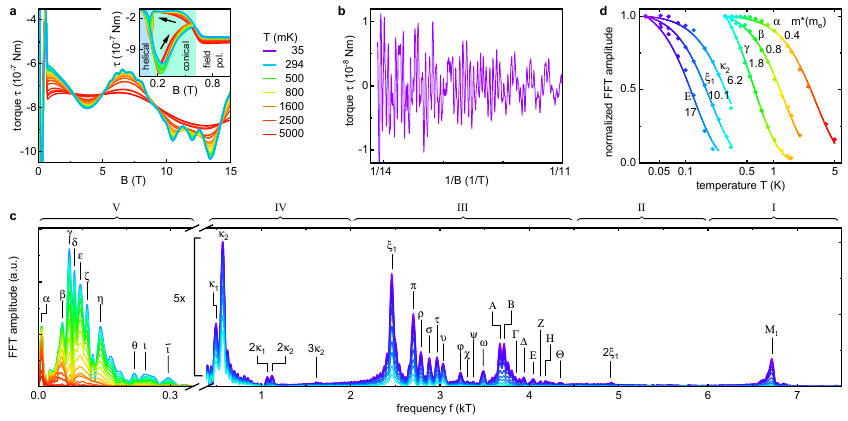}}
\linespread{1.0}\selectfont{}
\caption{\raggedright
{\bf $|$ Typical de Haas -- van Alphen data of ferromagnetic MnSi.} 
{\bf a}, 
De Haas -- van Alphen oscillations detected in the magnetic torque $\tau$ as a function of magnetic field for fixed field direction $\varphi=82.5^{\circ}$. Different colors represent different temperatures within the range  $0.035\,{\rm K}$ and $5\,{\rm K}$. The inset displays the hysteretic behavior in the regime of the helical and conical phases at low fields. 
{\bf b}, High-field part of the magnetic torque $\tau (1/B)$ at T$=35$~mK with low-frequency components removed. 
{\bf c} FFT spectra of $\tau (1/B)$ for the same field angle and temperature range as in {\bf a}.  The spectra naturally group into five regimes (labelled by I-V), each of which exhibits a number of pronounced dHvA frequencies (greek letters).
{\bf d} Normalized FFT amplitudes of six selected dHvA frequencies as a function of temperature. The lines represent fits to the Lifshitz-Kosevich formula, from which we obtain the effective masses $m^*$ for the corresponding extremal FS orbits.  
} \label{fig:2}
\end{figure}

\clearpage \thispagestyle{empty}

\begin{figure}[h]
\centerline{\includegraphics[width=8.6cm,clip=]{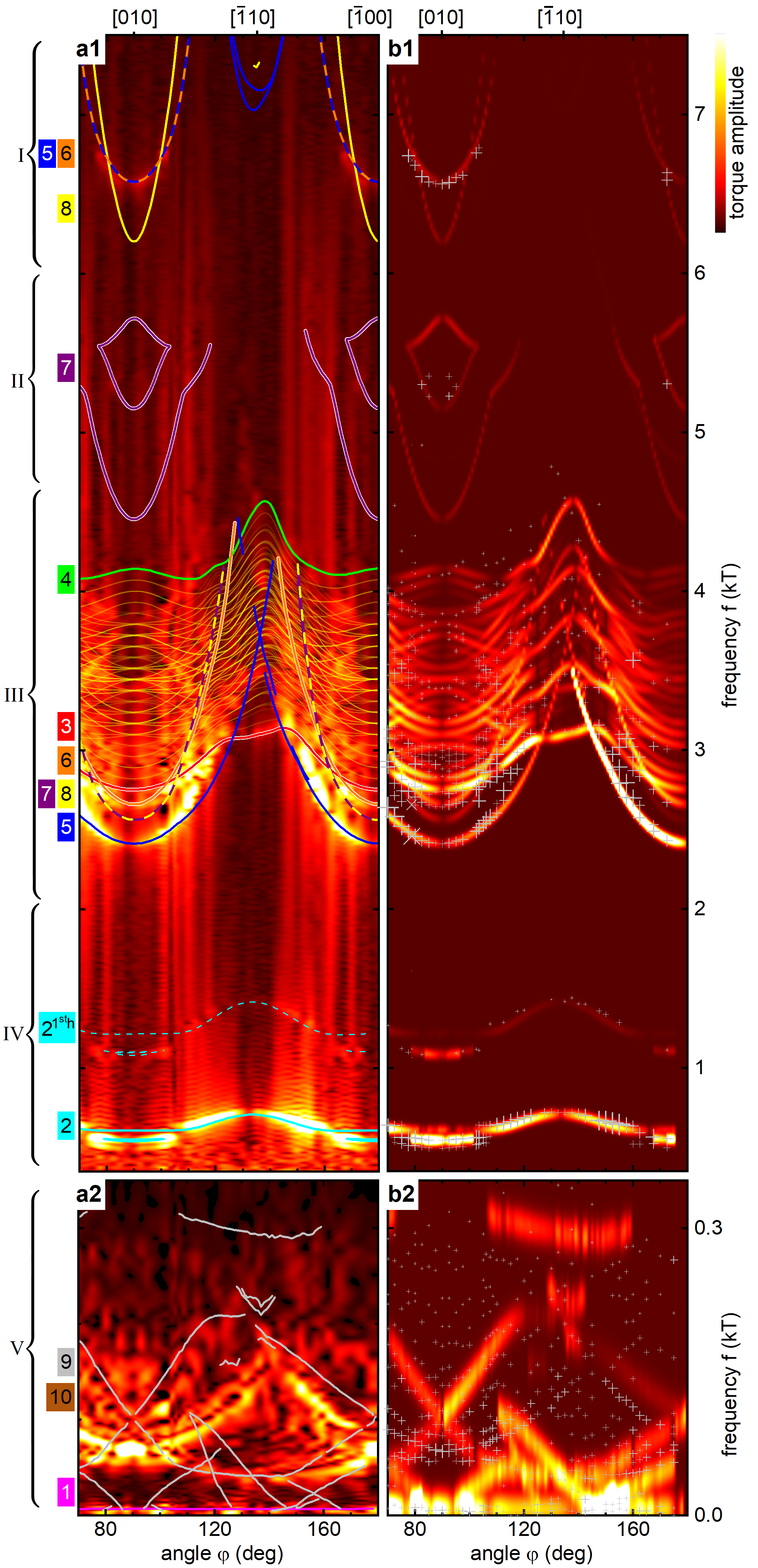}}
\linespread{1.0}\selectfont{}
\caption{\raggedright
\textbf{$|$
Experimental and theoretical dHvA spectra in the $\mathbf{(001)}$ plane as a function of field angle $\varphi$.} 
{\bf a}, 
FFT amplitudes of the experimentally observed dHvA spectra at $T=280$~mK as a function of frequency $f$ and field angle $\varphi$. The thin colored lines represent the theoretical dHvA branches, calculated from the \textit{ab-inito} band structure, where the color indicates the FS sheet  (cf.~\ Fig.~\ref{fig:1}f) from which the dHvA branch originates. The first harmonic (1$^{\mathrm{st}}$h) of the branches originating from sheet 2 is also labelled for clarity. A line cut of this color map for fixed  field angle $\varphi=82.5^{\circ}$ is shown in Fig.~\ref{fig:2}\,\textbf{c}. More than 40 dHvA branches were observed as listed in Extended Data Table~\ref{EDI:Exp:tab2}. 
{\bf b}, 
Torque amplitudes of the dHvA spectra inferred from the \textit{ab-initio} band structure (Methods), as a function of $f$ and $\varphi$, with the experimental frequencies of the dHvA branches indicated by crosses. To obtain a quantitative matching between theoretical and experimental dHvA branches, small rigid energy shifts to the \textit{ab-initio} bands were applied, as summarized in Extended Data Table~\ref{EDI:Exp:tab1} and Supplementary Note~S4. The detailed procedure how the experimental and theoretical dHvA branches were matched is described in the main text and in Supplementary Note~\ref{SI_sec_fs_determination}. }
\label{fig:3}
\end{figure}

\clearpage \thispagestyle{empty}

\begin{figure}
\centering
\centerline{\includegraphics[width=7.0cm,clip=]{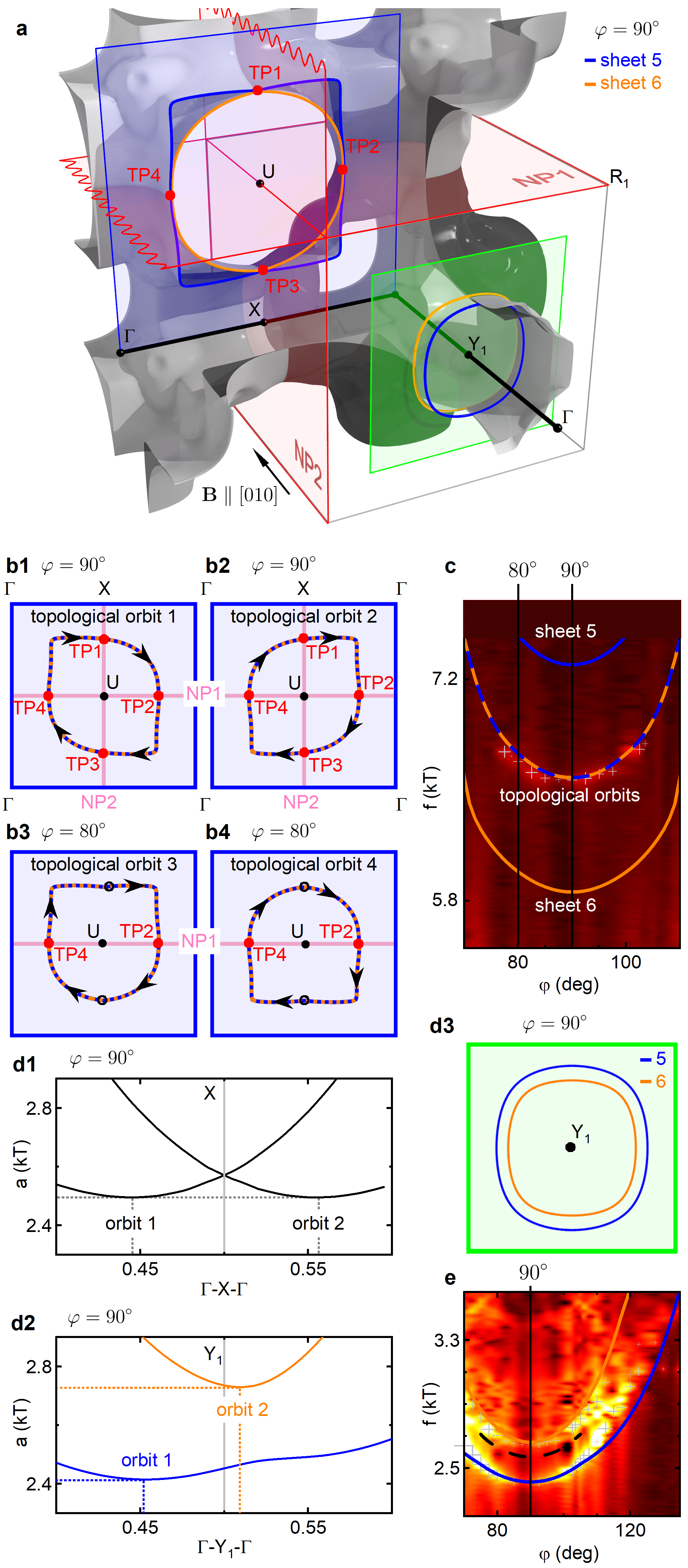}}
\linespread{0.8}\selectfont{}
\caption{\raggedright
\textbf{$|$
Extremal orbits and spectroscopic signatures of nodal planes and topological protectorates. }
Identical features presented here for sheet pair (5,6) are also observed for FS sheet pair (7,8) (Extended Data Fig.\,\ref{EDI:Exp:Fig10} and Suppl. Note\,S5).
{\bf a},
FS sheet pair (5,6) for a field $\mathbf{B}$ along the $\left[ 010 \right]$ direction (for an alternative color shading see Extended Data Fig.\,\ref{EDI:Exp:Fig9}d1). Planes illustrating loop- and neck-type orbits around the U point and the $\Gamma$-Y$_1$-$\Gamma$ line are indicated by blue and green shading, respectively. Loop orbits with respect to the U point intersect at TP1 to TP4 with the nodal planes on the $k_x = \pi$ and $k_z = \pi$ BZ boundaries. The nodal planes enforce degeneracies at TP1 to TP4, where the wave functions are orthogonal. Instead of anticrossing and magnetic breakdown topological orbits stabilize as illustrated in panels b1 to b4. 
\textbf{b1} and {\bf b2} Cross-sectional areas under field along $[010]$ at $\varphi=90^{\circ}$. 
\textbf{b3} and {\bf b4} Schematic cross-sectional areas under rotated field for $\varphi=80^{\circ}$.
{\bf c},
Intensity map of dHvA spectra in the regime of loop- and neck-type orbits around the U point (cf. Fig.\,\ref{fig:3}). Spectra are in excellent agreement with the topological orbits. No evidence for independent orbits of FS sheets 5 and 6 are observed.
\textbf{d1},  Symmetrical position of extremal FS cross-sections along $\Gamma$-X-$\Gamma$ with respect to the nodal plane at the X point. The associated orbits give rise to identical dHvA frequencies. Note that these orbits are not accessible experimentally.
{\bf d2}, and {\bf d3} Asymmetrical position of extremal FS cross-sections along $\Gamma$-Y$_1$-$\Gamma$, where the BZ boundary at $k_y=\pm\pi$ is no NP. The associated orbits, sketched in d3, give rise to different dHvA frequencies.
{\bf e},
Intensity map of dHvA spectra in the regime of neck-type orbits around the Y$_1$ point (cf. Fig.\,\ref{fig:3}). Spectra are in excellent agreement with two orbits as shown in panels d2 and d3, i.e., no NP at the BZ boundary at $k_y=\pm\pi$ containing Y$_1$. 
}
\label{fig:4}
\end{figure} 

\clearpage \thispagestyle{empty}

\newpage
\section*{Methods}

\textbf{Sample preparation}: 
For our study, two MnSi samples were prepared from a high-quality single crystalline ingot obtained by optical float-zoning \cite{Neubauer2011}. The samples were oriented by x-ray Laue diffraction and cut into 1x1x1~mm$^3$ cubes with faces perpendicular to $[100]$, $[110]$, and $[110]$ and $[110]$, and $[111]$ and $[112]$ cubic equivalent directions, respectively. Both samples exhibited a residual resistivity ratio (RRR) close to $300$.

\textbf{Experimental methods}: 
Quantum oscillations of the magnetization, i.e., the de Haas-van Alphen (dHvA) effect was measured by means of cantilever magnetometry measuring the magnetic torque $\mbox{\boldmath $ \tau$}=\mathbf{m} \times \mathbf{B}$. The double-beam type cantilevers sketched in Extended Data Fig.~\ref{EDI:Exp:Fig6}\,e were obtained from CuBe foil by standard optical lithography and wet-chemical etching. The cantilever position was read out in terms of the capacitance between the cantilever and a fixed counter electrode using an Andeen-Hagerling AH2700A capacitance bridge, similar to the design described in Refs.~\onlinecite{Wilde2008,Wilde2010}.

Angular rotation studies were performed in a $^3$He insert with a manual rotation stage at a base temperature $T=280$~mK under magnetic fields up to 15\,T. In addition, the effective charge carrier mass was determined using a dilution refrigerator insert with fixed sample stage under magnetic fields up to 14\,T (16\,T using a Lambda-stage) at temperatures down to 35\,mK.

We discuss partial rotations in the $(001)$ and $(\bar{1} \bar{1}0)$ crystallographic planes. The angle $\varphi$ is measured from $[100]$ in the $(001)$ plane and the angle $\theta$ is measured from $[001]$ in the $(\bar{1} \bar{1}0)$ plane. Corresponding data are shown in Fig.\ref{fig:3}\,a and Extended Data Fig.\ref{EDI:Exp:Fig6}\,g. Due to the topology of the FS and the simple cubic BZ, the $(001)$ plane rotation exhibits most of the extremal orbits and is already sufficient for an assignment to the FS sheets. For this reason, the discussion of the dHvA data in the main text focuses on the rotation in the $(001)$ plane.

The response of the cantilever was calibrated by means of the electrostatic displacement taking into account the cantilever bending line obtained from an Euler-Bernoulli approach \cite{WildeDiss2004}. Applying a DC voltage, $U$, to the capacitance $C_0=\epsilon_0A/d_0$ defined by the area $A$ and the plate distance $d_0$ leads to an electrostatic force $F=C_0U^2/2d_0$. This force is equivalent to a torque $\tau =\beta F L$, where $L$ is the effective beam length and $\beta=0.78$ is a geometry-dependent prefactor accounting for the different mechanical response of a bending beam to a torque and force, respectively. From this, the calibration constant $K(C)=\tau / \Delta C$ quantifying the capacitance change $\Delta C$ in response to the torque was obtained for different values of $C$. Changes in $K(C)$ up to 10$\%$ were recorded during magnetic field sweeps. The torque was calculated using
\begin{equation}
\tau (C)=\int_{C_0}^C K(C')dC' \mbox{ .}
\end{equation}\label{eq:torquecal}

\textbf{Evaluation of the dHvA signal}:
The magnetic field dependence of the capacitance, $C(B_{\text{ext}})$, was converted into torque and corrected as described below. An exemplary torque curve obtained at $T=280$\,mK and $\varphi=82.5^{\circ}$ is shown in Fig.~\ref{fig:2}\,a. In the regime below $B\sim0.7$~T the transitions from helical to conical and field polarized state generated a strongly hysteretic behavior. At higher fields, magnetic quantum oscillations on different amplitude and frequency scales could be readily resolved. The first low-frequency components appeared at magnetic fields as low as $B\sim4$~T, whereas several high-frequency components, corresponding to larger extremal cross sections, could only be resolved in high fields (Fig.~\ref{fig:2}\,b). Consequently, the data acquisition and evaluation was optimized by treating low- and high-frequency components separately.

In order to eliminate the non-oscillatory component of the signal, low-order polynomial fits or curves obtained by adjacent averaging over suitable field intervals were subtracted from the data, producing consistent results. Fast Fourier transforms (FFTs) of $\tau (1/B)$ were used to determine the frequency components contained in the signal. Field sweeps were performed from 0 to 15\,T at 0.03-0.04\,T/min and from 15\,T to 10\,T at 0.008\,T/min. FFTs over the range 4 to 15\,T (10\,T to 15\,T) were performed to evaluate frequency components below (above) $f=350$\,T for measurements in the $^3$He insert and from  10\,T to 14\,T (11\,T to 16\,T with Lambda-stage) in the dilution refrigerator. The values correspond to the applied field before taking into account demagnetization. Rectangular FFT windows were chosen in order to maximize the ability to resolve closely spaced frequency peaks. See Supplementary Note~\ref{SI_sec_experimental_details} for details.

\textbf{Internal magnetic field and dHvA frequency f(B) in a weak itinerant magnet}:
MnSi is a weak ferromagnet with an unsaturated magnetization up to the largest magnetic fields studied. This results in two different peculiarities concerning the observed dHvA frequencies. (i) The field governing the quantum oscillations is the internal field \cite{Shoenberg1984} $\mathbf{B}_{int}=\mu_0 \mathbf{H}_{ext}+\mu_0 (1-N_d)\mathbf{M}$. Taking into account the demagnization factor\cite{Aharoni1998} $N_d=\frac{1}{3}$ for a cubic sample to first order yields a field correction $\Delta B=B_{int}-B_{ext}=\frac{2}{3}\mu_0 M_{exp} \approx 0.131$~T, taking the low-field value of $M_{exp}$ in the field-polarized phase. The applied field was corrected by this value. The field dependence of the magnetic moment yields only a minor correction of the internal field that may be neglected. (ii) The effect of the unsaturated magnetization on the Fermi surface is more prominent and may be described in a good approximation as a rigid Stoner exchange splitting that scales with the magnitude of the magnetization. Consequently, FS cross-sectional areas are enlarged with increasing $B$ for majority electron orbits and minority hole orbits. Cross-sectional areas shift downwards for majority hole and minority electron orbits.

This change in cross-sectional area is not directly proportional to the change in the observed dHvA frequencies $f$, i.e., the dHvA frequencies deviate from the field-dependent frequency $f_B(B)=\frac{\hbar}{2\pi e}A_k(B)$ obeying the Onsager relation (Here $A_k$ is the extremal cross-sectional area in k-space, $\hbar$ is the reduced Planck constant and $e$ is the electron charge). The frequency $f$ observed may be inferred\cite{Ruitenbeek_1982} from the derivative of the dHvA phase factor $2\pi \left( \frac{f_B(B)}{B}-\gamma \right)\pm \frac{1}{4}$ with respect to $1/B$:
\begin{equation}
f(B)=\frac{d}{dB^{-1}} \left(\frac{f_B(B)}{B} \right) = f_B(B) -B\frac{df_B(B)}{dB} \mbox{ .}
\label{eq_fB}
\end{equation}
Thus, a linear relation $f_B(B)$ results in a \emph{constant} $f(B)$. This may be understood intuitively, because a linear term in $f_B(B)$ leads only to a phase shift since the oscillations are periodic in $1/B$. Equation \eqref{eq_fB} shows that $f(B)$ is the zero-field intercept of the tangent to $f_B(B)$.

In the Stoner picture of rigidly split bands $f_B(B)$ may be related to the magnetization\cite{Ruitenbeek_1982,Kimura_2004} using
\begin{equation}
f_B(B)-f_0=\pm \frac{m_b}{m_e} \frac{Is}{4\mu_B^2} M(B) \mbox{ ,}
\end{equation}
where $I$ is the Stoner exchange parameter, $m_b$ is the band mass, the $\pm$ is for electron and hole orbits, respectively, $s=\pm 1$ is the spin index and $f_0$ is the hypothetical frequency without exchange splitting. Note, that this model is only meaningful in the field-polarized regime $B\gtrsim 0.7$~T. Using the experimental $M(B)$ curve of MnSi \cite{Bauer2010}, we estimate that the frequencies $f(B)$ in the windows used for $f>350$~T defined above with center fields $B_{average}=2B_{high}B_{low}/(B_{low}+B_{high})$ ranging from $11.8$~T to $13.2$~T correspond to the extremal cross sections at $B\approx 1.7-1.9$~T (see Extended Data Fig.~\ref{EDI:Exp:Fig6}\,f). For the window used for frequencies $f<350$~T it is $B_{average}=6.5$~T and $f(B)$ corresponds to the extremal cross sections at $B\approx 0.7$~T. Thus, even under large magnetic fields the experimental frequency values correspond to a field-polarized state in a low field.

\textbf{Quantum oscillatory torque and Lifshitz-Kosevich equation}:
Evaluation and interpretation of the quantum oscillatory torque magnetization was performed using the Lifshitz-Kosevich formalism \cite{Shoenberg1984}.The components of $\mathbf{M}$ $\parallel$ and $\perp$ to the field are given by:
\begin{equation}\label{eq:MoscPara}
	M_{osc,\parallel}=-\left( \frac{e}{ \hbar} \right)^{3/2} \frac{e\hbar f B^{1/2} V}{m^* 2^{1/2} \pi^{5/2} \sqrt{A^{\prime \prime}}} \sum_{p=1}^{\infty}\frac{R_TR_D}{p^{3/2}}\sin(2\pi p\left( \frac{f}{B}-\gamma \right) \pm \frac{\pi}{4}) \mbox{ ,}
\end{equation}
and
\begin{equation}\label{eq:MoscPerp}
M_{osc,\perp}=-\frac{1}{f}\frac{\partial f}{\partial \varphi} M_{osc,\parallel} \mbox{ .}
\end{equation} 
Here, $V$ is the sample volume, $A^{\prime \prime}$ is the curvature of the cross sectional area parallel to $\mathbf{B}$ and $f$ is the dHvA frequency observed (see comments above). The phase $\gamma=\frac{1}{2}$ corresponds to a parabolic band. In general, the phase includes also contributions due to Berry phases when the orbit encloses topologically non-trivial structures in $k$-space. The $\pm$ holds for maximal and minimal cross sections, respectively. The torque amplitude is given by $\tau_{osc}=M_{osc,\perp}B$. The torque thus vanishes in high-symmetry directions where $f(\varphi)$ is stationary. This feature of $\tau$ may be used to infer additional information about the symmetry properties of a dHvA branch. $R_T$ describes the temperature dependence of the oscillations
\begin{equation}
R_T=\frac{X}{\sinh(X)} \mbox{ with } X=\frac{2\pi^2p m^* k_BT}{e\hbar B} \mbox{ ,}
\label{eq:LK-fit}
\end{equation}
from which the effective mass $m^*$ including renormalization effects can be extracted, where $k_{\mathrm{B}}$ is the Boltzmann constant. Equation \eqref{eq:LK-fit} was fitted to the temperature dependence of the FFT peaks using the average fields $B_{average}$ defined above. No systematic changes in the mass values were observed within the standard deviation of the fits when different window sizes were chosen. See Supplementary Note~\ref{SI_sec_experimental_details} for details.
The Dingle factor
\begin{equation}
R_D=\exp(-\frac{\pi p m^*}{e B \tau})=\exp(-\frac{\pi p}{\omega_c\tau})
\end{equation}
describes the influence of a finite scattering time $\tau$. Here, $\omega_c$ is the cyclotron frequency.

\textbf{DFT calculations}:
The band structure and FS sheets of MnSi in the field-polarized phase  were calculated using DFT. The calculations included the effect of spin-orbit coupling. In all calculations, the magnetic part of the exchange-correlation terms was scaled \cite{Hoshino1993} to match the experimental magnetic moment of $~0.41\mu_B$ per Mn atom at low fields.  As input for the DFT calculations the experimental crystal structure of MnSi was used, i.e., space group P$2_13$ (198) with an experimental lattice constant $a=4.558$~\AA. Both Mn and Si occupy Wyckoff positions $4a$ with $u_{Mn}=0.137$ and $u_{Si}=0.845$ (Extended Data Fig.~\ref{EDI:Exp:Fig6}\,a). 

Calculations were carried out using WIEN2k \cite{WIEN2k}, ELK \cite{ELK}, and VASP~\cite{kresse_comp_mat_sci_96,kresse_PRB_96} using different versions of the local spin density approximation (LSDA). The results are consistent within the expected reproducibility of current DFT codes\cite{Lejaeghere2016}. The remaining uncertainties motivate a comprehensive experimental FS determination as reported in this study. In the main text, we focus on the results obtained with WIEN2k, using the LSDA  parametrization of Perdew and Wang\cite{Perdew1992} and a sampling of the full Brillouin zone (BZ) with a $23 \times 23 \times 23$ $\Gamma$-centered grid. The results of Extended Data Figs.~\ref{EDI:Theory:Fig1}, \ref{EDI:Theory:Fig2}, and~\ref{EDI:Theory:Fig4} were obtained using VASP with the PBE functional~\cite{PBE_functional} and  a BZ sampling with a $15 \times 15 \times 15$ k-mesh centered around $\Gamma$. 

Bands used for the determination of the Fermi surface were calculated with WIEN2k on a $50 \times 50 \times 50$ k-mesh. Due to the presence  of spin-orbit coupling, but the absence of both inversion and time-reversal symmetry, band structure data had to be calculated for different directions of the spin quantization axis. For a given experimental plane of rotation, calculations were performed in angular steps of $10^{\circ}$. The bands were then interpolated k-point-wise using third order splines to obtain band structure information in $1^{\circ}$ steps.

For the prediction of the dHvA branches from the DFT results, the Supercell k-space Extremal Area Finder (SKEAF) \cite{Rourke2012} was used on interpolated data corresponding to $150 \times 150 \times 150$ k-points in the full Brillouin zone. The theoretical torque amplitudes shown in Fig.~\ref{fig:3}b1 and~\ref{fig:3}b2 were calculated directly from the prefactors in Eqs.~\eqref{eq:MoscPara} and \eqref{eq:MoscPerp} convoluted with a suitable distribution function.

In order to compute the surface states of MnSi in the field-polarized phase (Extended Data Fig.~\ref{EDI:Theory:Fig4}), we first constructed a DFT-derived tight-binding model using the maximally localized Wannier function method as implemented in Wannier90~\cite{wannier90}. Using this tight-binding model, we computed the momentum-resolved surface density of states by means of an iterative Green's function method, using WannierTools~\cite{wannierTools}. The symmetry eigenvalues of the DFT bands were computed from expectation values  using VASP pseudo wavefunctions, as described in Ref.~\onlinecite{gao2020irvsp}.

\textbf{Magnetic breakdown}:
The probabilities for magnetic breakdown at a junction $i$ is given by $p_i=e^{-\frac{B_0}{B}}$. The probability for no breakdown to occur is thus $q_i=1-p_i$. The breakdown fields $B_0$ were calculated from Chamber's formula
\begin{equation}
B_0=\frac{\pi \hbar}{2e} \sqrt{\frac{k_g^3}{a+b}} \mbox{ ,}
\end{equation}\label{eq:chambers}
where $k_g$ is the gap in k-space and $a$ and $b$ are the curvatures of the trajectories at the breakdown junction\cite{Shoenberg1984}. In our study of MnSi we observed magnetic breakdown in particular between sheets 3 and 4, which exhibit up to eight junctions depending on the magnetic field direction and between FS sheet pairs touching the BZ surfaces on which the nodal-plane degeneracy is lifted. Only breakdown orbits that are closed after one cycle are considered in the analysis. Further details may be found in the Supplementary Note \ref{SI_sec_fs_determination}.

\textbf{Assignment of dHvA orbits and rigid band shifts}:
The assignment of the experimental dHvA branches to the corresponding extremal FS cross sections was based on the following criteria
\begin{compactitem}
\item dHvA frequency -- determining sheet size in terms of the cross-sectional area
\item angular dispersion -- relating to sheet shape, topology and symmetry
\item torque signal strength -- relating to sheet shape and symmetry
\item direction of f(B) shift -- relating to spin orientation and charge carrier type
\item effective mass -- relating to the temperature dependence
\item magnetic breakdown behavior -- relating to proximity of neighboring sheets
\end{compactitem}
The majority of the observed dHvA branches could be related directly to the FS as calculated. In addition we used the well-established procedure of small rigid band shifts to optimize the matching. While this procedure is, in general, neither charge nor spin conserving, it results in a very clear picture of the experimental FS. One has to bear in mind, however, that the deviations between the true FS and the calculated FS are not due to a rigid band shift (this might be justified, e.g., in case of unintentional doping, which we rule out here). Rather, it may be attributed to differences in the band dispersions that originate in limitations of our DFT calculations (e.g., neglecting electronic correlations and the coupling to the spin fluctuation spectrum).

The dHvA orbits, the assignments to a specific extremal cross section, the observed and predicted frequencies, the observed and predicted masses and mass enhancements are listed in Extended Data Table~\ref{EDI:Exp:tab2}.
Extended Data Table~\ref{EDI:Exp:tab1} summarizes the resulting characteristic properties of the FS sheets including their contribution to the density of states at the Fermi level.

\textbf{Symmetry analysis}:
The symmetry-enforced band crossings and the band topology follow from the non-trivial winding of the symmetry eigenvalues through the BZ. This winding of the eigenvalues is derived in the Supplementary Note~\ref{SI_sec_band_topology}, both for the paramagnetic and ferromagnetic phases of MnSi. Supplementary Note~\ref{SI_sec_band_topology} also contains the derivation of the topological charges of the nodal planes, Weyl points, and fourfold points, which are obtained from generalizations of the Nielsen-Ninomiya theorem~\cite{nielsen_no_go}. To illustrate the band topology for ferromagnets in SG 19.27 and SG 4.9 two tight-binding models are derived in the Supplementary Note~\ref{sec_tight_binding_model}, which includes also a discussion of the Berry curvature and the surface states. The classification of nodal planes in magnetic materials is given in Supplementary Note~\ref{classification_nodal_surfaces}. It is found that among the 1651 magnetic SGs, 254 exhibit symmetry-enforced nodal planes. We find that (at least) 33 out of these have nodal planes whose topological charge is guaranteed to be nonzero due to symmetry alone.


\newpage
\section*{References}

\newpage

\section*{Extended Data}

\setcounter{figure}{0}
\captionsetup[figure]{labelfont={bf},name={Extended Data Fig.},labelsep=space}

\clearpage \thispagestyle{empty}

\begin{figure}
\centerline{\includegraphics[width=0.9\textwidth,clip=]{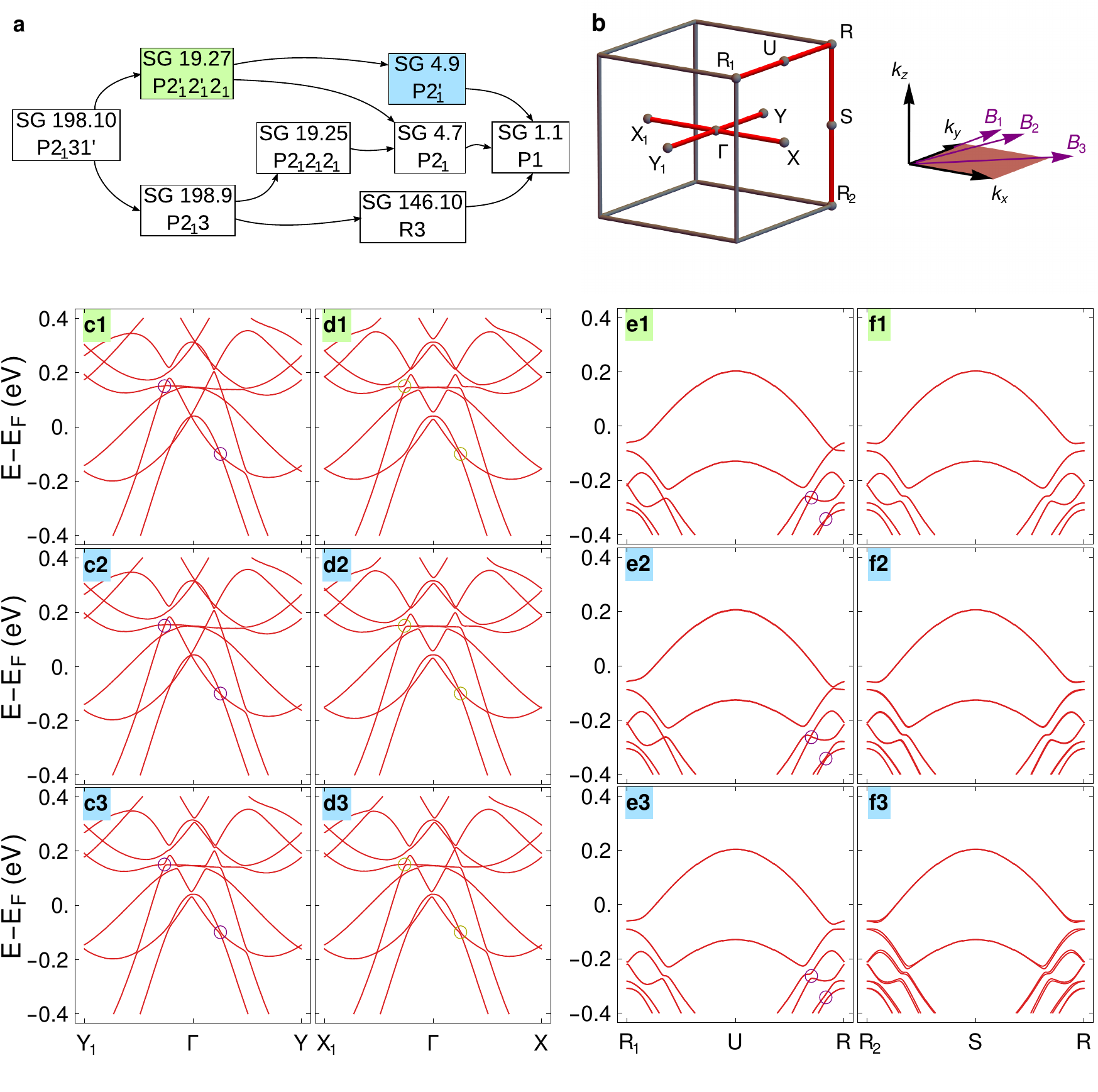}}
\linespread{1.0}\selectfont{}
\caption{\raggedright
{\bf $|$Magnetic space groups and electronic band structures for different directions of the magnetization.}
{\bf a}, Magnetic subgroups of space group 198 (P2$_1$3) and their group-subgroup relations. The magnetic space groups describing the symmetries for magnetizations along $[010]$ within the $xy$-plane are highlighted in green and blue, respectively.
{\bf b}, Orthorhombic BZ for ferromagnetic MnSi (left) and magnetization directions used for the ab-initio calculations in panels {\bf c}-{\bf f} (right). 
{\bf c}-{\bf f}, Ab-initio electronic band structure of ferromagnetic MnSi along the four high-symmetry paths indicated in panel {\bf b}. In the first, second, and third rows the magnetization is oriented along $[010]$, 10$^\circ$  rotated into the $xy$-plane, and along $[110]$, respectively. Some of the Weyl points (WPs) and fourfold degenerate points (FPs) at (or near) the high-symmetry lines are highlighted by violet and brown circles.
}
\label{EDI:Theory:Fig1}
\end{figure}

\clearpage \thispagestyle{empty}

\begin{figure}
\centerline{\includegraphics[width=0.9\textwidth,clip=]{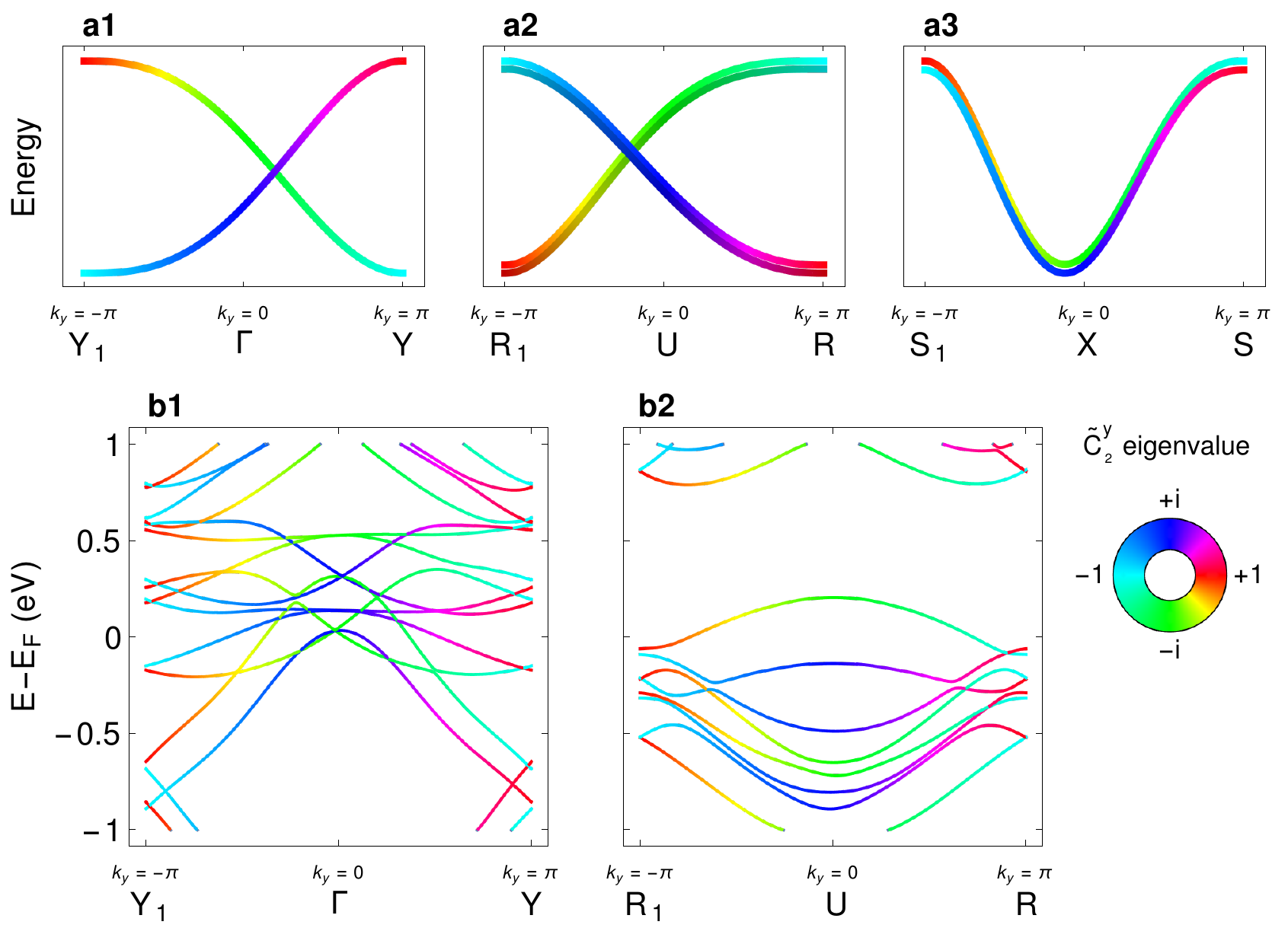}}
\linespread{1.0}\selectfont{}
\caption{\raggedright
{\bf $|$Momentum dependence of the screw rotation eigenvalues.} {\bf a1}-{\bf a3}, Schematic band connectivity diagrams for a minimal set of bands along the Y$_1-$$\Gamma$$-$Y line, the  R$_1-$U$-$R line, and the S$_1-$X$-$S line, respectively. The eigenvalues of the screw rotation symmetry $\tilde{C}^y_2$ are indicated by color. 
{\bf b1}-{\bf b2} Ab-intio electronic band structure of MnSi in the [010] FM phase with the $\tilde{C}^y_2$ eigenvalues indicated by color. The crossings of bands with different color on the  Y$_1-$$\Gamma$$-$Y line in {\bf b1} and on the R$_1-$U$-$R line in {\bf b2} are Weyl points (WPs) and fourfold degenerate points (FPs), respectively.
}
\label{EDI:Theory:Fig2}
\end{figure}

\clearpage \thispagestyle{empty}

\begin{figure}
\centerline{\includegraphics[width=1\textwidth,clip=]{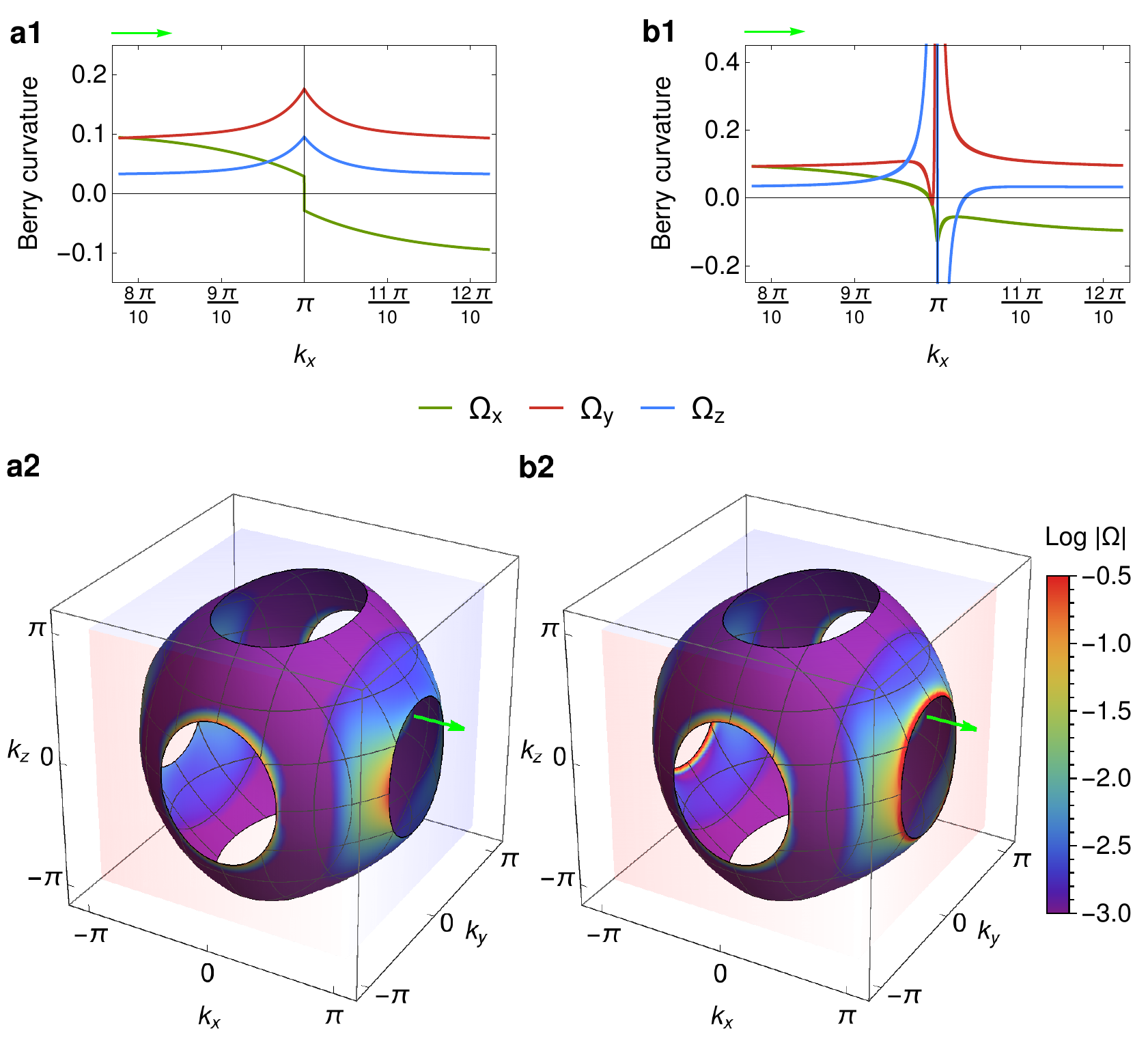}}
\linespread{1.0}\selectfont{}
\caption{\raggedright
{\bf $|$~Berry curvature on the Fermi surface.}  
{\bf a1}-{\bf a2}, Berry curvature ${\Omega}_{\mu} ({\bf k})$ on one of the Fermi surfaces of a tight-binding model in SG 19.27, corresponding to ferromagnetic MnSi with the magnetization pointing along [010]. 
{\bf a1} shows the three components of ${\Omega}_{\mu}$ as a function of $k_x$, along the direction indicated by the green arrow in {\bf a2}.  The absolute value of the Berry curvature $| {\bf \Omega} ({\bf k}) |$ is indicated in {\bf a2} by a logarithmic color code.
{\bf b1}-{\bf b2}, Same as {\bf a} but for a tight-binding model in SG 4.9, corresponding to ferromagnetic MnSi with the magnetization rotated into the $xy$-plane. 
}
\label{EDI:Theory:Fig3}
\end{figure}

\clearpage \thispagestyle{empty}

\begin{figure}
\centering
\centerline{\includegraphics[width=1.0\textwidth,clip=]{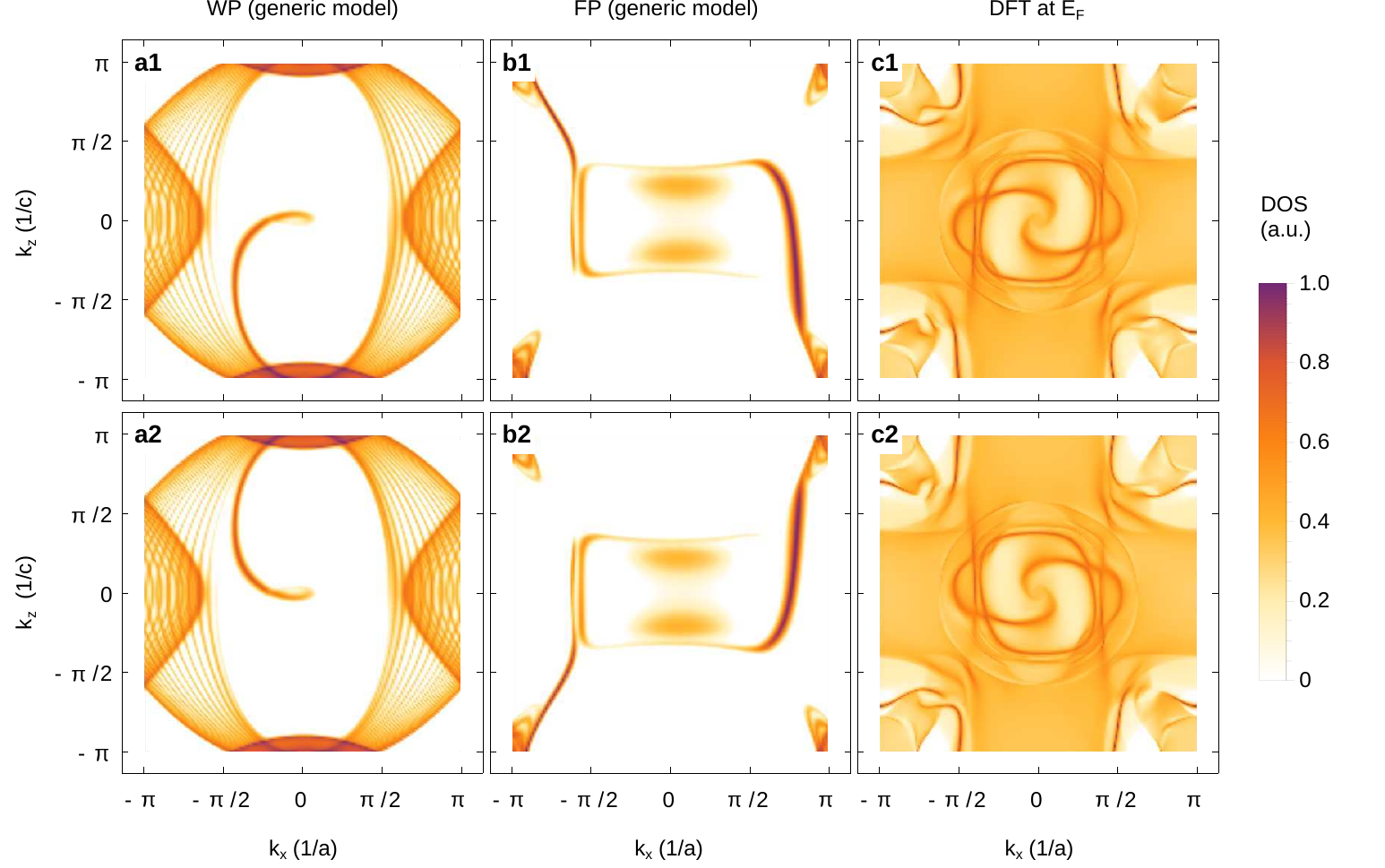}}
\linespread{1.0}\selectfont{}
\caption{\raggedright
\textbf{$|$~Topological surface states.} 
Density of states (DOS) at the (010) surface of ({\bf a}, {\bf b}) the tight-binding model with SG 19.27 and (){\bf c}) ferromagnetic MnSi with the magnetization aligned along $[010]$. The first and second rows display the DOS at the top and bottom surfaces, respectively.  
In panel {\bf a} the surface DOS is shown at an energy $E=-1.2$ of the single Weyl point on the  Y$_1-$$\Gamma $$-$Y line. A single Fermi arc emanates from the projected Weyl point and connects to the $k_z = \pi$ nodal plane. In panel {\bf b} the surface DOS is shown at the energy $E=+1.4$ of the fourfold point on the R$_1-$U$-$R line, whose chirality $\nu = -2$ is compensated by two accidental Weyl points in the bulk. Two Fermi arcs emanate from the projected fourfold point and connect to the accidental Weyl points in the bulk. 
In panel {\bf c} the DFT-derived surface DOS of ferromagnetic MnSi is shown at the Fermi level $E=E_{\textrm{F}}$. Fermi arcs emanate from the projected Weyl points on the Y$_1-$$\Gamma $$-$Y line and connect with the bulk bands forming nodal planes on the BZ boundaries.
} 
\label{EDI:Theory:Fig4}
\end{figure} 

\clearpage \thispagestyle{empty}

\begin{figure}
\centering
\centerline{\includegraphics[width=\textwidth,clip=]{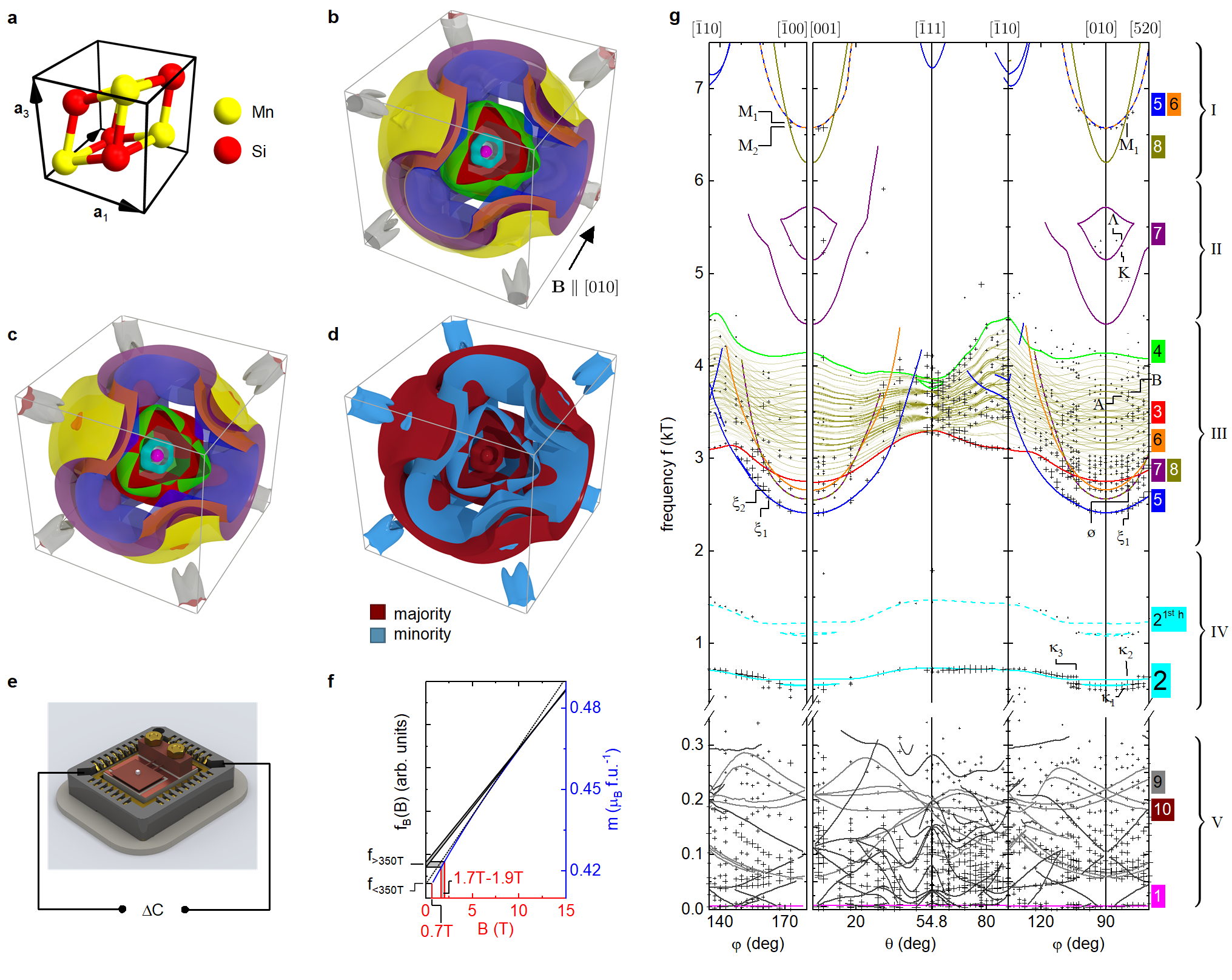}}
\linespread{1.0}\selectfont{}
\caption{\raggedright
{\bf $|$ Crystal structure, calculated Fermi surfaces, experimental methods, and dHvA spectra in the $\mathbf{(001)}$ and $\mathbf{(\bar 1 \bar 1 0)}$ planes.} {\bf a}, Crystal structure of MnSi. {\bf b}, Fermi surface as calculated within LDSA without rigid band shifts. {\bf c}, Calculated Fermi surface neglecting spin-orbit coupling. {\bf d}, Calculated Fermi surface neglecting spin-orbit coupling and highlighting majority and minority spin. {\bf e}, Sketch of the cantilever magnetometer chip with capacitive readout. {\bf f}, Magnetic field-dependence of the frequency $f_B(B)$ tracking the magnetic-field dependence of the unsaturated magnetization in the field-polarized phase. The frequency $f(B)$ observed corresponds to the zero-field intercept of the tangent to $f_B(B)$.  {\bf g}, Experimental dHvA frequency branches (crosses) for rotation in the $(001)$ and $(\bar 1 \bar1 0)$ planes together with the theory (lines) matched to the experiment. See Supplementary Note \ref{SI_sec_fs_determination} for details.
}
\label{EDI:Exp:Fig6}
\end{figure} 

\clearpage \thispagestyle{empty}

\begin{figure}
\centering
\centerline{\includegraphics[width=0.8\textwidth,clip=]{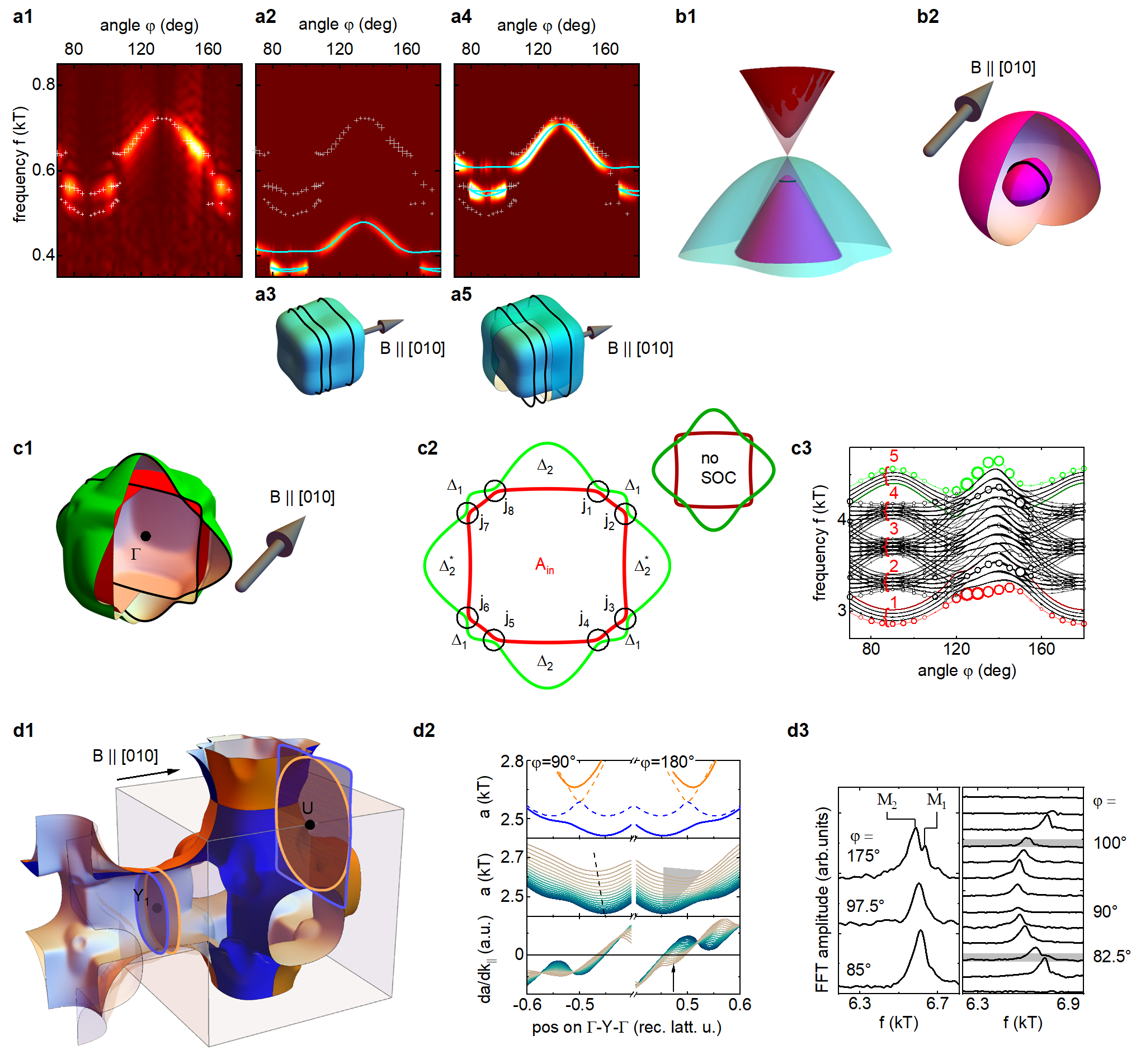}}
\linespread{0.78}\selectfont{}
\caption{\raggedright
\textbf{$|$Details of the assignment of experimental dHvA orbits to FS sheets 1 to 6.}
\textbf{a1} Experimental signature of sheet 2. Color scale corresponds to experimental FFT amplitude and crosses show positions of maxima.
\textbf{a2} Torque signal predicted from DFT as calculated. Lines show theoretical branches, crosses show experimental positions.
\textbf{a3} FS sheet 2 as calculated. Three extremal orbits $2\Gamma$, $2\Gamma Y(1,2)$ are present for $\mathbf{B}$ close to $[010]$ which are assigned to $\kappa_{1,2,3}$.
\textbf{a4} Calculated dHvA branch including a small upward band shift of 20~meV, yielding a good match with experiment (crosses).
\textbf{a5} Comparison of as-calculated (inner) and matched (outer) FS sheet 2.
\textbf{b1} Dispersions of bands 1, 2 and 3 in the $k_x$-$k_z$-plane without (transparent) and with SOC (solid), showing a spin-1 excitation-like 3-fold degeneracy that is lifted by SOC. Since band 2 (cyan) crosses the Fermi level, the $\alpha$ branch must originate from band 1 and not band 3. The Fermi level matching the experimental frequency is shown in black.
\textbf{b2} FS sheet 1 as calculated (outer) and matched to experiment (inner). The $\alpha$ branch is assigned to orbit $1\Gamma$.
\textbf{c1} FS sheets 3 and 4 exhibit extremal orbits with 8 breakdown junctions $j_1$ to $j_8$ for $\mathbf{B}$ close to $[010]$ as shown in \textbf{c2}. The inset shows the two extremal orbits that arise when SOC is neglected. 
\textbf{c3} 256 breakdown orbit branches originating from sheets 3 and 4. Symbol size reflects orbit probability. The torque amplitude is not considered in this graph. The breakdown orbits group into five sets labelled in red. The branches $\rho$ and $H$ are assigned to the inner and outer orbits $3\Gamma$ and $4\Gamma$, respectively.
\textbf{d1} FS sheets 5 and 6 as in Fig.~\ref{fig:4}\,a with two neck orbits $5\Gamma Y$, $6\Gamma Y$ and the loop orbits $5U6U$ assigned to ($\xi_1$,$\xi_2$), $\pi$ and $M_1$, respectively.
\textbf{d2} Upper panel: neck cross-sectional areas of sheet 5 (blue) and 6 (orange) vs $\mathbf{k}_{\parallel}$ neglecting (dashed) and including (solid) SOC for $\varphi=90^{\circ}$ and $\varphi=180^{\circ}$. Middle panel: Cross-sectional area $a$ of sheet 5 vs $\mathbf{k}_{\parallel}$ for field directions $70^{\circ}-90^{\circ}$ and $160^{\circ}-180^{\circ}$. Dashed line: position of single extremal area around $\varphi=90^{\circ}$. Shaded grey area: neck being on the verge of developing a second minimum close to $180^{\circ}$ but not around $90^{\circ}$ that could give rise to $\xi_2$. Lower panel: derivative $\frac{da}{dk_{\parallel}}$, where zero-crossings correspond to extremal orbits.
\textbf{d3} FFT amplitude of the loop orbits around $\varphi=90^{\circ}$ and $\varphi=180^{\circ}$. Left panel: a distinct splitting of the $M_1$-branch into  $M_1$ and  $M_2$ is observed close to $\varphi=180^{\circ}$ but not around $\varphi=90^{\circ}$. Right panel: The FFT amplitude of the $M_1$-branch shows unexpected secondary minima (shaded areas) close to $\varphi=90^{\circ}$ on both sides. Both effects may be connected to either the quasi-degeneracy of the $U5U6$ orbits shown in Fig.~\ref{fig:4}\,b3,b4 or to a crossing with the $8\Gamma Y$ branch.
}
\label{EDI:Exp:Fig9}
\end{figure} 

\clearpage \thispagestyle{empty}

\begin{figure}[!]
\centering
\centerline{\includegraphics[width=1.0\textwidth,clip=]{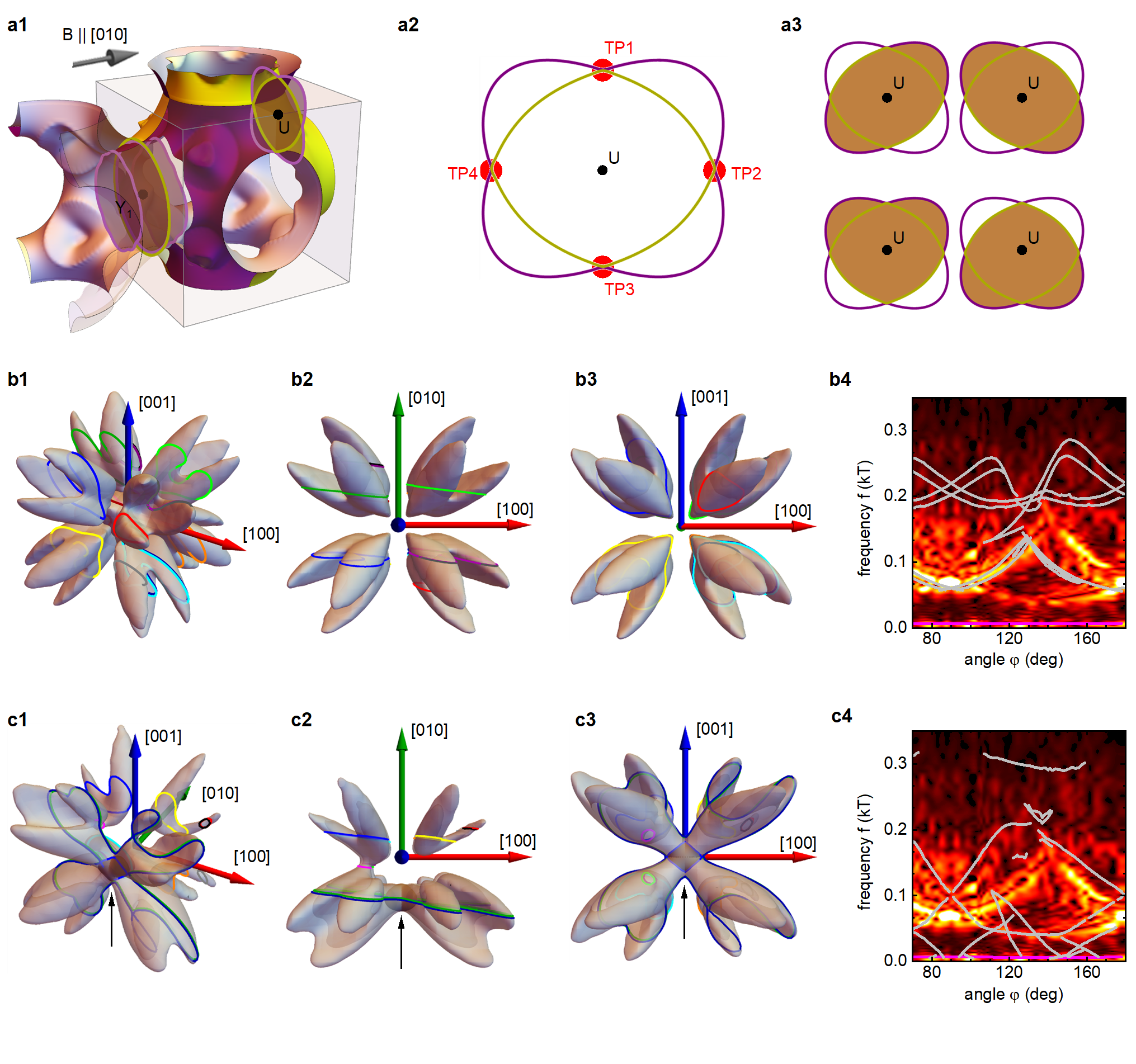}}
\linespread{1.0}\selectfont{}
\caption{\raggedright
\textbf{$|$Assignment of experimental dHvA orbits to FS sheets 7 to 10.} \textbf{a1}, Along $\Gamma$-Y$_1$-$\Gamma$ three neck orbits on sheet 7 (purple), one neck orbit on sheet 8 (yellow), and two loop orbits around U are predicted for $\mathbf{B} \parallel \left[ 010 \right]$. \textbf{a2}, Loop orbits are shared between the sheets at TP1 to TP4 in analogy to the sheet pair (5,6). \textbf{a3}, For $\mathbf{B} \parallel [010]$, the upper two lentil-shaped orbits exist, while for $\mathbf{B}$ in the $(001)$-plane away from $[010]$ the lower two heart-shaped orbits are allowed in addition. \textbf{b1}, Sheet 9 neglecting SOC in perspective, top (\textbf{b2}) and back (\textbf{b3}) view for $\varphi=83^{\circ}$, i.e., $\mathbf{B}$ sligthly off the $\left[ 010 \right]$-direction. Bands 9 and 10 are shifted upward by $10$~meV for an optimal match to the low frequencies as shown in \textbf{b4}, where gray lines correspond to the calculations. In total $15$ orbits are predicted for this specific field direction alone. Without SOC, band 10 does not cross the Fermi level for this shift.		
\textbf{c1}-{\bf c3} Sheets 9 and 10 and predicted dHvA orbits including SOC for $\varphi=83^{\circ}$ and a shift of $11$~meV yielding a good match as shown in \textbf{c4}. Sheet 10 resides inside sheet 9 as highlighted by black arrows. It occurs only for field directions where two or more "banana bunches" cross the BZ surface and connect. In the situation depicted here, $(001)$ is a nodal plane, thereby connecting parts of sheet 9 and 10 in such a way that extremal orbits cross from one sheet to the other.
}
\label{EDI:Exp:Fig10}
\end{figure} 

\clearpage \thispagestyle{empty}

\renewcommand{\thetable}{\arabic{table}}
\captionsetup[table]{labelfont={bf},name={Extended Data Table},labelsep=space}

\begin{table}[htbp]
\renewcommand*{\arraystretch}{1.0}
\linespread{1.0}\selectfont{}
\caption{\raggedright
\textbf{$|$ Key properties of the Fermi surface sheets}. Information as calculated and matched to experiment. Upper part: sheet number, topology and location, carrier type and spin refer to the dominant properties of carriers on the corresponding dHvA orbits. There are sheets with, both, strongly mixed electron/hole and mixed spin character. The column labelled $f(B)$-shift states the direction of the expected frequency shift with increasing magnetic field. Lower part: $D(E_{\text{F}}$) states the density of states as calculated. $E_{\text{F}}$-shift states the shift of the Fermi level used to achieve an optimal match to experiment. $D(E_{\text{F}}+E_{\text{F}}$-shift) states the density of states following the shift of $E_{\text{F}}$. $\gamma$ corresponds to the contribution of the shifted band to the Sommerfeld coefficient. $m^*/m_{\text{b}}$ is the mass enhancement factor, where $m^*$ is determined from the Lifshitz-Kosevich-behavior of the dHvA amplitude and $m_{\text{b}}$ is the bare band mass obtained in DFT. $\gamma^*$ is the Sommerfeld coefficient scaled with the mass enhancement. The Sommerfeld coefficient of $28.15$~mJ/(mol K$^2$) inferred from the dHvA data matches the value of the experimentally determined specific heat at $B=12$~T within a few percent, confirming that all thermodynamically significant parts of the Fermi surface were observed.}
\label{EDI:Exp:tab1}
\vspace{-5mm}
\begin{flushleft}
\begin{tabular}{|c|c|c|c|c|c|}
\hline
 sheet no. & topology & location & carrier type & spin character  & $f(B)$-shift \\  
\hline
\hline
 & & & e/h & &  \\
\hline
1 & pocket & $\Gamma$-centered & h & majority & $\searrow$ \\
2 & pocket & $\Gamma$-centered & h & majority & $\searrow$ \\
3 & pocket & $\Gamma$-centered & h & mixed & - \\
4 & pocket & $\Gamma$-centered & h & mixed & - \\
5 & jungle-gym & - & necks: h & minority & necks: $\nearrow$ \\
& & & loops: e & minority & loops: $\searrow$ \\
6 & jungle-gym & - & necks: h & minority & necks: $\nearrow$ \\
& & & loops: e & minority & loops: $\searrow$ \\
7 & jungle-gym & - & necks: h & majority & necks: $\searrow$ \\
& & & loops: e & majority & loops: $\nearrow$ \\
8 & jungle-gym & - & necks: h & majority & necks: $\searrow$ \\
& & & loops: e & majority & loops: $\nearrow$ \\
9 & pockets & $\Gamma$-$R$ & e & minority & $\nearrow$ \\
10 & pocket/none & $R$ & h & minority & $\searrow$ \\                              
\hline
\end{tabular}
\end{flushleft}
\begin{flushleft}
\begin{tabular}{|c|c|c|c|c|c|c|}
\hline
 sheet no. & $D(E_{\text{F}}$) & $E_{\text{F}}$-shift & $D(E_{\text{F}}+E_{\text{F}}$-shift) & $\gamma$ & $m^*/m_{\text{b}}$ & $\gamma^*$ \\  
\hline
\hline
 & (states/(eV u.c.)) & (mJ/(mol K$^2$)) & & (mJ/(mol K$^2$)) & & \\
\hline
1 & $0.011$ & $27$ & $0.001$ & $10^{-4}$ & $\sim 5$ & $0.004$ \\
2 & $0.084$ & $-20$ & $0.11$ & $0.07$ & $5.9$ & $0.41$ \\
3 & $0.36$ & $8.5$ & $0.36$ & $0.21$ & $7.3$ & $1.53$ \\
4 & $0.62$ & $9.5$ & $0.62$ & $0.36$ & $5.0$ & $1.80$ \\
5 & $1.66$ & $4$ & $1.66$ & $0.98$ & $5.5$ & $5.39$ \\
6 & $1.74$ & $4$ & $1.74$ & $1.03$ & $5.6$ & $5.77$ \\
7 & $2.44$ & $-4$ & $2.47$ & $1.46$ & $5$   & $7.3$ \\
8 & $1.74$ & $-4$ & $1.77$ & $1.04$ & $\sim 5$ & $5.2$ \\
9 & $1.09$ & $-11$ & $0.84$  & $0.5$  & $\sim 1.5$ & $0.75$ \\
10 & $0.36$ & $-11$ & $0-0.08$ & $0-0.05$ & - & - \\                              
\hline
\bf{sum} & $\mathbf{10.1}$ & & $\mathbf{9.65}$ & $\mathbf{5.65}$ & $\mathbf{5.1}$ & $\mathbf{28.15}$ \\
\hline
\multicolumn{6}{|r|}{ \bf{specific heat experiment\cite{Bauer2010}:}} & $\mathbf{28}$ \\
\hline
\end{tabular}
\end{flushleft}

\end{table}

\clearpage \thispagestyle{empty}

\begin{table}
	\renewcommand*{\arraystretch}{0.8}
	\linespread{1.0}\selectfont{}
	\centering
	\caption{\raggedright
	\textbf{$|$Assignment of observed dHvA branches to orbits on the calculated FS.} Arrows denote the direction of the frequency shift with increasing magnitude of $B$. Error bars of effective masses reflect the standard deviation of the Lifshitz-Kosevich fits. Peaks at frequencies marked with a prime were only observed in magnetic field sweeps up to $16$~T. Frequency values in brackets are given at a second angle $\varphi$ in the (001)-plane measured from the [100]-direction. }
	\vspace{5mm}
	\begin{tabular}{lllrrccl}
	\hline 
	\hline 
		Branch \hspace{1mm} & Orbit \hspace{3mm} & $f_{\rm exp.} [\rm{kT}]$  & $f_{\rm pred.} [\rm{k}T] $	\hspace{2mm} & $m^*\, [m_e]$ \hspace{1mm} & $m_b\, [m_e]$ \hspace{3mm} & $\frac{m^*}{m_b}$\hspace{2mm}& $\varphi$~$[\rm{deg}]$	\\
	\hline 
	\hline
	
		$\alpha$	& $1\Gamma$		& 0.007$\searrow$	& 0.068$\searrow$		& 0.4$\pm$0.1		& 0.1				& 4			& 82.5		\\
		$\beta$		& $9\Gamma$R(1)	& 0.054				& $-$ $\searrow$		& 0.8$\pm$0.1		& 0.6				& 1.4		& 82.5		\\
		$\gamma$	& $9\Gamma$R(2)	& 0.070				& $-$ $\searrow$		& 2.7$\pm$0.3		& 1.9				& 1.4		& 82.5		\\
		$\delta$	& $9\Gamma$R(3)	& 0.082				& $-$ $\searrow$		& 2.3$\pm$0.6		& 1.9				& 1.2		& 82.5		\\
		$\epsilon$	& 9$\Gamma$R(4)	& 0.095				& $-$ $\searrow$		& 2.0$\pm$0.5		& $-$				& $-$		& 82.5		\\
		$\zeta$		& 9$\Gamma$R(5)	& 0.110				& $-$ $\searrow$		& 2.4$\pm$0.4		& $-$				& $-$		& 82.5		\\
		$\eta$		& 9$\Gamma$R(6)	& 0.141				& $-$ $\searrow$		& 2.5$\pm$0.5		& $-$				& $-$		& 82.5		\\
		$\mu$		& 9$\Gamma$R(7)	& 0.130				& $-$ $\searrow$		& $-$				& 1.9				& $-$		& 152.5		\\
		$\theta$	& 9$\Gamma$R10R(1)	& 0.225			& $-$ $\searrow$		& 3.5$\pm$0.6 		& $-$				& $-$		& 82.5		\\
		$\iota$		& 9$\Gamma$R10R(2)	& 0.248			& $-$ $\searrow$		& 5.4$\pm$0.6		& $-$				& $-$		& 82.5		\\
		$\tilde{\iota}$	& 9$\Gamma$R10R(3))& 0.290		& $-$$\searrow$			& 5.4$\pm$0.6		& $\sim2.0$			& $\sim2.7$	& 82.5		\\
		$\kappa_1$	& 2$\Gamma$		& 0.488$\searrow$ (0.523)	& 0.369$\searrow$	& 6.3$\pm$0.6	& 1.1				& 6.0		& 82.5 (106)\\
		$\kappa_2$	& 2$\Gamma$Y(1)	& 0.566$\searrow$ (0.564)	& 0.371$\searrow$	& 6.2$\pm$0.1	& 1.1				& 5.9		& 82.5 (106)\\
		$\kappa_3$	& 2$\Gamma$Y(2)	& 0.641$\searrow$ 	& 0.411$\searrow$		& 6.5$\pm$0.5		& 1.2				& 5.6		& 106		\\
		2$\kappa_1$	& 2$\kappa_1$	& 1.065				& 2f$_{\kappa_1}$		& 14.2$\pm$0.8		& 2m$_{\kappa_1}$	& $-$		& 82.5		\\
		2$\kappa_2$	& 2$\kappa_2$	& 1.120				& 2f$_{\kappa_2}$		& 14.0$\pm$0.6		& 2m$_{\kappa_2}$	& $-$		& 82.5		\\
		3$\kappa_2$	& 3$\kappa_2$	& 1.610				& 3f$_{\kappa_2}$		& 16$\pm$6			& 3m$_{\kappa_2}$	& $-$		& 82.5		\\
		$\xi_1$		& 5$\Gamma$Y(1)	& 2.459$ \nearrow$ (2.576)	& 2.532$\nearrow$	& 10.3$\pm$0.1	& 2.0				& 5.4		& 82.5 (165)\\
		$\xi_2$		& 5$\Gamma$Y(2)	& 2.653				& $-$					& $-$				& $-$				& $-$		& 165		\\
		$\o^{\prime}$	& 7U8U		& 2.658				& 2.765$\nearrow$		& 10.0$\pm$0.3		& 2.0				& 5.0		& 82.5		\\
		$\pi$		& 6$\Gamma$Y	& 2.701$\nearrow$	& 2.822$\nearrow$		& 11.1$\pm$0.3		& 2.0				& 5.6		& 82.5		\\
		$\rho$		& 3$\Gamma$		& 2.786$\searrow$	& 2.891$\rightarrow$	& 10.9$\pm$0.4		& 1.5				& 7.1		& 82.5		\\
		$\rho^{\prime}$	& 3$\Gamma$4$\Gamma$(1)& 2.833	& 2.934$\rightarrow$	& $-$				& 1.5				& $-$		& 82.5		\\
		$\sigma$	& 3$\Gamma$4$\Gamma$(2)& 2.879		& 2.976$\rightarrow$	& 11.2$\pm$0.4		& 1.5				& 7.5		& 82.5		\\
		$\sigma^{\prime}$	& 3$\Gamma$4$\Gamma$(3)& 2.918	& 3.021$\rightarrow$	& $-$			& $-$				& $-$		& 82.5		\\
		$\tau$		& 3$\Gamma$4$\Gamma$(4)& 2.966		& 3.019$\rightarrow$	& 10.2$\pm$0.4		& 1.5				& 6.8		& 82.5		\\
		$\upsilon$	& 3$\Gamma$4$\Gamma$(5)& 3.034$\searrow$	& 3.061$\searrow$	& 8.7$\pm$0.3	& 1.5				& 5.9		& 82.5		\\
		$\upsilon^{\prime}$	& 3$\Gamma$4$\Gamma$(6)& 3.105	& 3.231$\rightarrow$	& $-$			& $-$				& $-$		& 82.5		\\
		$\varphi$	& 3$\Gamma$4$\Gamma$(7)& 3.229	& 3.453$\rightarrow$		& 13.2$\pm$0.4		& $-$				& $-$		& 82.5		\\
		$\chi^{\prime}$	& 3$\Gamma$4$\Gamma$(8)& 3.350	& 3.583$\rightarrow$	& $-$				& $-$				& $-$		& 82.5		\\
		$\psi$		& 3$\Gamma$4$\Gamma$(9)& 3.450	& 3.715$\rightarrow$		& 11.6$\pm$0.6		& $-$				& $-$		& 82.5		\\
		$\omega$	& 3$\Gamma$4$\Gamma$(10)& 3.485$\nearrow$	& 3.626$\rightarrow$	& 11.6$\pm$0.6	& $-$			& $-$		& 82.5		\\
		$A$			& 3$\Gamma$4$\Gamma$(11)& 3.671	& 3.931$\rightarrow$		& 13.6$\pm$0.1		& $-$				& $-$		& 82.5		\\
		$B$			& 3$\Gamma$4$\Gamma$(12)& 3.717$\searrow$& 4.017$\rightarrow$	& 13.7$\pm$0.3	& $-$				& $-$		& 82.5		\\
		$\Gamma^{\prime}$	& 3$\Gamma$4$\Gamma$(13)& 3.840	& 4.323$\nearrow$	& $-$				& 3.2				& $-$		& 82.5		\\
		$\Delta$	& 3$\Gamma$4$\Gamma$(14)& 3.940	& 4.366$\rightarrow$		& 14$\pm$1			& 3.2				& 4.4		& 82.5		\\
		$E$			& 3$\Gamma$4$\Gamma$(15)& 4.040	& 4.409$\rightarrow$		& 17$\pm$4			& 3.2				& 5.5		& 82.5		\\
		$Z$			& 3$\Gamma$4$\Gamma$(16)& 4.120	& 4.451$\rightarrow$		& 15$\pm$3			& 3.2				& 4.7		& 82.5		\\
		$H$			& 4$\Gamma$		& 4.180				& 4.493$\rightarrow$	& 16$\pm$5			& 3.1				& 5.1		& 82.5		\\
		$\Theta$	& 7$\Gamma$Y(1)	& 4.350				& 4.569					& 15$\pm$5			& 4.0				& 3.8		& 82.5		\\
		2$\xi_1$	& 2$\xi_1$		& 4.920				& 2f$_{\xi_1}$ $\searrow$	& 24$\pm$5		& 2m$_{\xi_1}$		& $-$		& 82.5		\\
		$K$			& 7$\Gamma$Y(2)	& 5.304				& 5.179$\searrow$		& $-$				& 4.2				& $-$		& 85		\\
		$\Lambda$	& 7Y			& 5.304				& 5.481$\searrow$		& $-$				& 3.4				& $-$		& 85		\\
		$M_1$ 		& 5U6U			& 6.715$\searrow$ (6.634)	& 6.627$\searrow$	& 15.1$\pm$0.2	& $\sim$2.8			& 5.4		& 82.5 (175)\\
		$M_2$		& 5U6U			& 6.587				& $-$					& $-$				&   $-$				& $-$		& 175		\\
		$-$			& 8$\Gamma$Y	& $-$				& 6.610$\searrow$		& $-$				& 	4.0				& $-$		& $-$		\\	
	\hline
	\hline
	\end{tabular}
	\label{EDI:Exp:tab2}
\end{table} 

\clearpage \thispagestyle{empty}

\begin{table}
	\centering
	\linespread{1.0}\selectfont{}
	\caption{\raggedright
	\textbf{$|$ Catalogue of space groups with symmetry-enforced nodal planes.} 
		Table listing all magnetic SGs with symmetry-enforced nodal planes. The list is grouped into three blocks: 32 SGs with time-reversal symmetry (describing nonmagnetic materials), 
		94 SGs without time-reversal symmetry (describing ferro- or ferri-magnets); and 129 SGs with a symmetry that combines time-reversal symmetry with a translation (describing
		antiferromagnets). In order for the nodal planes to have nonzero topological charge, the SG must be chiral (labelled by ``[t]" or ``[\textbf{T}]").      
		The 33 SGs  labelled by ``[\textbf{T}]" have nodal planes whose topological charge is enforced to be nonzero by symmetry, as discussed in 
		Supplementary Note~\ref{classification_nodal_surfaces}.
	}
	\label{EDI:Theory:tab1}
	\vspace{5mm}
	\renewcommand{\arraystretch}{1.1} 
	\begin{tabular}{lllllllll}
		\hline 
		\hline
 4.8 [t]  & 17.8 [t]  & 18.17 [t]  & 19.26 [\textbf{T}]  & 20.32 [t]  & 26.67  & 29.100  & 31.124  & 33.145  \\
 36.173  & 76.8 [t]  & 78.20 [t]  & 90.96 [t]  & 91.104 [t]  & 92.112 [\textbf{T}]  & 94.128 [\textbf{T}]  & 95.136 [t]  & 96.144 [\textbf{T}]  \\
 113.268  & 114.276  & 169.114 [t]  & 170.118 [t]  & 173.130 [t]  & 178.156 [t]  & 179.162 [t]  & 182.180 [t]  & 185.198 
   \\
 186.204  & 198.10 [\textbf{T}]  & 212.60 [\textbf{T}]  & 213.64 [\textbf{T}] & & & & & \\
 \hline
 4.9 [t]  & 11.54  & 14.79  & 17.10 [t]  & 18.18 [t]  & 18.19 [t]  & 19.27 [\textbf{T}]  & 20.34 [t]  & 26.68  \\
 26.69  & 29.101  & 29.102  & 31.125  & 31.126  & 33.146  & 33.147  & 36.174  & 36.175  \\
 51.294  & 51.296  & 52.310  & 52.311  & 53.327  & 53.328  & 54.342  & 54.344  & 55.357  \\
 55.358  & 56.369  & 56.370  & 57.382  & 57.383  & 57.384  & 58.397  & 58.398  & 59.409  \\
 59.410  & 60.422  & 60.423  & 60.424  & 61.436  & 62.446  & 62.447  & 62.448  & 63.463  \\
 63.464  & 64.475  & 64.476  & 90.98 [t]  & 90.99 [t]  & 92.114 [\textbf{T}]  & 92.115 [\textbf{T}]  & 94.130 [\textbf{T}]  & 94.131 [t]  \\
 96.146 [\textbf{T}]  & 96.147 [\textbf{T}]  & 113.269  & 113.271 [t]  & 114.277  & 114.279 [t]  & 127.390  & 127.393  & 128.402  \\
 128.405  & 129.414  & 129.417  & 130.426  & 130.429  & 135.486  & 135.489  & 136.498  & 136.501  \\
 137.510  & 137.513  & 138.522  & 138.525  & 169.115 [t]  & 170.119 [t]  & 173.131 [t]  & 176.147  & 178.157 [t]  \\
 178.158 [t]  & 179.163 [t]  & 179.164 [t]  & 182.181 [t]  & 182.182 [t]  & 185.199  & 185.200  & 186.205  & 186.206  \\
 193.258  & 193.259  & 194.268  & 194.269  & & & & & \\
 \hline
 3.5 [t]  & 3.6 [t]  & 4.10 [t]  & 16.4 [t]  & 16.5 [t]  & 16.6 [\textbf{T}]  & 17.11 [t]  & 17.13 [t]  & 17.14 [\textbf{T}]  \\
 17.15 [\textbf{T}]  & 18.20 [t]  & 18.21 [\textbf{T}]  & 18.22 [\textbf{T}]  & 18.24 [t]  & 19.28 [\textbf{T}]  & 19.29 [t]  & 20.36 [t]  & 21.42 [t]  \\
 21.44 [t]  & 25.61  & 25.64  & 25.65  & 26.71  & 26.72  & 26.76  & 27.82  & 27.85  \\
 27.86  & 28.94  & 28.95  & 28.96  & 28.98  & 29.104  & 29.105  & 29.109  & 30.118  \\
 30.119  & 30.120  & 30.122  & 31.128  & 31.129  & 31.133  & 32.139  & 32.142  & 32.143  \\
 33.149  & 33.150  & 33.154  & 34.161  & 34.162  & 34.164  & 35.169  & 35.171  & 36.178  \\
 37.184  & 37.186  & 75.4 [t]  & 75.6 [t]  & 76.11 [t]  & 77.16 [t]  & 77.18 [t]  & 78.23 [t]  & 81.36 [t]  \\
 81.38 [t]  & 89.92 [t]  & 89.93 [t]  & 89.94 [\textbf{T}]  & 90.100 [\textbf{T}]  & 90.102 [t]  & 91.109 [\textbf{T}]  & 91.110 [\textbf{T}]  & 92.116 [\textbf{T}] 
   \\
 92.117 [t]  & 93.124 [t]  & 93.125 [\textbf{T}]  & 93.126 [\textbf{T}]  & 94.132 [\textbf{T}]  & 94.134 [t]  & 95.141 [\textbf{T}]  & 95.142 [\textbf{T}]  & 96.148
   [\textbf{T}]  \\
 96.149 [t]  & 99.168  & 99.170  & 100.176  & 100.178  & 101.184  & 101.186  & 102.192  & 102.194  \\
 103.200  & 103.202  & 104.208  & 104.210  & 105.216  & 105.218  & 106.224  & 106.226  & 111.256  \\
 111.257  & 111.258  & 112.264  & 112.265  & 112.266  & 113.272  & 113.274  & 114.280  & 114.282  \\
 115.288  & 115.290  & 116.296  & 116.298  & 117.304  & 117.306  & 118.312  & 118.314  & 168.112 [t]  \\
 171.124 [t]  & 172.128 [t]  & 177.154 [t]  & 180.172 [t]  & 181.178 [t]  & 183.190  & 184.196  & 195.3 [\textbf{T}]  & 207.43 [\textbf{T}]
    \\
 208.47 [\textbf{T}]  & 215.73  & 218.84  & & & & & & \\
		\hline
		\hline
	\end{tabular}
\end{table} 

\clearpage \thispagestyle{empty}

\setcounter{page}{1}

\newpage
\newpage
\section*{Supplementary Information}

\setcounter{section}{0}
\renewcommand{\thesection}{S\arabic{section}}

In these Supplementary Notes we present  additional information on the theoretical derivations and experimental results. The details reported are intended to be rather pedagogical and thus rather detailed.
In Supplementary Note~\ref{SI_sec_band_topology} the band topology of MnSi, notably SG~198   and its magnetic subgroups,  is derived and discussed both for the paramagnetic and ferromagnetic phases. In Supplementary Note~\ref{sec_tight_binding_model} two tight-binding models for SG~19.27 and SG~4.9 are introduced, corresponding to the magnetization in MnSi  aligned along [010] or within the $xy$ plane, respectively. This section contains also a discussion of the Berry curvature and  surface states, notably the formation of large Fermi arcs. The magnetic SGs that allow for  nodal planes and topological  protectorates 
of the FS are catalogued in Supplementary Note~\ref{classification_nodal_surfaces}.
We use this catalogue to identify a number of candidate materials with topological protectorates, such as CoNb$_3$S$_6$ and Nd$_5$Si$_3$. In Supplementary Note~\ref{SI_sec_experimental_details} we discuss additional details of the experimental data treatment and analysis. In Supplementary Note~\ref{SI_sec_fs_determination} we present the detailed considerations underlying the identification of the experimentally observed de Haas-van Alphen (dHvA) frequencies with the calculated extremal FS cross-sections. In this Note we also discuss additional experimental data and corresponding DFT calculations that complement the results reported in the main part of the paper.
For ease of presentation we denote the BZ boundaries at $k_i=\pm\pi$ ($i=x,y,z$) as $k_i=\pi$. We note further, that in the spirit of itinerant-electron magnetism we use the expressions magnetization and magnetic moments equivalently.


\section{Band Topology of $\textrm{MnSi}$}
\label{SI_sec_band_topology}

An important topological feature of MnSi in its paramagnetic (PM) and ferromagnetic (FM) phases are the  nodal planes. These are two-dimensional, topologically charged band crossings at the boundary of the Brillouin zone (BZ). Besides these, MnSi also exhibits  Weyl points and fourfold degenerate points  in its band structure. The existence of the  nodal planes, the  fourfold degenerate points, and some of the Weyl points  is enforced by the non-symmorphic symmetries of MnSi alone. For this reason, these topological band crossings occur in all bands of MnSi, irrespective of the specific band dispersions.  
(In fact, any material with the same symmetries as MnSi exhibits these topological features.)
In this section we derive the existence of these topological band crossings in the PM and FM phases using the symmetry eigenvalues of the space group (SG) 198 (and its magnetic subgroups),
taking into account the effects of spin-orbit coupling. In Sec.~\ref{supp_paramag_MnSi} we first discuss the topological band crossings in the PM phase. Section~\ref{supp_ferro_MnSi} is concerned with the topology of the FM phase with different orientations of the ordered moment.	


\subsection{Paramagnetic phase of $\textrm{MnSi}$}
\label{supp_paramag_MnSi}

The crystal symmetries of the PM phase of MnSi are described by SG 198 (P2$_1$3). This SG  is chiral, due to the absence of inversion, mirror or roto-inversion symmetries. The BZ of SG 198 is cubic as shown in Fig.~\ref{fig:1}\,c. SG 198 is generated by the following symmetry operations
\begin{subequations}
	\begin{eqnarray}
	\tilde{C_2^z} \equiv \{ C_2^z \, | \, \tfrac{1}{2} 0 \tfrac{1}{2} \}
	&:&
	(x,y,z) \, \to (-x + \tfrac{1}{2}, - y, z + \tfrac{1}{2} )  \otimes (   i \sigma_z ),
	\\
	\tilde{C_2^y} \equiv\{ C_2^y \, | \, 0 \tfrac{1}{2}  \tfrac{1}{2} \}
	&:&
	(x,y,z) \, \to (-x, y + \tfrac{1}{2}, - z + \tfrac{1}{2} ) \otimes (   i \sigma_y ),
	\\
	\{ C_3^{xyz} \, | \, 0 \}   
	&:&
	(x,y,z) \, \to (z, x, y) \otimes \left[  \tfrac{1}{2} \sigma_0 + \tfrac{i}{2}  ( \sigma_x + \sigma_y + \sigma_z) \right],
	\label{three_fold_rotation_EQ_198}
	\end{eqnarray}
\end{subequations}
the mappings (indicated by the arrows)  specify how the symmetries act on the three real-space coordinates $(x,y,z)$.  The last brackets on each line describe  how the symmetries transform the spin parts of the Bloch wave functions $| \psi ( {\bf k}) \rangle$. Here, $\{ \sigma_x, \sigma_y, \sigma_z \}$ are the three Pauli matrices. 


\subsubsection{Nodal planes}
\label{SG_198_nodal_planes}

The existence of nodal planes follows from the combination of the two-fold screw rotations with time-reversal symmetry $\theta = i \sigma_y \mathcal{K} \, $ ~\cite{FURUSAKI2017788}, where $\mathcal{K}$ denotes the complex conjugation operator. For concreteness, let us focus on the screw rotation $\tilde{C_2^z} \equiv \{ C_2^z \, | \, \tfrac{1}{2} 0 \tfrac{1}{2} \}$  combined with the time-reversal symmetry $\theta $, i.e., $\theta \,\tilde{C_2^z} $. This combined symmetry acts like an effective mirror symmetry. That is, it relates the wave function at $(k_x, k_y, k_z)$ to the wave function  at $(k_x, k_y, -k_z)$, as illustrated in Fig.~\ref{fig:1}a. In particular, $\theta \,\tilde{C_2^z} $ leaves each point within the $k_z=0$ and $k_z = \pi $ planes of the BZ invariant. We note, however, that this effective mirror symmetry is anti-unitary and and acts upon the Berry curvature by mirroring its $z$ component [i.e., $\Omega^z (k_x, k_y, k_z) \to - \Omega^z (k_x, k_y, -k_z)$]. This is unlike an actual mirror symmetry, which leaves the Berry curvature, a pseudovector, invariant. For this reason $\theta \,\tilde{C_2^z} $ acts quite differently on the wave functions and topological charges than a regular mirror symmetry, see below. Since  $\theta \,\tilde{C_2^z} $ is anti-unitary, we can use it to derive the existence of degenerate Kramers pairs. For this purpose, we compute the square of $\theta \,\tilde{C_2^z} $. Since $\theta$ and $\tilde{C_2^z} $ commute, we find that
\begin{eqnarray} \label{square_magnetic}
\left[  \theta \, \{ C_2^z \, | \, \tfrac{1}{2} 0 \tfrac{1}{2} \} \right]^2
= 
\theta^2   \{ C_2^z \, | \, \tfrac{1}{2} 0 \tfrac{1}{2} \}^2 
=
T_{(0,0,1)} 
=  e^{i k_z } 
=
\left\{
\begin{array}{l}
+1 , \;  \textrm{for}  \; k_z = 0 \cr
-1 ,  \;  \textrm{for} \; k_z = \pi
\end{array}  \right. .
\end{eqnarray}
Here, $T_{(0,0,1)}$ denotes the translation by the vector $(0,0,1)$. The last two equations are obtained by acting on a Bloch state $| \psi ( {\bf k}) \rangle$. From Eq.~\eqref{square_magnetic} it follows that $\theta \,\tilde{C_2^z} $ acts on the Bloch states within the invariant plane $k_z = 0$ ($k_z= \pi$) like a time-reversal symmetry which squares to $+1$ ($-1$). Hence, we can use Kramers theorem~\cite{kramers_theorem_paper} to show that the Boch states $| \psi ( {\bf k}) \rangle$ and $\theta \,\tilde{C_2^z}  | \psi ( {\bf k}) \rangle$ within the $k_z = \pi$ plane have the same energy and are orthogonal to each other. In other words, all  Bloch states $| \psi ( {\bf k}) \rangle$  within the $k_z = \pi$ plane  are (at least) two-fold degenerate. As we move away from the $k_z= \pi$ plane in the BZ, the ${\bf k}$ points are no longer invariant under $\theta\,\tilde{C_2^z} $, such that the Bloch states $| \psi ( {\bf k}) \rangle$ are in general singly degenerate. (Note that SG 198 does not contain inversion, which together with $\theta$ would lead to a two-fold degeneracy at every ${\bf k}$ point.) Therefore, the bands in materials with SG 198 form  band-crossing planes (i.e., nodal planes) that are pinned at $k_z = \pi$. 

Similar arguments also hold for the combined symmetries $\theta \,\tilde{C_2^x}$ and $\theta \,\tilde{C_2^y}$ which lead to nodal planes that are pinned at the $k_x = \pi$ and $k_y = \pi$ planes, respectively. These three nodal planes are mapped onto each other by the three-fold rotation $C_3^{xyz}$, Eq.~\eqref{three_fold_rotation_EQ_198}. Taking into account the periodicity of the BZ, we observe that the nodal planes actually form three 2-tori, which intersect each other.  We conclude that the bands of MnSi cross each other on all three boundary planes of the cubic BZ. That is, the band structure of MnSi is grouped into pairs of bands that form  three nodal planes on the BZ boundary.

From the above arguments it is not clear whether or not this trio of nodal planes carries a topological charge, i.e., a nonzero Chern number. To infer the topology of the nodal plane trio, we study  in the next section the multiplicity and chirality of the Weyl points in the interior of the BZ cube. The Chern number (i.e., the chirality) of the nodal plane trio can then be inferred from the fermion doubling theorem by Nielsen and Ninomiya~\cite{nielsen_no_go}, wich states that the chiralities of all band crossings formed by a given pair of bands must add up to zero.   


\subsubsection{Topological charge of the nodal plane trio}
\label{PM_phase_Weyl_points}

The topological charge of a Weyl point (i.e., its chirality) is given by the Berry curvature integrated over a surface that encloses the Weyl point. Similar to Weyl points, also the nodal planes of the previous section can carry   nonzero topological charges. This topological charge is given by the Berry curvature integrated over the surface of a cube that is just a little bit smaller than the BZ cube, thereby enclosing the BZ boundary from within. According to the Nielsen-Ninomiya theorem~\cite{nielsen_no_go}, the sum over the topological charges of all band crossings (i.e., Weyl points and nodal planes) formed by a given pair of bands within the first BZ must vanish. We now use this theorem to show that the nodal planes   of MnSi carry   nonzero topological charges. 

To do so, we must first study the possible Weyl points in the interior of the BZ cube. Kramers theorem guarantees that there exists always a  Weyl point at $\Gamma$, i.e., at the center of the BZ cube, because $\Gamma$ is a time-reversal invariant momentum. Since the Weyl point at $\Gamma$ is left invariant by all the symmetries of SG 198, it has multiplicity one, i.e., it is not related by symmetry to any other Weyl point in the BZ. Besides the Weyl point at $\Gamma$, the band structure of MnSi can, in principle, also exhibit Weyl points on the two-fold and three-fold rotation axes, on the invariant planes of the effective mirror symmetries, or somewhere else in the BZ. These Weyl points away from $\Gamma$ must have multiplicity larger than one, since they transform nontrivially under the symmetries. That is, these Weyl points are part of a group of Weyl points that are mapped onto each other by the symmetries. These symmetry-related Weyl points have all the same topological charge, because the symmetries of SG 198 act on the Berry curvature in reciprocal space either with a proper rotation or contain time-reversal.

Hence, in both cases the chirality of a Weyl point is preserved (in contrast, inversion or mirror symmetry would flip the chirality of the Weyl points.) We find that the Weyl points on the two-fold and three-fold rotation axes have multiplicity 6 and 8, respectively. The Weyl points on the invariant planes of the effective mirror symmetries have multiplicity 12. Finally, Weyl points away from any high-symmetry lines or planes have multiplicity 24. The last ingredient for our argument is that the  chirality of all Weyl points  must be $\nu = \pm 1$, since higher-order chiralities are only stable in the presence of symmetries other than the two-fold and three-fold rotations of SG 198~\cite{tsirkin_vanderbilt_PRB_17}. With this, we can sum up the topological charges form all the Weyl points as
\begin{eqnarray} \label{sum_chiralities_PM_198}
6 \nu_6 + 8 \nu_8 + 12 \nu_{12} + 24 \nu_{24} + \nu_{\Gamma} + \nu_{\text{npt}} = 2\mu +  \nu_{\Gamma} + \nu_{\text{npt}} = 0,
\end{eqnarray}
where $\nu_j \in \mathbb{Z}$, $\mu \in \mathbb{Z}$, and $\nu_\Gamma = \pm 1$. Here, $j v_j$ is the sum of the topological charges of all Weyl points with multiplicity $j$, $v_\Gamma$ is the chirality of the Weyl point at $\Gamma$, and $\nu_{\text{npt}}$ is the topological charge of the nodal plane trio. The sum in Eq.~\eqref{sum_chiralities_PM_198}  must vanish due to the Nielsen-Ninomiya theorem, which is only possible for $\nu_{\text{npt}}$ odd. Hence, the topological charge of the nodal plane  trio is  nonzero~\cite{2019_PRB_Yuxin,PRB_Yuxin}. 

As an aside, we note that at $\Gamma$ also four-fold band degeneracies are stable, since the little group at $\Gamma$ has irreducible representations of both dimension two and four. These four-fold degeneracies are two copies of conventional Weyl points, which are related by time-reversal symmetry. They are referred to as doubled spin-$\tfrac{1}{2}$ Weyl points~\cite{FURUSAKI2017788,grushin_multifold_fermions_PRB_18}, or sometimes also as spin-3/2 fermions~\cite{CoSi_PRL_17}, and have been discussed in the context of CoSi and other  transition metal silicides~\cite{CoSi_PRL_17,RhSi_Nature_hasan_19,CoSi_Nature_hong_19}. But the existence of these doubled Weyl points does not invalidate our argument, because they enter  in the sum of topological charges only for an even number of occupied bands. For the evaluation of the topological charges of the nodal planes, however,  an odd number of occupied bands is considered. Hence, we concluded that the topological charges $\nu_\Gamma = \pm 1$ of the crossings at $\Gamma$ can only be compensated by the nodal plane trio, since   all other band crossings have even topological charge.


\subsection{Ferromagnetic phase of $\textrm{MnSi}$}
\label{supp_ferro_MnSi}

In the ferromagnetically ordered  phase of MnSi the symmetry is lowered to a magnetic subgroup of SG 198. These subgroups contain a subset of the symmetry elements of SG 198, possibly combined with time-reversal symmetry. Which of the elements of SG 198 remain as   good symmetries, depends on the orientation of the magnetic moment. That is, for different orientations of the moment the symmetries are described by different magnetic subgroups. In Extended Data Fig.~\ref{EDI:Theory:Fig1}\,a we plot all possible magnetic subgroups of SG 198, as obtained from the {\tt k-subgroupsmag} tool of the Bilbao Crystallographic Server~\cite{bilbao_server_k_subgroups_mag}. For the moments along $[010]$,   the subgroup is SG 19.27, for the moments within the $xy$-plane it is SG 4.9, while for the moments along $[111]$ it is SG R3. Here and in the following we use the BNS convention to label the magnetic space groups. Let us now study the band topology for the different moment directions separately. 


\subsubsection{Moments pointing along $[010]$ direction}

For the moments oriented along   $[010]$, the three-fold rotation $C_3^{xyz}$ and all symmetries generated by it  no longer leave the FM  phase invariant. However, the screw rotation $\tilde{C_2^y}$ is preserved in the $[010]$ FM phase. The other two screw rotations, $\tilde{C_2^x}$ and $\tilde{C_2^z}$, combined with time-reversal  $\theta$ also remain as good symmetries. Hence, the symmetries that leave the $[010]$ FM phase invariant are
\begin{eqnarray} \label{symmetries_FM_phase}
\tilde{C}_2^y \equiv \{ C_2^y \, | \, 0  \tfrac{1}{2}   \tfrac{1}{2} \} ,
\quad
\theta \, \tilde{C}_2^z \equiv \theta \{ C_2^z \, | \, \tfrac{1}{2} 0 \tfrac{1}{2} \} , 
\quad
\textrm{and}
\quad
\theta \, \tilde{C}_2^x \equiv \theta \{ C_2^x \, | \, \tfrac{1}{2}  \tfrac{1}{2} 0  \} ,
\end{eqnarray} 
which generate the magnetic SG 19.27 (P$2_1 2'_1 2'_1$), see green box in Extended Data Fig.~\ref{EDI:Theory:Fig1}\,a. Similar to $ \theta \, \tilde{C}_2^z$ of the PM phase (SG~198), the two combined symmetries  $ \theta  \, \tilde{C}_2^x$ and $\theta \, \tilde{C}_2^z $ act like effective mirror symmetries, since they relate the wave functions and energy bands at $(k_x,k_y,k_z)$ to those at $(-k_x, k_y, k_z)$ and $(k_x, k_y, - k_z)$, respectively. Using the reasoning of Sec.~\ref{SG_198_nodal_planes}, we now explain how these two combined symmetries enforce the existence of nodal planes at the BZ boundary. 


\paragraph{\textbf{Nodal planes.}} 
\label{sec_010_nodal_planes}
The effective mirror symmetry $\theta \, \tilde{C}_2^x $  ( $ \theta \, \tilde{C}_2^z $ )  leaves the $k_x=0$ and $k_x = \pi$ planes (the $k_z=0$ and $k_z= \pi$ planes) invariant. Similar to Eq.~\eqref{square_magnetic}, we find that $\theta \, \tilde{C}_2^x $  ( $ \theta \, \tilde{C}_2^z $ ) squares to $+1$ on the $k_x=0$ plane ($k_z=0$ plane), while it square to $-1$ on the $k_x = \pi$ plane ($k_z = \pi$ plane). Hence, we can use Kramers theorem to show that all bands are two-fold degenerate on the  $k_x = \pi$  and $k_z = \pi$ planes. Away from the $k_x = \pi$  and $k_z = \pi$ planes the bands are, in general, singly degenerate, since time-reversal combined with inversion is not a good symmetry of the $[010]$ FM phase. Therefore, all bands cross each other pairwise on the $k_x = \pi$ and $k_z = \pi$ boundary of the BZ, forming a duo of nodal planes (Fig.~\ref{fig:1}). Taking into account the periodicity of the BZ these  two nodal planes are actually two 2-tori, which cross each other. We note that in contrast to the PM phase (SG~198), the bands are not Kramers degenerate on the $k_y = \pi$ BZ boundary, since  $\theta \,  \tilde{C}_2^y$ is broken in the $[010]$ FM phase. Using the fermion doubling theorem by Nielsen and Ninomiya, we will show in Sec.~\ref{top_charge_nodal_plane_duo} that this duo of nodal planes carries a topological charge. But before doing so, we first discuss how the symmetries Eq.~\eqref{symmetries_FM_phase}, enforce the existence of Weyl points and  fourfold degenerate points. 


\paragraph{\textbf{Weyl points.}}
\label{weyl_point_subsec}
The screw rotation $\tilde{C}_2^y$, which leaves the  $(0,k_y,0)$ axis of the BZ invariant, is a good symmetry of the $[010]$ FM phase. Therefore, the Bloch states $| \psi ( {\bf k}) \rangle$ on the  $(0,k_y,0)$  axis can be labelled by the eigenvalues of  $\tilde{C}_2^y$. Due to the translation part, the eigenvalues of $\tilde{C}_2^y$ are momentum dependent. Taking the square of  $\tilde{C}_2^y$ [cf.~Eq.~\eqref{square_magnetic}]
\begin{eqnarray}
\left[ \{ C_2^y \, | \, 0  \tfrac{1}{2}   \tfrac{1}{2} \} \right]^2 
= 
T_{(0,1,0)} ( i \sigma_y) ^2 = - e^{i k_y } ,
\end{eqnarray}
we find that the eigenvalues of $\tilde{C}_2^y$ are
\begin{eqnarray} \label{EQ:EVs_srew_rot_Y}
\tilde{C}_2^y  | \psi_{\pm} ( {\bf k} )\rangle 
=
\pm i e^{i k_y /2}
| \psi_{\pm}  ( {\bf k} )\rangle .
\end{eqnarray}
Thus, the $\tilde{C}_2^y$ eigenvalues live on the complex unit circle and wind in a non-trivial way as $k_y$ is changed from $-\pi \to 0  \to + \pi$ on the Y$_1-$$\Gamma $$-$Y line of the BZ. For example, at $k_y = - \pi$ the eigenvalues are either $+1$ or $-1$, see the two bands in Extended Data Fig.~\ref{EDI:Theory:Fig2}\,a1. As we move from $k_y = -\pi$ through $k_y=0$ to $k_y = + \pi$ the eigenvalues move counterclockwise on the complex unit circle form $+1$ ($-1$) through $+i$ ($-i$) to $-1$ ($+1$). That is, the eigenvalues of the two bands in Extended Data Fig.~\ref{EDI:Theory:Fig2}\,a1 are interchanged. Since  the BZ is periodic and the bands and their eigenvalues are continuous functions of ${\bf k}$, this interchange of eigenvalues must be accompanied by an odd number of band crossings~\cite{zhao_schnyder_PRB_16}.  That is,   on the Y$_1-$$\Gamma $$-$Y line the bands have a nontrivial connectivity with at least one Weyl point. Since the two bands that form these Weyl points have different $\tilde{C}_2^y$ eigenvalues, a gap cannot be opened, as long as the screw rotation  $\tilde{C}_2^y$ is preserved. 

Let us now extend these arguments to more than two bands. To that end, we consider a collection of $2N$ bands   ($N \in \mathbb{N}$)  and label the bands at each ${\bf k}$ in order of increasing energy. That is, the band with lowest energy is labeled $n=1$, the band with next higher energy has label $n=2$, and so on. We note that with this convention the nodal planes on the $k_x = \pi$ and $k_z =\pi$ BZ boundaries are formed between bands $n=2m-1$ and $n=2m$, but not between bands $n=2m$ and $n=2m+1$, where $m \in \mathbb{N}$. We now follow the $n=1$ band along the Y$_1-$$\Gamma $$-$Y line and count the number of crossings with the band $n=2$. Note that if a crossing appears, we follow the energetically lower band, as the band label $n$ is defined at each ${\bf k}$ in order of increasing energy. At a crossing the $\tilde{C}_2^y$ eigenvalue of the $n=1$ band must flip by the factor $-1$ (i.e., $\pm i e^{i k_y /2} \to  \mp i e^{i k_y /2}$), since stable crossings are only possible between bands with opposite eigenvalues. Due to the continuity of the eigenvalues and the periodicity of the BZ, there must be an odd number of crossings between the $n=1$ and the $n=2$ bands. Next, we follow the $n=2$ band along the Y$_1-$$\Gamma $$-$Y line and count the number of crossings with the $n=3$ band. As for the $n=1$ band, there must be an odd number of crossings, where the $\tilde{C}_2^y$ eigenvalue flips, as otherwise continuity would be violated. Since the $n=2$ band already has an odd number of crossings with the $n=1$ band, there must be an even number of crossings with the $n=3$ band. Continuing to the $n=3$ band, we can repeat   the logic employed for the $n=1$ band, as both have an even number of crossings with their lower band (i.e., no crossing in the case of $n=1$). Repeating this process iteratively, we find that there are odd number of crossings between bands $n=2m-1$ and $n=2m$, while there is an even number of crossings between bands $n=2m$ and $n=2m+1$, where $m \in \mathbb{N}$. \label{argument_with_more_than_two_bands}

The above reasoning holds for all of the bands of any material crystallizing in SG 19.27,  and hence, in particular, also for the bands of MnSi in the $[010]$ FM phase. In Extended Data Fig.~\ref{EDI:Theory:Fig2}\,b1 the  $\tilde{C}_2^y$ eigenvalues of the MnSi band structure in the  $[010]$ FM phase are indicated by color. We observe  that the eigenvalues of all bands wind half way around the complex unit circle as we move along the Y$_1-$$\Gamma $$-$Y line. The band structure can be decomposed into pairs of nearby bands with opposite eigenvalues, which exhibit an odd number of Weyl crossings, as expected from the above analysis. The number of crossings of bands from different band pairs is even. We note that in Extended Data Fig.~\ref{EDI:Theory:Fig2}\,b1 (apparent) crossings of bands with the same eigenvalues are non-topological, i.e., they can be gapped out by small perturbations or are actually already avoided crossings with a small gap that is not resolved.  


\paragraph{\textbf{Fourfold points (double spin-1/2 fermions).}}
\label{dirac_point_subsec}

The screw rotation $\tilde{C}_2^y$ leaves, besides the  $(0,k_y,0)$ axis, also the $(0, k_y, \pi)$ axis, the $(\pi, k_y, 0)$ axis, and the $(\pi, k_y, \pi)$ axis invariant. The latter three axes are part of the nodal planes, as discussed in Sec.~\ref{sec_010_nodal_planes}. That is, the magnetic symmetries $\theta \, \tilde{C}_2^x$ and $\theta \, \tilde{C}_2^z$ enforce the existence of Kramers pairs on the $(0, k_y, \pi)$, $(\pi, k_y, 0)$, and $(\pi, k_y, \pi)$ axes. The screw rotation $\tilde{C}_2^y$ could increase the degeneracy further, giving rise to four-fold degenerate points, also known as fourfold fermions. In order to check for this possibility, we need to derive the commutation relations among the symmetry operations. For the commutation relation between  $ \tilde{C}_2^y$ and $\theta \, \tilde{C}_2^x$ we find 
\begin{eqnarray} \label{sec_dirac_communation_eins}
\{ C_2^y \, | \, 0  \tfrac{1}{2}   \tfrac{1}{2} \} \,
\theta \{ C_2^x \, | \, \tfrac{1}{2}  \tfrac{1}{2} 0  \}
&=&
- T_{(-1,1,1)} \, \theta \{ C_2^x \, | \, \tfrac{1}{2}  \tfrac{1}{2} 0  \} \,\{ C_2^y \, | \, 0  \tfrac{1}{2}   \tfrac{1}{2} \}
\nonumber\\
&=&
-  e^{i ( - k_x + k_y - k_z ) } \theta \{ C_2^x \, | \, \tfrac{1}{2}  \tfrac{1}{2} 0  \} \, \{ C_2^y \, | \, 0  \tfrac{1}{2}   \tfrac{1}{2} \},
\end{eqnarray}
where we moved the translation through the operators $\{ C_2^x \, | \, \tfrac{1}{2}  \tfrac{1}{2} 0  \} \{ C_2^y \, | \, 0  \tfrac{1}{2}   \tfrac{1}{2} \}$ changing it to $T_{(1,-1,1)}$ before replacing it with its eigenvalue, cf.~Eq. (\ref{square_magnetic}). Letting Eq.~\eqref{sec_dirac_communation_eins} act on a Bloch state $| \psi ( {\bf k}) \rangle$ on one of the two axes $(\pi, k_y, 0)$ or $(\pi, k_y, \pi)$, which are left invariant by both $  \tilde{C}_2^y $ and $\theta \,  \tilde{C}_2^x $, we obtain using Eq.~\eqref{EQ:EVs_srew_rot_Y}  
\begin{eqnarray} \label{eq_EV_kramers_pairs_1}
\tilde{C}_2^y   \,
\theta \,  \tilde{C}_2^x \, 
| \psi_\pm ( {\bf k} ) \rangle
=
\pm i e^{i (  - k_x + \tfrac{k_y}{2} - k_z )} \theta \, \tilde{C}_2^x \, 
| \psi_\pm ( {\bf k} ) \rangle .
\end{eqnarray}
Repeating this calculation for the symmetry operations $\tilde{C}_2^y$ and $\theta \, \tilde{C}_2^z$ we find
\begin{eqnarray} \label{eq_EV_kramers_pairs_2}
\tilde{C}_2^y  \,
\theta \, \tilde{C}_2^z \, 
| \psi_\pm ( {\bf k} ) \rangle
=
\pm i e^{i (  + k_x + \tfrac{k_y}{2} - k_z )} 
\theta \,  \tilde{C}_2^z  \,
| \psi_\pm ( {\bf k} ) \rangle ,
\end{eqnarray}
where here $| \psi_\pm ( {\bf k} ) \rangle$ are Bloch states on the two axes $(0, k_y, \pi)$ or $(\pi, k_y, \pi)$, which are left invariant by both $ \tilde{C}_2^y$ and $\theta \, \tilde{C}_2^z$. Equations~\eqref{eq_EV_kramers_pairs_1} and~\eqref{eq_EV_kramers_pairs_2} tell us  the  $\tilde{C_2^y}$ eigenvalues of the  Kramers pairs
$\left( | \psi_{\pm} ( {\bf k} ) \rangle , 
\; \theta \,
\tilde{C_2^x} | \psi_\pm ( {\bf k} ) \rangle \right)$
and
$\left( | \psi_{\pm} ( {\bf k} ) \rangle  , 
\; \theta \,
\tilde{C_2^z} | \psi_\pm ( {\bf k} ) \rangle \right)$, respectively. 
That is, we find that on the $(\pi, k_y, \pi)$ line  the two states of a  Kramers pair have the same $\tilde{C_2^y}$ eigenvalues, 
\begin{eqnarray} \label{EQ:PairingRUR}
\pm i e^{i k_y / 2} \quad \textrm{and} \quad \pm i e^{i k_y /2}   ,
\end{eqnarray}
while on the $(0, k_y, \pi)$ and $(\pi, k_y, 0)$  lines  they have opposite $\tilde{C_2^y}$ eigenvalues, 
\begin{eqnarray}  \label{kramers_pairs_oposite_EVs}
\pm i e^{i k_y / 2} \quad \textrm{and} \quad \mp i e^{i k_y /2}  .
\end{eqnarray}

\emph{(i) \; R$_1-$U$-$R line.} From Eq.~\eqref{EQ:PairingRUR} we see that the two eigenvalues of the Kramers pair on the $(\pi, k_y, \pi)$ axis rotate in lock-step and counterclockwise by a factor of $-1$, as we move along $k_y$ from, say, $-\pi$ $\to$ $0$ $\to$ $\pi$ on the R$_1-$U$-$R line of the BZ. This is similar to the winding of the  $\tilde{C_2^y}$ eigenvalue of the nondegenerate Bloch states on the $(0,k_y,0)$ axis. Hence, we can use the same argument as in Sec.~\ref{weyl_point_subsec} to infer the band connectivity, but now for   Kramers pairs instead of  single Bloch states. We therefore find that the $\tilde{C_2^y}$ eigenvalues of two Kramers pairs of bands are interchanged, as we go from $k_y = -\pi$ to $k_y =0$ to $k_y = \pi$, see Extended Data Fig.~\ref{EDI:Theory:Fig2}\,a2. This interchange of eigenvalues is accompanied by an odd number of crossings. That is, there exists at least one four-fold degenerate point on the R$_1-$U$-$R line. 
This fourfold point, which is also known as a fourfold fermion (or double spin-1/2 fermion)~\cite{grushin_multifold_fermions_PRB_18},    cannot be gapped out,
because the four mutually orthogonal states at the crossing point are either Kramers partners or have opposite $\tilde{C_2^y}$ eigenvalues. 

If we have more than two Kramers pairs, it follows from the argument on page~\pageref{argument_with_more_than_two_bands} that the Kramers pairs $( 4 m-3, 4m-2)$ and $(4m-1, 4m)$ have an odd number of crossings, while the Kramers pairs $(4m-1, 4m)$ and $(4m+1, 4m+2)$ have an even number of crossings, where $m \in \mathbb{N}$.

\emph{(ii) \; S$_1-$X$-$S and T$_1-$Z$-$T lines.} From Eq.~\eqref{kramers_pairs_oposite_EVs} it follows that the two eigenvalues of the Kramers pairs on the  $(0, k_y, \pi)$ and $(\pi, k_y, 0)$  axes are opposite  for any value of $k_y$. They rotate counterclockwise   from $-1$ and $+1$ $\to$ $-i$ and $+i$ $\to$ $+1$ and $-1$, as we move along $k_y$ from $-\pi$ $\to$ $0$ $\to$ $+\pi$, see Extended Data Fig.~\ref{EDI:Theory:Fig2}\,a3. Hence, the Kramers pair can connect with itself in a continuous manner at the BZ boundary $k_y = \pm \pi$. Therefore, as opposed to the $(\pi, k_y, \pi)$ axis, there is no need to introduce a second Kramers pair to satisfy the nontrivial winding of the eigenvalues. Moreover, any crossing between two Kramers pairs on the $(0, k_y, \pi)$ and $(\pi, k_y, 0)$  axes is unstable, since every Kramers pair has the same pair of $\tilde{C_2^y}$ eigenvalues. 

\vspace{0.2 cm}

The above arguments apply to all bands of any material in SG 19.27, specifically also to MnSi in the $[010]$ FM phase. In Extended Data Fig.~\ref{EDI:Theory:Fig2}\,b2 we show the band structure of MnSi along the R$_1-$U$-$R line with their $\tilde{C_2^y}$ eigenvalues indicated by  color. All bands are Kramers degenerate and their eigenvalues wind by a factor $-1$ as we go through the BZ along R$_1-$U$-$R. We can see from this figure that the bands can be grouped into pairs of Kramers partners with opposite $\tilde{C_2^y}$ eigenvalues that cross each other an odd number of times. Hence the connectivity of the bands  agrees with the above discussion and is of the same type as in Extended Data Fig.~\ref{EDI:Theory:Fig2}\,a2. 


\paragraph{\textbf{Topological charge of the nodal plane duo.}} \label{top_charge_nodal_plane_duo}

Having found all symmetry-enforced band crossings in SG 19.27, we now use the Nielsen-Ninomiya theorem~\cite{nielsen_no_go}, as in Sec.~\ref{PM_phase_Weyl_points}, to show that the nodal plane duo at the $k_x = \pi$ and $k_z = \pi$ BZ boundaries has a nonzero topological charge. For this purpose, we must first consider the possible existence of accidental Weyl points, i.e., Weyl points that are not symmetry enforced. These accidental Weyl points can appear either within one of the two effective mirror planes $k_x=0$ or $k_z=0$, or at a generic position in the interior of the BZ cube. Due to the effective mirror symmetries, accidental Weyl points within the effective mirror planes appear in pairs with same Chern number (i.e., they have multiplicity two). Similarly, Weyl points at generic positions appear in quartets with same Chern number (i.e., they have multiplicity four). Hence, according to the Nielsen-Ninomiya theorem, the topological charges of all Weyl points formed by the bands $n=2m-1$ and $n=2m$, which form nodal plane duos,  must add up to zero as
\begin{eqnarray} \label{sum_top_charge_nodal_plane_duo}
2 \nu_2 + 4 \nu_4  + \nu_{\text{Y$_1 \Gamma$Y}}^{2m-1,2m} + \nu_{\text{npd}} 
= 0,
\end{eqnarray}
where $\nu_j \in \mathbb{Z}$ and $j \nu_j$ is the sum of the topological charges of all accidental Weyl points with multiplicity~$j$. $\nu_{\text{npd}} $ is the topological charge of the nodal plane duo and $ \nu_{\text{Y$_1 \Gamma$Y}}^{2m-1,2m}$ is the sum of the topological charges on the  Y$_1-$$\Gamma $$-$Y line. As discussed in Sec.~\ref{weyl_point_subsec},  there is an odd number of Weyl crossings between bands $n=2m-1$ and $n=2m$ on the Y$_1-$$\Gamma $$-$Y line. Since the chiralities of the Weyl points is in general $\nu = \pm 1$, $\nu_{\text{Y$_1 \Gamma$Y}}^{2m-1,2m}$ is therefore an odd number. (Note that Weyl points with higher chiralities are only stable in the presence of additional symmetries besides the two-fold rotations of SG 19.27.)  Thus, in order for Eq.~\eqref{sum_top_charge_nodal_plane_duo} to be satisfied, also $ \nu_{\text{npd}} $ must be an odd number. That is, the nodal plane duo must carry an odd Chern number (i.e., $\nu_{\text{npd}} \ne 0$), such that the total topological charge of all Weyl crossings between bands $n=2m-1$ and $n=2m$ sums up to zero in the first BZ. 

Next, we turn to the crossings between the Kramers pairs on the R$_1-$U$-$R line. As discussed in Sec.~\ref{dirac_point_subsec}, the number of crossings on the R$_1-$U$-$R line  between the Kramers pairs $(4m-3, 4m-2)$ and $(4m-1,4m)$ is odd, while it is even between the pairs $(4m-1,4m)$ and $(4m+1,4m+2)$. The fourfold degenerate points on the R$_1-$U$-$R line have even topological charge, i.e., in general $\pm 2$, since both Kramers partners have the same  $\tilde{C_2^y}$ eigenvalues, see Eq.~(\ref{EQ:PairingRUR})   (for higher even numbers additional symmetries would be needed.). Hence, the sum of the topological charges of the fourfold points is either 
\begin{eqnarray}
\nu^{\text{\tiny{$(4m-3, 4m-2)$,$(4m-1,4m)$}}}_{\text{R$_1$U$$R}}  = 4 \mu + 2 
\quad
\textrm{or}
\quad 
\nu^{\text{\tiny{$(4m-1,4m)$,$(4m+1,4m+2)$}}}_{\text{R$_1$U$$R}} = 4 \mu, 
\end{eqnarray}
where $\mu \in \mathbb{Z}$, depending on which two Kramers pairs we are looking at. These nonzero topological charges of the fourfold points are in general compensated by the topological charges of   accidental Weyl points. For example, the topological charge $\nu^{\text{\tiny{$(4m-3, 4m-2)$,$(4m-1,4m)$}}}_{\text{R$_1$U$ $R}}  = 4 \mu + 2 $ can be compensated by an even number of Weyl points on one of the effective mirror planes with total topological charge $ \nu_{\text{mir}}^{(4m-2,4m-1)}  = - 4 \mu -2$.	

As an example, let us discuss the simplest configuration of topological charges for a single set of four connected bands, as described by the tight-binding model of Sec.~\ref{sec_tight_binding_model}, see Fig.~\ref{fig:1}d. This tight-binding model realizes only the minimum number of required Weyl points and fourfold    points. That is, on  the Y$_1-$$\Gamma $$-$Y line there is only one Weyl point between the lower two and upper two bands. Likewise, on the R$_1-$U$-$R line there exists only one fourfold degenerate  point. Using the above arguments, we find that for this tight-binding model the simplest way to assign the Weyl point chiralities is
\begin{eqnarray} \label{Chern_nos_TB_model}
&&
\nu_{\text{Y$_1 \Gamma$Y}}^{1,2} = \pm 1 , 
\quad
\nu_{\text{Y$_1 \Gamma$Y}}^{2,3} = 0 , 
\quad
\nu_{\text{Y$_1 \Gamma$Y}}^{3,4} = \pm 1 , 
\quad
\nu_{\text{npd}}^{ 1,2} = \mp 1 ,
\quad
\nu_{\text{npd}}^{ 3,4} = \mp 1 ,
\\
&&
\nu_{\text{R$_1$UR}} =+2,
\quad
\textrm{and} \quad
\nu_{\text{mir}}^{2,3} = -2 ,
\nonumber
\end{eqnarray}
where $\nu_{\text{mir}}^{2,3}$ denotes the sum of the topological charges of the two Weyl points on the effective mirror planes, which compensate the topological charge of the fourfold   point. 


\subsubsection{Moments pointing away from $[010]$ direction}

So far we focused on the magnetic moment direction with highest symmetry, namely the  $[010]$ direction. Let us now consider lower symmetry situations. When the moments are within the $xy$-plane (but not along $[010]$), only the combined symmetry  $ \theta \, \tilde{C}_2^z$ remains as a good symmetry, which corresponds to SG 4.9 (P$2'_1$), see Extended Data Fig.~\ref{EDI:Theory:Fig1}\,a. Similar to Eq.~\eqref{square_magnetic}, we find that this symmetry squares to $-1$ on the $k_z = \pi$ plane, i.e., $\left.  (   \theta \, \tilde{C}_2^z )^2 \right|_{k_z=\pi} = -1$. Hence, the bands on $k_z=\pi$ are doubly degenerate forming a single nodal plane of Kramers pairs. In general, this nodal plane at $k_z=\pi$ does not carry a topological charge. However, there may exist accidental crossings in the interior of the BZ, whose topological charges are compensated by a nonzero topological charge of the nodal plane. Hence, the nodal plane at $k_z=\pi$ may carry, in principle, a nonzero topological charge, but this requires some fine-tuning and is not guaranteed by the symmetries alone as in the previous section.

When the moments are along $[111]$, only the thee-fold rotation $C^{xyz}_3$ remains as a good symmetry, corresponding to SG 146.10 (R3), see Extended Data Fig.~\ref{EDI:Theory:Fig1}\,a. Since this is a symmorphic group, there are no  symmetry-enforced band crossings on high-symmetry lines or planes, but only accidental Weyl points somewhere at generic positions in the interior of the BZ. This is also the case when the moments point along some arbitrary direction, since then no symmetries remain at all.

It is interesting to study how the nodal plane duo of the previous section evolves, as the magnetic moments are rotated away from $[010]$, for example, into the $xy$-plane. For this purpose, we present in Extended Data Figs.~\ref{EDI:Theory:Fig1}c1-f3  the ab-initio electronic structure of ferromagnetic MnSi with the moments pointing along [010], 10$^\circ$ rotated into the $xy$-plane, and along [110], respectively, for four high-symmetry directions in the BZ. We observe that, as the moments are rotated into the $xy$-plane, the Weyl points Y$_1-$$\Gamma $$-$Y are slightly shifted away from the high-symmetry line, see Extended Data Figs.~\ref{EDI:Theory:Fig1}c2-c3. Likewise, the fourfold  points on R$_1-$U$-$R are split into Weyl points that are located slightly away from R$_1-$U$-$R, see Extended Data Figs.~\ref{EDI:Theory:Fig1}e1-e3. Furthermore,  as the moments are rotated away from $[010]$, the Kramers degeneracies on the $k_x = \pi$ plane are lifted and a small gap of about 10~meV opens, see Extended Data Figs.~\ref{EDI:Theory:Fig1}f1-f3. In general, new Weyl points are formed in this process, somewhere in the interior of the BZ but close to the $k_x = \pi$ and $k_z=\pi$ planes, which inherit the topological charges from the nodal plane duos. Since these new Weyl points are formed by bands which cross each other at a small angle, their Berry curvature ${\bf \Omega}_n ( {\bf k} )$ radiates out over a large volume near the $k_x = \pi$ and $k_z = \pi$ planes. Hence, the bands and FSs near the BZ boundaries have large Berry curvatures, even when the moments point away from the $[010]$ directions with no topological planes present. In fact, as discussed in  Sec.~\ref{supp_sec_berry_curvature}, the cusp- and jump-like features of ${\bf \Omega}_n ( {\bf k} )$ at the nodal planes turn into   quasi divergences, when a small gap is opened, such that ${\bf \Omega}_n ( {\bf k} )$  is enhanced (Extended Data Fig.~\ref{EDI:Theory:Fig3}).  

In conclusion, rotating the magnetic moments away from $[010]$ represents a relatively small perturbation, as far as topology is concerned. It only leads to small gaps of the nodal plane duo and slight changes of the Berry curvature profile. Also the Weyl points on Y$_1-$$\Gamma $$-$Y are only slightly shifted, but remain otherwise intact and far away from their opposite chirality partners near the BZ boundary. Hence, the Fermi arc surface states remain largely unchanged, still spanning almost half of the surface BZ, see Sec.~\ref{supp_surface_states}.  

\newpage 

\section{Tight-binding models,  Berry curvature, and surface states}
\label{sec_tight_binding_model}

To illustrate the band topologies discussed in the previous section, we construct here two tight-binding models: One which shares the same symmetries as MnSi in the $[010]$ FM phase, i.e., which is invariant under the symmetries of SG 19.27, and a second one which has the same symmetries as MnSi with the magnetic moments in the $xy$ plane, i.e., which is invariant under SG 4.9.

We note that for a generic tight-binding model to have symmetry-enforced nodal planes controlling the low-energy physics, only two conditions must be satisfied:
(i) the tight-binding model must satisfy the symmetries of one of the space groups listed in Extended Data Table~\ref{EDI:Theory:tab1}, and (ii) there must be Fermi surfaces that extend across the BZ boundary. That is, stringent filling conditions are not necessary for the nodal planes to control the low-energy physics. 
This is in stark contrast to topological semimetals supporting point nodes, i.e., Weyl or Dirac systems, where one needs to invoke arguments about filling enforcement to ensure
that the point nodes are at the Fermi level (see, e.g., Refs.~\onlinecite{PhysRevLett.118.186401,chen_nat_phys_2018}).


\subsection{Tight-binding model for SG 19.27}
\label{sec_model_SG19_27}

Due to the fourfold  point on the $R_1 - U -R$ line, the minimum number of distinct bands is four. Hence, we consider a tight-binding model with four atoms in the unit cell, which are located at the four Wyckoff positions $4a$ of SG 19.27, and which each contribute one $s$ orbital to the band structure. Taking into account only nearest-neighbor hopping terms, the tight-binding Hamiltonian reads 
\begin{eqnarray} \label{ham_model_SG19_27}
H(\mathbf{k}) &= \begin{pmatrix}
\mu  & H_{12} & H_{13} & H_{14}
\\
H_{12}^* & \mu  & H_{23} & H_{24}
\\
H_{13}^*  &  H_{23}^* & \mu  &  H_{34}
\\
H_{14}^* &  H_{24}^*  &  H_{34}^*  & \mu 
\end{pmatrix},
\end{eqnarray}
with
\begin{eqnarray}
H_{12} &= a_1 + a_1^* e^{i k_y} + a_2 e^{i k_x} + a_2^* e^{i (k_x + k_y)}, 
\qquad
H_{13} &= a_3 (1+e^{i k_x}) + a_4 e^{i k_z}(1+e^{i k_x}),  \\
H_{14} &= a_5 (1+e^{i k_z}) + a_6 e^{i k_y}(1+ e^{i k_z}),
\qquad
H_{23} &= a_6 (1+e^{i k_z}) + a_5 e^{-i k_y}(1+ e^{i k_z}),  \\
H_{24} &= a_4 (1 + e^{-i k_x}) + a_3 e^{i k_z}(1 + e^{-i k_x}),
\qquad
H_{34} &= a_2 + a_2^* e^{i k_y} + a_1 e^{-i k_x} + a_1^* e^{i (-k_x+k_y)},
\end{eqnarray}
where the hopping amplitudes $a_i$ are complex and are arbitrarily chosen to be
$a_1 = -0.612421-0.386933i$,
$a_2 = -0.205746 + 0.0929439 i$,
$a_3 = -0.940811-0.0412139 i$, 
$a_4 =  0.133936 +0.344822$, 
$a_5 = -0.0649457-0.529284 i$, and
$a_6 = -1.051 - 0.111635 i$. The chemical potential $\mu$ is set to 
$\mu = 1.06929$.

The band structure of this model is shown in Fig.~\ref{fig:1}d. We observe that there is a Weyl point along the Y$_1-$$\Gamma $$-$Y path and a four-fold crossing point along  the R$_1-$U$-$R path, as predicted in the previous sections. Along the R$_1-$U$-$R path the bands are doubly degenerate, since this path is part of the nodal plane duo $k_x = \pi$ ($k_z = \pi$).


\subsection{Tight-binding model for SG 4.9}

In order to obtain a tight-binding model for SG 4.9, we start from a model in SG 19, which explicitly contains the spin degree of freedom. We then add to this model a Zeeman term $H_{\text{Z}} = \mathbf{B} \cdot\mathbf{\sigma}  $, which lowers the symmetry from SG 19 down to SG 4.9. This allows us to study the effects of Zeeman splitting as a function of  field direction $\mathbf{B}$. The model constructed in this way contains twelve nearest-neighbor hopping terms with twelve complex hopping parameters. As in Sec.~\ref{sec_model_SG19_27} we consider four atoms in the unit cell, located at the four Wykoff positions $4a$ of SG 19, i.e., at $(0, 0, 0)$, $(1/2, 1/2, 0)$, $(1/2, 0, 1/2)$, and  $(0, 1/2, 1/2)$. Each atom at these four Wykoff positions contributes two $s$ orbitals to the band structure, i.e., one for spin up and one for spin down. For convenience, the spin quantization is chosen in the $z$ direction.  
With this, the Hamiltonian $H (\mathbf{k})$ for SG 4.9 can be written as
\begin{eqnarray} \label{ham_model_SG19_4_9}
H (\mathbf{k}) &= 
\begin{pmatrix}
H_\uparrow(\mathbf{k}) & H^\text{SOC}(\mathbf{k})\\
H^\text{SOC}(\mathbf{k})^\dagger & H_\downarrow(\mathbf{k}) 
\end{pmatrix}
+ 
\mathbf{B} \cdot\mathbf{\sigma} \otimes \mathbb{1}_{4 \times 4} 
,
\end{eqnarray}
where $H_\downarrow(\mathbf{k}) = H_\uparrow(-\mathbf{k})^*$,  $\mathbf{\sigma}$ is the vector of Pauli matrices operating in spin space, and $\mathbf{B}$ is the magnetic field vector. $H_\uparrow(\mathbf{k})$ has a similar structure as Eq.~\eqref{ham_model_SG19_27}, i.e., 
\begin{eqnarray}
H_\uparrow(\mathbf{k}) &= 
\begin{pmatrix}
0 & H^{\uparrow}_{12}  & H^{\uparrow}_{13}  & H^{\uparrow}_{14}  \\ 
H^{\uparrow\ast}_{12} & 0 & H^{\uparrow}_{23}  & H^{\uparrow}_{24}  \\ 
H^{\uparrow\ast}_{13}  & H^{\uparrow\ast}_{23}  & 0 & H^{\uparrow}_{34}  \\ 
H^{\uparrow\ast}_{14}  & H^{\uparrow\ast}_{24}  & H^{\uparrow\ast}_{34}  & 0 \\
\end{pmatrix},
\end{eqnarray}
with
\begin{eqnarray}
H^{\uparrow}_{12}  &= b_1 + b_1 e^{i k_x} + b_2 e^{i k_y} + b_2 e^{i (k_x + k_y)},  \qquad
H^{\uparrow}_{13}  &= b_3 + b_3^* e^{i k_z} + b_4 e^{i k_x} + b_4^* e^{ i (k_x + k_z)} , \\
H^{\uparrow}_{14}  &= b_5 + b_5 e^{i k_y}  + b_6 e^{i k_z} + b_6 e^{i (k_y + k_z)}, \qquad
H^{\uparrow}_{23}  &= b_6^* + b_6^* e^{-i k_y} + b_5^* e^{i k_z} + b_5^* e^{-i (k_y - k_z)} , \\
H^{\uparrow}_{24} &= b_4 + b_4^*e^{i k_z}+ b_3 e^{-i k_x} + b_3^* e^{-i (k_x - k_z)} , \qquad 
H^{\uparrow}_{34}  &= b_2 + b_2 e^{-i k_x} + b_1 e^{i k_y} + b_1 e^{-i (k_x - k_y)},  
\end{eqnarray}
where $b_i$ are complex hopping amplitudes, which we arbitrarily set to 
$ 
b_1 = -0.430005 - 0.818889 i, \,
b_2 = 0.139629 - 0.587502 i, \,
b_3 = 1.00815 - 0.17464 i, \,
b_4 = 0.0620589 - 0.0593171 i, \,
b_5 = -0.226315 + 0.64551 i, \,\text{ and }  
b_6 = -0.326725 + 0.76938 i  $.
The two spin sectors are coupled to each other via spin-orbit coupling  
\begin{eqnarray}
H_{\text{SOC}}(\mathbf{k}) &= 
\begin{pmatrix}
0 & H^{\text{SOC}}_{12} (\mathbf{k}) & H^{\text{SOC}}_{13} (\mathbf{k}) & H^{\text{SOC}}_{14} (\mathbf{k}) \\ 
- H^{\text{SOC}}_{12} (- \mathbf{k}) & 0 & H^{\text{SOC}}_{23} (\mathbf{k}) & H^{\text{SOC}}_{24} (\mathbf{k}) \\ 
- H^{\text{SOC}}_{13} (- \mathbf{k}) & - H^{\text{SOC}}_{23} (- \mathbf{k}) & 0 & H^{\text{SOC}}_{34} (\mathbf{k}) \\ 
- H^{\text{SOC}}_{14} (- \mathbf{k}) & - H^{\text{SOC}}_{24} (- \mathbf{k}) & - H^{\text{SOC}}_{34} (- \mathbf{k}) & 0 \\
\end{pmatrix} ,
\end{eqnarray}
with
\begin{eqnarray}
H^{\text{SOC}}_{12}  &= c_1 + c_1^* e^{i k_x } + c_2^* e^{i k_y} + c_2 e^{i (k_x + k_y)} , \qquad 
H^{\text{SOC}}_{13}  &= c_3 + c_3 e^{i k_z} + c_4 e^{i k_x} + c_4 e^{i (k_x + k_z)}, \\
H^{\text{SOC}}_{14}  &= c_5 - c_5^* e^{i k_y} - c_6^* e^{i k_z} + c_6 e^{i (k_y + k_z)}, \qquad
H^{\text{SOC}}_{23}  &= c_6 - c_6^* e^{-i k_y} - c_5^* e^{i k_z} + c_5 e^{-i (k_y - k_z)}, \\
H^{\text{SOC}}_{24}  &= - c_4^* - c_4^* e^{i k_z} - c_3^* e^{-i k_x}  - c_3^* e^{-i (k_x - k_z)}, \qquad
H^{\text{SOC}}_{34}  &= - c_2^* - c_2 e^{-i k_x}  - c_1 e^{i k_y} - c_1^*e^{-i (k_x - k_y)},  
\end{eqnarray}
where $c_i$ are spinflip hopping parameters, which we arbitrarily set to
$
c_1 = -0.0673999 - 0.0654125 i, \,
c_2 = 0.895971 + 0.223187 i, \,
c_3 = 0.397634 + 0.222339 i, \,
c_4 = -0.0167943 + 0.100792 i, \,
c_5 = 0.388337 - 0.727044 i,  \,\text{ and }  
c_6 = 0.180885 - 0.0104037 i$.


\subsection{Berry curvature and topological charges}
\label{supp_sec_berry_curvature}

Using the above  tight-binding models, we compute the Berry curvature by the formula
\begin{eqnarray}
{\bf \Omega}_n ( {\bf k} ) =
i \sum_{n' \ne n} 
\frac{   
	\langle \psi_n |  \nabla_{\bf k} H ( {\bf k} ) | \psi_{n'} \rangle 
	\times
	\langle \psi_{n'} |  \nabla_{\bf k}  H ( {\bf k} )   | \psi_{n} \rangle 
}
{( E_n ({\bf k} ) - E_{n'} ({\bf k} ) )^2} ,
\end{eqnarray}
where  $| \psi_{n} \rangle $ denotes the $n$th Bloch state with energy $E_n ( {\bf k} ) $.
By integrating up the Berry curvature on closed surfaces that surround the band crossing points and planes, we compute the topological charges (chiralities) of the Weyl points, fourfold points, and nodal planes~\cite{fukui_JPSJ_05}. For the tight-binding model~\eqref{ham_model_SG19_27} in SG 19.27 (see Fig.~\ref{fig:1}d) we find that there  are two symmetry-enforced Weyl points on $Y_1-\Gamma-Y$ with charges $\nu^{1,2}_{Y_1 \Gamma Y}=+1$ and $\nu^{3,4}_{Y_1 \Gamma Y}=+1$. These Weyl points are the topological partners of two nodal plane duos, which have opposite topological charges $\nu^{1,2}_{\text{npd}}=-1$ and  $\nu^{3,4}_{\text{npd}}=-1$. We note that the two nodal planes, forming the nodal plane duo, carry together this topological charge. It is not possible to assign topological charges to each of the two nodal planes individually, since there does not exist a two-dimensional closed integration contour, which encloses only a single nodal plane and on which there exists everywhere a band gap. Any such contour must cross the other nodal plane, where the band gap closes. Hence the topological charge can only be defined for the two nodal planes together, i.e., for the nodal plane duo.

Finally, for the fourfold point on $R_1-U-R$ we find that it has topological charge $\nu_{R_1UR}=-2$. This charge is compensated by four accidental Weyl points in the bulk, whose total topological charge is $\nu^{2,3}_{\text{bulk}}=+4$, and two accidental  Weyl points on the $k_z=0$ plane, whose total topological charge is  $\nu^{2,3}_{\text{mir}}=-2$.

To investigate how the Berry curvature changes as the magnetization direction is rotated away from  high-symmetry axes, we use the tight-binding model defined in Eq.~\eqref{ham_model_SG19_4_9}. Extended Data Figures~\ref{EDI:Theory:Fig3}a and~\ref{EDI:Theory:Fig3}b show the Berry curvature projected onto the jungle-gym FS at $E=-4.9$ for the magnetization pointing along [010] and 10$^\circ$ rotated into the $xy$ plane, respectively. For the magnetization along [010] (corresponding to SG 19.27) the three components of the Berry curvature exhibit cusp or jump singularities at the BZ boundary $k_x = \pi$ (Extended Data Fig.~\ref{EDI:Theory:Fig3}a1), while for the magnetization within the $xy$ plane (corresponding to SG 4.9) these singularities are turned into quasi divergences (Extended Data Fig.~\ref{EDI:Theory:Fig3}b1). 
These singularities arise as soon as the magnetic screw rotation protecting the nodal plane is broken. Hence a small change in the field direction leads to a large change in the Berry curvature, 
i.e., a finite jump / cusp is turned into a singularity. This could potentially be exploited for applications, such as ultrasensitive detectors.

Interestingly, the jump in the Berry curvature is directly related to the topological charge of the NP. This is because the topological charge of the NP is given by the integral of the Berry curvature over two surfaces that enclose the NP with an  infinitesimal distance from either side. Hence, the topological charge of the NP is just simply the integral of the jump over the relevant BZ boundaries. Furthermore, we note that the texture of the Berry curvature on the FS sheets is restricted by the symmetries. Specifically, the effective mirror symmetry $\theta \, \tilde{C}^x_2$ of SG 19.27 requires that 
\begin{eqnarray}
{\bf \Omega}_n ( k_x, k_y, k_z )
=  
\begin{pmatrix}
-1 & 0 & 0 \cr
0 & +1 & 0 \cr
0 & 0 & +1 
\end{pmatrix}
{\bf \Omega}_n ( - k_x, k_y, k_z ),
\end{eqnarray}
and similarly for $\theta \,  \tilde{C}^z_2$, see Extended Data Fig.~\ref{EDI:Theory:Fig3}a1.
 
From the above reasoning it follows that materials with topological Fermi surface protectorates, such as MnSi, should exhibit large Berry curvatures, leading to an enhancement of the anomalous Hall effect. Indeed, this is consistent with the findings of Ref.~\onlinecite{2014_Franz_PRL}, which reports   large Berry curvatures contributing  to the anomalous Hall effect in MnSi.

\subsection{Topological surface states}
\label{supp_surface_states}

Here, we discuss the surface states of both the tight-binding model in SG 19.27, Eq.~\eqref{ham_model_SG19_27}, and ferromagnetic MnSi with magnetization along [010]. 
First of all, we note that MnSi with magnetization along [010] has the simplest type of topological nodal planes, namely nodal plane duos (two nodal planes) instead of nodal plane trios (three nodal planes). The latter occur in many paramagnetic and antiferromagnetic space groups (see Supplementary Note~\ref{classification_nodal_surfaces}). Fermi arcs exist only for the nodal plane duos, but not for the nodal plane trios. That is, by the bulk-boundary correspondence~\cite{chiu_RMP_16,armitage_mele_vishwanath_review}, the nontrivial topology of the band crossings gives rise to surface states, i.e., Fermi arcs that connect the projected band crossings in the surface BZ. In particular, for the [010] surface we expect a Fermi arc that connects the Weyl point  on the  $Y_1-\Gamma-Y$ line to the nodal plane duo. Due to the continuity of the band structure, this Fermi arc must attach smoothly (i.e., tangentially) to the nodal plane duo at the BZ boundary. Therefore, it cannot connect to the nodal plane duo in a straight line, but rather with a spiral shape. 
The handedness of this spiral is determined by the Chern number of the Weyl point~\cite{schroeter_Nat_Phys_2019}. Topology, however, does not seem to determine to which of the two NPs the Fermi arc connects. This may depend on the surface termination and other details.

Our calculations of the surface DOS confirm these considerations, both for the tight-binding model~\eqref{ham_model_SG19_27} (Extended Data Fig.~\ref{EDI:Theory:Fig4}a) and ferromagnetic MnSi (Extended Data Fig.~\ref{EDI:Theory:Fig4}c). In Extended Data Fig.~\ref{EDI:Theory:Fig4}a we observe a single Fermi arc that spirals outwards from the center of the surface BZ towards the BZ boundary, where it attaches smoothly to the nodal plane duo. In Extended Data Fig.~\ref{EDI:Theory:Fig4}b we see  two Fermi arcs that emanate from two unpaired Weyl points on $Y_1-\Gamma-Y$ close to the Fermi level (Fig.~\ref{fig:1}d) and connect to the bulk bands forming nodal planes on the BZ boundaries. Even though these Fermi arcs merge with bulk states before they reach the nodal plane duos, their existence proves the topology of the nodal plane duos. This is because all Weyl points away from $Y_1-\Gamma-Y$ have multiplicity two, and hence the single Fermi arcs emanating from the unpaired Weyl points on $Y_1-\Gamma-Y$ must connect to the nodal plane duos.
 
The topological partners of the fourfold point on R$_1-$U$-$R are accidental Weyl points located somewhere on the effective mirror planes or in the bulk of the BZ. Since the fourfold point of Hamiltonian~\eqref{ham_model_SG19_27} has topological charge $\nu = -2$, we expect two Fermi arcs that connect the fourfold point to accidental Weyl points in the bulk. Again, this agrees with our calculations of the surface DOS for the tight-binding model~\eqref{ham_model_SG19_27}. Extended Data Fig.~\ref{EDI:Theory:Fig4}b shows two Fermi arcs emanating from the projected fourfold point on R$_1-$U$-$R and connecting to accidental Weyl points in the bulk. 


\newpage
\section{Catalogue of space groups with symmetry-enforced nodal planes}
\label{classification_nodal_surfaces}

The topological analysis of SG 198 and its magnetic subgroups, presented in the previous sections,  can be extended to 254 of the 1651 magnetic SGs (Shubnikov groups). The 1651 magnetic space groups describe all possible symmetries of spin-full, periodic systems in three dimensions with and without time-reversal symmetry~\cite{miller_love_book}. Here, as in the entire article, we use for the magnetic space groups  the BNS setting, rather than the OG setting. This has the advantage that in the BNS setting the magnetic unit cell is used, such that the symmetry operations capture the unit cell size more clearly than in the OG setting. In Sec.~\ref{sub_sec_symm_enforced_NPs} we show that 254 magnetic SGs have the necessary symmetries that enforce the existence of nodal planes. In Sec.~\ref{sub_sec_top_nodal_planes} we give conditions on the topological charges of the nodal planes in these 254 SGs. We find that 33 SGs have nodal planes whose topological charges are enforced to be nonzero by symmetry alone. 

Using the database MAGNDATA~\cite{magndata_paper_database}, we find a number of magnetic materials that crystalize in one of these 33 SGs. We note that for these materials to have nodal planes at the Fermi level (leading to topological Fermi surface protectorates) the only other condition needed is that the material is metallic with Fermi surfaces extending across the BZ boundary. That is, no fine tuning is needed for the Fermi level to cross parts of the nodal planes, in stark contrast to, e.g., Weyl or Dirac semimetals. Moreover, even a considerable detuning of the chemical potential or change in the band structure (by, e.g., impurity doping or other deformations) are generally not able to shift the nodal plane away from the Fermi level or remove it altogether. In turn, the topological character of the Fermi surface (i.e., the topological protectorate) remains robust under perturbations, as long as the magnetic screw rotation symmetry is not broken.

\subsection{Symmetry-enforced nodal planes}  \label{sub_sec_symm_enforced_NPs}

The conditions that need to be satisfied for the existence of symmetry-enforced nodal planes are:
\begin{compactitem}
\item Presence of at least one magnetic screw rotation  $\theta \, \tilde{C}_2^q$  (with $q \in \{ x,y,z \}$), i.e., a screw rotation combined with time-reversal $\theta$. 
\item Presence of a reciprocal lattice vector that relates $k_q = -\pi$ to $k_q = \pi$ in the first BZ.
\item Absence of $PT$-symmetry, i.e, the absence of any symmetry that can be written as the combination of time-reversal $\theta$ with inversion and possibly some translation part.
\end{compactitem}
These three conditions for symmetry-enforced nodal planes are both sufficient and necessary. Conditions 1 and 2 ensure that there exist Kramers pairs on the BZ boundary $k_q = \pi$. We note that condition 2 is required, since some SGs with non-primitive unit cells (e.g., body-centered or base-centered unit cells), may not contain a reciprocal lattice vector that relates $k_q = -\pi$ to $k_q = \pi$, such that the plane $k_q = \pi$ is not left invariant by the effective mirror symmetry  $\theta \, \tilde{C}_2^q$ (see the discussion of  Fig.~\ref{fig:1} in the main text). This excludes any SG containing a noninteger translation parallel to the rotation axis of $\theta \, \tilde{C}_2^q$ (within the BNS setting). The third condition is necessary, since the presence of a  $PT$-symmetry leads to Kramers pairs in the entire BZ, such that all bands in the entire BZ are Kramers degenerate. 

As an aside, we note that there exist also accidental nodal planes which are not symmetry-enforced and only protected by a topological invariant, but not by a non-symmorphic symmetry~\cite{turker_moroz_PRB_18,PhysRevB.96.155105}. 
The perhaps most natural examples of accidental nodal planes are Bogoliubov Fermi surfaces in unconventional superconductors~\cite{PhysRevB.98.224509}. Classifications of these accidental nodal planes have been discussed in, e.g., Refs.~\onlinecite{chiu_RMP_16,PhysRevLett.116.156402}.

We retrieve the properties of the magnetic SGs from the ISO-MAG database~\cite{iso_mag_database} and find that among the 1651 magnetic SGs 254 satisfy the above three criteria. These are listed in Extended Data Table~\ref{EDI:Theory:tab1}, which contains the three blocks: 
\begin{compactitem}
\item
32 grey SGs with an onsite time-reversal symmetry of the form $\theta = i \sigma_y \mathcal{K}$, describing nonmagnetic or paramagnetic materials.
\item 94 black-white SGs without time-reversal symmetry, describing ferro- or ferri-magnetic materials. 
\item 129 black-white SGs with a symmetry that combines onsite time-reversal with a translation, describing antiferromagnetic materials. 
\end{compactitem} 
We note that this catalogue of magnetic space groups with nodal planes  is consistent with the irreducible co-representations that have recently been posted on the Bilbao Crystallographic Server~\cite{elcoro2020magnetic}. That is, all space groups with symmetry-enforced nodal planes have two-dimensional co-representations of the little groups on one (or more) BZ boundary planes. 
We note, however, that the topology of the nodal planes cannot be determined from the co-representations alone, but must be inferred from the global topology in the entire BZ, i.e., from the chirality and multiplicity of all band crossings in the entire BZ, as discussed in the section below.

\subsection{Topological charge of nodal planes} \label{sub_sec_top_nodal_planes}

Having found all magnetic SGs with symmetry-enforced nodal planes, we now study the topological charges of these nodal planes. In general, the symmetries of the SGs put constraints on the possible values of the topological charges carried by the nodal planes.  A necessary condition for a nonzero topological charge is that the SG is chiral, i.e., that it does not contain any inversion, mirror or roto-inversion symmetries. This is because mirror or (roto-)inversion symmetries relate Weyl points with opposite chiralities to each other, thereby guaranteeing that the sum over all topological charges of all Weyl points in the interior (bulk) of the BZ is zero. In other words, there cannot exist a single Weyl point in the bulk whose chirality is compensated by a nonzero topological charge of the nodal planes. In Extended Data Table~\ref{EDI:Theory:tab1} we have marked all 101 chiral SGs   with ``[t]" or ``[\textbf{T}]". The nodal planes of these SGs can have nonzero topological charges, while those of   all other SGs  are ensured to be zero by symmetry. 

Remarkably, we find that among the 101 chiral SGs there are 33 SGs with nodal planes whose topological charges are guaranteed to be nonzero by symmetry.   These are labelled by ``[\textbf{T}]" in Extended Data Table~\ref{EDI:Theory:tab1}. We now discuss each of these 33 SGs, first focusing on those with nodal plane trios and then on those with nodal plane duos. (We note that for the chiral SGs labelled by ``[t]", there could, in principle, also exist symmetry conditions that enforce nonzero topological charges. But we have not yet been able to infer these.)


\subsubsection{Trios of nodal planes}

We first focus on nodal plane trios, whose existence is enforced by three different magnetic screw rotations $\theta \, \tilde{C}_2^q$. In this case there are three mutually intersecting nodal planes in the BZ torus, i.e., on all boundaries of the BZ at $k_x=\pi$, $k_y=\pi$, and $k_z=\pi$. Since $\theta \, \tilde{C}_2^q$ act like effective mirror symmetries, all Weyl points in the interior of the BZ (with the exception of $\Gamma$) must have a partner with same chirality at a symmetry-related position in the BZ (see discussion in Sec.~\ref{PM_phase_Weyl_points}). As a consequence, the topological charges of all Weyl points in the interior of the BZ (but away from $\Gamma$) must add up to an even number. If, due to time-reversal symmetry, there exists a single Weyl point at the time-reversal invariant momentum $\Gamma$, its chirality can only be compensated by the nodal plane trio (which contains all other time-reversal invariant momenta).

Using these arguments, we find that there are six grey SGs with nodal plane trios whose nonzero topological charge is compensated by a single Weyl point at $\Gamma$, namely\cite{2019_PRB_Yuxin}: 19.26, 92.112, 96.144, 198.10, 212.60, and 213.64. These grey SGs describe nonmagnetic materials with an onsite time-reversal symmetry of the form $\theta = i \sigma_y \mathcal{K}$. (We note that the first number of these grey SGs gives the corresponding nonmagnetic SG number.) The SG symmetries of ferro- or ferri-magnets, on the other hand, cannot give rise to nodal plane trios, since the finite magnetization of ferro- or ferri-magnets breaks at least one of the three required magnetic screw rotations  $\theta \, \tilde{C}_2^q$. Moreover, these ferro-/ferri-magnetic SGs do not contain an onsite time-reversal  symmetry $\theta$ that could pin a Weyl point at $\Gamma$. In contrast, chiral SGs describing antiferromagnets can possess three magnetic screw rotations and a combined symmetry consisting of $\theta$ with a translation, which enforces the existence of a single Weyl point a $\Gamma$. If so, then the nodal plane trio of the antiferromagnets must have a nonzero topological charge, wich is the case in the 13 SGs: 16.6, 17.14, 18.21,  19.28, 89.94, 90.100, 91.109, 93.126, 94.132, 95.141, 195.3, 207.43, and 208.47. 

So, in total we have found 19 chiral SGs with nodal plane trios that have a nonzero topological charge. For all of these 19 SGs we have constructed generic tight-binding models with randomly chosen hopping parameters, including up to fifth-nearest-neighbor hopping terms. For these generic tight-binding models we have computed the topological charges of the nodal plane trio and have found that it is always nonzero (i.e., and odd number), independent of the chosen hopping parameters. This provides a numerical confirmation of the above symmetry arguments.  Using the database MAGNDATA~\cite{magndata_paper_database} we identify two antiferromagnets with nodal plane trios of nonzero topological charge: CoNb$_2$O$_6$ (SG 19.28) and Ba(TiO)Cu$_4$(PO$_4$)$_4$ (SG 94.132). We note that materials with nodal plane trios, as opposed to those with nodal plane duos,  in general do not exhibit surface Fermi arcs that emanate from the nodal planes~\cite{2019_PRB_Yuxin}.


\subsubsection{Duos of nodal planes}

Second, we study nodal plane duos whose existence is enforced by two different magnetic screw rotations $\theta \, \tilde{C}^q_2$. These nodal plane duos are located at two of the three BZ boundaries $k_x=\pi$, $k_y=\pi$, or $k_z=\pi$, such that  all bands are doubly degenerate at two BZ boundaries, but non-degenerate at the third BZ boundary. The two screw rotations $\theta \, \tilde{C}^q_2$ represent effective mirror symmetries that leave two BZ boundaries invariant, as well as two planes in the interior of the BZ (i.e., two of the three planes $k_x=0$, $k_y=0$, and $k_z=0$), see discussion in Sec.~\ref{top_charge_nodal_plane_duo}. Due to these effective mirror symmetries, any Weyl point in the interior of the BZ (but away from the intersection of the mirror planes) must have multiplicity two or four. That is, these Weyl points come in pairs or quartets with the same topological charge. The exception to this are Weyl points located on the main axis $k_\alpha \in \{ k_x, k_y, k_z\}$ that is left invariant by both  effective mirror symmetries. These Weyl points on $k_\alpha$ have in general multiplicity one. Now, in order to generalize the argument of Sec.~\ref{top_charge_nodal_plane_duo}, we need a nonsymmrophic symmetry that enforces an odd number of Weyl points on $k_\alpha$.

Below we show that also even, but nonzero, charged nodal planes exist in the presence of four-fold rotation symmetry.  Several ways to attribute topological charges to nodal planes have to be distinguished. 

Consider the on-site time-reversal as broken and a nonsymmorphic rotation axis parallel to the nodal planes. This requires an odd number of band exchanges on the rotation axis. The nodal plane must be topological for candidate space groups fulfilling these and the general conditions subjected to Extended Data Table~\ref{EDI:Theory:tab1}. Such space groups are 17.15, 18.22, 19.27, 91.110, 92.114, 92.115, 92.116, 95.142, 96.146, 96.147, and 96.148. This list includes systems with nonzero total magnetic moment as well as antiferromagnetic systems, where the translation associated to time-reversal symmetry fulfills Kramers theorem only at 4 out of 8 time-reversal invariant momenta. Besides MnSi there are other metallic material realizations of above space groups:   Cu$_3$Mo$_2$O$_9$ (SG 19.27),  Nd$_5$Si$_4$ (SG 92.114), AgNiO$_2$ (SG 18.22), and CoNb$_3$S$_6$ (SG 18.22) as well as the likely nonmetallic magnetic phases of TbFeO$_3$ (SG 19.27) and BaCrF$_5$ (SG 19.27). 

So far paramagnetic phases have only been discussed with trios of nodal planes. We like to point out that time-reversal may pair the bands at TRIMs such that their chiralities add up, yet cannot be compensated by accidental crossing points within the Brillouin zone. To be more specific consider a space group containing a four-fold screw rotation $\tilde{C}^z_4 \equiv 4_{001}(0,\:0,\:1/2)$ as well as time-reversal $\mathcal{\theta}$ additional to $\theta\tilde{C}^x_2$ and $\theta\tilde{C}^y_2$. Here, for the simplest case the Weyl points at $\Gamma$ and Z are enforced to have the same chirality. The total chirality contribution $\nu_{\Gamma, \text{Z}} = \pm 2$ cannot be compensated by any Weyl point away from the four-fold rotation axis, as its minimal multiplicity is 4. We show that this behavior cannot be avoided with additional crossings on the line $(0,\,0,\,k_z )$ and consider for clarity only for the lowest band. Central to the argument is the relation between the chirality of a band crossing and the phase change of the corresponding $\tilde{C}^z_4$ eigenvalues \cite{tsirkin_vanderbilt_PRB_17}. On the $\tilde{C}^z_4$ rotation axis accidental crossings can only be introduced in pairs due to time-reversal symmetry, which carry therefore the same chirality and thus give the same phase change for the symmetry eigenvalue of the lowest band. 

Following the lowest band from $(0,\,0,\,-\pi )$ to $(0,\,0,\,\pi )$ the total phase accumulated must be equal to $\pi \mod 2 \pi$, which is required by the nonsymmorphic winding of the eigenvalues of $\tilde{C}^z_4$. Band crossings on the rotation axis away from the time-reversal invariant momenta change the eigenvalue of the lowest band by a phase of $\pm \pi/2$ or $\pi$. The latter changes the total Chern number of crossings on the line by $4$, because each crossing and its time-reversal partner carry a chirality of 2. Therefore they cannot cancel the phase and the chirality of the enforced crossings at $\Gamma$ and Z. The second possibility is adding two time-reversal related Weyl points, which each change the phase for the lowest band by $\pm \pi/2$. 

Note that the Weyl points at the TRIMs for the given symmetry also change the phase by $\pm \pi/2$, albeit for them the signs are not related by time-reversal symmetry. Since the summed phase changes must add up to $\pm\pi$, when traversing the BZ along $(0,\,0,\,-\pi )$ to $(0,\,0,\,\pi )$, the Weyl points at $\Gamma$ and Z must necessarily have opposite phase changes. To be specific if there are two accidental crossings on the rotation axis, their  chiralities are for example both $\nu_{\text{acc}} = 1$ and the chiralities at the TRIMs cancel, i.e., $\nu_{\Gamma} = \pm 1 $ and $\nu_{\text{Z}} = \mp 1 $. Should there be no accidental crossings their chiralities are the same, $\nu_{\Gamma} = \nu_{\text{Z}} $, as they must accumulate the necessary phase.

We conclude that the accidental crossings on the rotation axis through $\Gamma$ and Z must change the chirality by 4 as would any accidental crossings away from it. Therefore the Chern number in the interior of the BZ may only be compensated by the nodal plane carrying an even, yet non-zero topological charge. The required symmetries are present in SG 94.128 extending the number of paramagnetic phases hosting necessarily topological nodal surfaces. For this space group it is possible to obverse the Fermi arcs emerging from a nodal plane without a magnetic field or otherwise broken time-reversal symmetry. Also the SG 93.125, which is anti-ferromagnetic, shows the discussed features. In our generic tight-binding models we find, as expected, even, non-zero Chern numbers for the nodal planes at any band index. 

It is actually possible that the number of crossings on the nonsymmorphic rotation axis parallel to the nodal planes is odd, whereas the Chern number of the nodal plane is even. Consider the ferromagnetic SG 94.130 it contains the 4-fold screw rotation $\tilde{C}^z_{4} \equiv 4_{001}(1/2,\:1/2,\:1/2)$ but no other symmetries that lead to enforced crossings on the rotation axis $(0,\,0,\,k_z)$. The 4 eigenvalues of $\tilde{C}^z_{4}$ divide into two pairs that must exchange if $k_z$ is varied from $k_z= -\pi$ to $k_z = \pi$. Therefore the number of crossings within a pair is odd and between pairs it must be even. Each pair of $\tilde{C}^z_{4}$ eigenvalues differs by a phase of $\pi$. Thus all Weyl points on the rotation axis have charge $\nu = \pm 2$ within an eigenvalue pair \cite{tsirkin_vanderbilt_PRB_17}, whereas crossings between pairs have charge $\nu = +1$ and occur with a second crossing of $\nu = -1$. Due to the 4-fold rotation symmetry Weyl points away from the axis occur in sets of four such that the nodal plane necessarily carries again an even, nonzero charge.

The cylindrical topological nodal planes with an even Chern number exhibit consequentially an even number of Fermi arcs for a suitably chosen termination.

\newpage
\section{Technical aspects of the analysis}
\label{SI_sec_experimental_details}
In the following, we elaborate on some of the technical details of the data analysis presented in a more compact form in the Methods section.

\subsection{Evaluation of experimental dHvA frequencies}

The dHvA spectra of MnSi in the field-polarized regime exhibit a number of characteristics that require special attention both in the data pretreatment ("background" removal) and the FFT windowing. These characteristics comprise (i) a very large range of dHvA amplitudes spanning several orders of magnitude, (ii) different onset fields of oscillations and (iii) frequency regimes where many dHvA frequencies are very closely spaced.

{\bf Background removal and filtering.} In order to obtain only the oscillatory part of $\tau$, the well-established techniques of subtracting either a moving average over a suitable field interval or a low-order polynomial fit have been employed. It is important to note here, that the exact choice made does not change the frequencies contained in the signal. However, it can alter our ability to identify dHvA frequencies in the FFT spectra. As an example, consider an imperfect subtraction of the semi-static torque offset due to the magnetic anisotropy that is apparent in Fig.\,\ref{fig:2}a. In the FFT, this would result in a very large signal centered at $f=0$. In this scenario, the dHvA peaks at $f>0$ would constitute only small wiggles of a decreasing signal by the $f=0$ contribution, making it difficult to identify them correctly. Similarly, the simultaneous presence of very large-amplitude dHvA oscillations with low frequencies and very small-amplitude dHvA oscillations with high frequencies can lead to an effective masking of the small-amplitude components. 

The method of subtracting a moving average from the data was found to give superior results if compared to the method of subtracting low-order polynomial fits. This can be understood by considering that the moving average follows low-frequency large-amplitude oscillations to an extent controlled by the size of the averaging interval. Subtracting a moving average thus acts as a smooth high-pass filter with a corner frequency scaling inversely with the size of the averaging interval. Thus, choosing a suitable averaging interval allows us to efficiently visualize the dHvA frequencies in a given regime.

{\bf FFT windows.} Both the size and the shape of FFT windows strongly influence the appearance of the FFT spectra. The window sizes have been discussed in the methods section and are mainly fixed by experimental parameters, i.e., by the maximum fields of the superconducting magnets on the high-field side and the onset fields of the quantum oscillations on the low-field side. In contrast, the shape of the FFT windows can - in principle - be chosen freely and has a strong influence on spectral leakage and FFT peak width. A large body of different FFT windows that are optimized for different purposes exist in the literature. The main parameters to be optimized for our purposes are (i) the FFT peak width and corresponding ability to individually resolve closely spaced peaks, (ii) the side lobe suppression influencing our ability to distinguish side lobes from genuine dHvA frequencies, and (iii) the overall spectral power influencing the signal-to-noise ratio. 

In our analysis, several different FFT window functions were tested in order to distinguish side lobes from genuine dHvA orbits. The main representatives we used included Rectangular, Hann, Hamming, Blackman-Harris and Tukey windows with different parameters $\alpha$.

Several windows yield excellent side lobe suppression as, e.g., the Blackman-Harris window. Such windows were used in the analysis in order to discriminate smaller peaks in close vicinity to large peaks from side lobes. However, adjacent frequency peaks in regimes III and V strongly overlap when using these windows. In addition, the spectral power is significantly lower, which is in particular detrimental to small peaks. Windows designed for small peak width in combination with reasonable side lobe suppression are, e.g., Hann and Tukey windows. However, the spectral resolution turned out to be not sufficient for the dHvA spectra under investigation. As a consequence, all FFT spectra shown in this work were performed with a Rectangular window in order to maximize the ability to resolve adjacent FFT peaks at the expense of the suppression of side lobes. However, use of different windowing functions played an important role in the identification of side lobes visible in these spectra.

{\bf FFT of synthesized quantum oscillations.} As an additional test to identify spurious effects due to the finite field range of our data we created FFT spectra with identical parameters using synthesized quantum oscillatory data generated by means of the LK formalism. That is, quantum oscillations were calculated using the LK formula with the measured frequencies and masses as input and the corresponding FFTs were compared with the FFTs of the experimental data. This method provides an additional impression of the side lobes, since it is known a priori which peaks correspond to actual dHvA orbits in the synthesized spectra. 

We finally also used the angular evolution of the spectra to discriminate between dHvA orbits and spurious signals. For instance, close-by dHvA frequencies may display a very different angular evolution and even cross as a function of angle, while a side lobe tends to track the main peak as a function of angle.

\subsection{Effective quasiparticle mass analysis}

Following convention, the effective masses were inferred from the $T$-dependence of the FFT intensity using the LK-formalism (Eq.\,\eqref{eq:LK-fit} in the Methods section), which is a function of $T/B$. The effective masses are hence subject to statistical uncertainties in the sample temperature and FFT intensity (e.g. due to numerical uncertainties), where we assume that uncertainties in the actual field values are vanishingly small. They are, further, subject to systematic variations due to the magnetic field range, the window of reciprocal fields analyzed, as well as the magnetic field dependence of the effective mass itself.

Starting with the error bars of the carrier masses reported in Extended Data Table\,\ref{EDI:Exp:tab2}, they represent the statistical uncertainty of the mass inferred from the Lifshitz-Kosevich fits of the temperature dependence observed experimentally with respect to the average magnetic field $B_{average}$ of the FFT window. They are thus well-defined, reflecting uncertainties in the sample temperature and FFT amplitude.

The systematic uncertainties are assumed to be dominated by FFT window boundaries as follows: The Lifshitz-Kosevich analysis yields accurate results, when the size of the FFT window in $1/B$ is small compared to $1/B_{average}$. It is thus important to assess the effect of the window size on the results. We tested our results for (i) different window sizes centered at the same $1/B_{average}$ and, (ii), smaller FFT windows centered at different $1/B_{average}$. This gives, in principle, an account of both, the dependence on the window size and a possible $B$-dependence of the quasiparticle masses themselves. 

For this type of analysis, we needed to narrow the FFT windows in comparison to the full-range rectangular window required to resolve most of the narrowly spaced frequency branches. In turn, a comprehensive analysis of the systematic dependence of the mass values on the window size was only possible for some branches. In particular, such an analysis was not possible for the low-frequency dHvA branches up to 350\,T which required the largest windows. Instead, the analysis was possible for FFT peaks that are well isolated on the frequency axis, such as the orbits $\kappa$ and $M$. While narrowing the window did have an effect on the values of the mass of $\kappa$ and $M$, it resulted also in much larger error bars. Within these larger error bars, however, there was no significant trend.

Keeping these limitations in mind, it has long been known that the quasiparticles in MnSi acquire a large part of their mass from coupling to the spectrum of spin fluctuations\cite{1988_Lonzarich_JMMM}. This is reflected in the magnetic field dependence of the Sommerfeld coefficient, which decreases by roughly 20\% up to 14\,T\cite{Bauer2010}. Such a reduction is known as quenching of spin fluctuations (a mode stiffening under applied fields here without discernible spin-wave contributions) that is well-established in many d- and f-electron compounds, notably the class of heavy fermion materials. For the comparison of the effective masses, the Sommerfeld coefficient close to the average field value was used.


\newpage
\section{Comprehensive Fermi surface determination}
\label{SI_sec_fs_determination}

In this section, details of the assignment of the experimental dHvA branches to the corresponding extremal FS cross sections are presented, using the criteria described in the Methods section.

\subsection{Sheet 2}
We start by considering the dHvA branches $\kappa_{1,2,3}$ and their harmonics $2\kappa_{1,2}$ and  $3\kappa_{2}$ in regime IV, since their assignment is most obvious. In Extended Data Fig.~\ref{EDI:Exp:Fig9}\,a1 the experimental angular dependence of the FFT amplitude is shown in the form of an intensity map, with peaks marked by crosses. The data exhibit an angular dispersion with maxima along $\left[ \bar 1 \bar 1 0 \right]$ and a threefold splitting of the branch close to the minima along $\left[ 010 \right]$. This behavior is characteristic of sheet 2 as highlighted in Extended Data Fig.~\ref{EDI:Exp:Fig9}\,a2 and a3 showing the predicted torque signature and the FS sheet with orbits, respectively. For $\mathbf{B}$ along edge directions of the cuboid FS sheet a single maximal dHvA orbit traverses the sheet diagonally. Under field rotation towards $\left[ 010 \right]$ this orbit splits into one minimal $\Gamma$-centered and two maximal $\Gamma$-Y-centered orbits. We may thus assign $\kappa_1$ to the orbit $2\Gamma$ and $\kappa_{2,3}$ to $2\Gamma Y(1,2)$. Note that the torque signal vanishes around $\varphi=90^{\circ}$ and $\varphi=135^{\circ}$ since the Lifshitz-Kosevich torque term $\frac{1}{f} \frac{df}{d\varphi}$ vanishes here. 

The high signal strength arises partly from the low curvature of this sheet together with the comparatively light effective mass. While the overall experimental behavior (crosses) is reproduced by the DFT calculation in Extended Data Fig.~\ref{EDI:Exp:Fig9}\,a2, the calculation does not predict the dimensions of the pocket accurately. In order to create a quantitatively accurate picture of the experimental FS we introduce a small rigid upward shift of band 2 by 20~meV, resulting in a good match (Extended Data Fig.~\ref{EDI:Exp:Fig9}\,a4).

This connects the $\kappa$-orbits with FS sheet 2, representing a majority hole sheet that is slightly larger in experiment than predicted by DFT as shown in Extended Data Fig.~\ref{EDI:Exp:Fig9}\,a5. The higher frequency FFT peaks in regime IV are the first harmonic components of the $\kappa$-branches (i.e., the term with $p=2$ in Eq.~\eqref{eq:MoscPara} in the Methods section). They do not correspond to additional extremal cross sections. This can be inferred from the angular dispersion, which tracks the doubled fundamental frequency over the whole angular regime together with the apparent doubling of the effective mass within the error bars. The latter is due to the $p=2$ term in Eq.~\eqref{eq:LK-fit} in the Methods section.

\newpage
\subsection{Sheet 1}
Branch $\alpha$ in regime V with $f\approx 7$~T exhibits are very large torque amplitude that dominates the raw data shown in Fig.~\ref{fig:2}\,a. The effective mass, $m^*=0.4m_e$ in Fig.~\ref{fig:2}\,d, is the smallest of all frequencies observed. There are in total $7$ band extrema within $\pm 50$~meV of $E_{\text{F}}$ that could give rise to such an ultra-low frequency. These belong to bands 7, 8, 9 and 10 around the $R$ point and 1, 2 and $3$ around the $\Gamma$ point. However, all except for two of the bands around $\Gamma$ exhibit DFT band masses that exceed the measured value of $m^*=0.4m_e$. 

Given that the experimental masses we observe are enhanced with respect to the band masses, only two choices remain, notably the light bands 1 and 3 around $\Gamma$ (magenta and red bands in Fig.~\ref{fig:1}\,e). Since we have unambiguously identified band $2$ (cyan) with the $\kappa$-branches above, we know that the minimum of band 3 (red) must be located above $E_{\text{F}}$. This is illustrated in Extended Data Fig.~\ref{EDI:Exp:Fig9}\,b1, where we show the dispersions of bands 1, 2 and 3 with and without SOC (transparent) together with the Fermi level (black) matching the experimental frequency. We can thus assign branch $\alpha$ to orbit $1\Gamma$ on the small light hole pocket of band 1, acquiring its small mass from the SOC-induced gap opening at $\Gamma$. 

This assignment is corroborated by the shift downward of $f(B)$ with increasing $B$ observed experimentally, which is expected for a majority hole sheet but would be reversed for the majority electrons of band 3. 

The FS sheet $1$ as calculated (large) and matched to the experiment (small) is shown in Extended Data Fig~\ref{EDI:Exp:Fig9}\,b2. A shift of $+27$meV of $E_{\text{F}}$ (or equivalently $-27$meV of the band) is required for a match.

\newpage
\subsection{Sheet 3 \& 4}
Neglecting SOC, the minority hole sheet 3 and majority hole sheet 4 are intersecting cuboid- and octaeder-shaped $\Gamma$-centered pockets with corners oriented in $\langle111\rangle$ and $\langle100\rangle]$ equivalent cubic directions, respectively. Including SOC the lines of intersection gap out, leading to an inner (red) and an outer (green) sheet of mixed spin character (Extended Data Fig.~\ref{EDI:Exp:Fig9}\,c1). 

Orbits on both sheets are connected by up to eight magnetic breakdown junctions $j_1$ to $j_8$ as depicted in Extended Data Fig.~\ref{EDI:Exp:Fig9}\,c2, where the normal of the rotation plane is pointing in vertical direction. We denote the areas of the orbit sections by $A_{in}$, $\Delta_1$, $\Delta_2$, $\Delta_2^*$, and $A_{out}$. Only orbits with an even number of breakdowns satisfy the conditions of being closed after one cycle. For 8 junctions, there are $2^8=256$ configurations, half of which have an even number of breakdowns. For each configuration of junctions, two orbits with the charge carrier travelling on an inner or an outer path on a given section have to be considered. This results in a total of 256 breakdown orbits. Calculation of the angular-dependent breakdown probabilities using the Chambers formula and of the orbit cross-sectional areas from the as-calculated DFT results yields the dispersion curves shown in Extended Data Fig.~\ref{EDI:Exp:Fig9}\,c3. 

The structure of the breakdown branches can be understood as follows. The lowest branch (red) encloses only $A_{in}$. The highest branch (green) encloses $A_{out}=A_{in}+4\Delta_1+4\Delta_2$. Both correspond to no breakdown. The set of the five lowest branches (in brackets) enclose in increasing order: $A_{in}$, $A_{in}+1\Delta_1$, $A_{in}+2\Delta_1$, $A_{in}+3\Delta_1$, $A_{in}+4\Delta_1$. The next higher set contains: $A_{in}+1\Delta_2$, $A_{in}+1\Delta_2+1\Delta_1$, $A_{in}+1\Delta_2+2\Delta_1$, etc.

Branches that are maximal at $\varphi=90^{\circ}$ enclose $\Delta_2^*$ while branches that are minimal at $\varphi=90^{\circ}$ enclose $\Delta_2$. There are five of these sets, ranging from $A_{in}$ to $A_{in}+4\Delta_2$. The symbol size in Extended Data Fig.~\ref{EDI:Exp:Fig9}\,c3 reflects the relative probability of a breakdown orbit and the line thickness reflects the degeneracy of a set of branches. The torque signal strength is not considered in this plot. The breakdown probabilities differ strongly between branches and exhibit strong angular dispersion for a given branch.

A comparison with experiment is shown in Fig.~\ref{fig:3} in the main text, displaying the simulated torque signals including rigid band shifts. The lowest branch ($A_{in}$) may be identified particularly well around $\varphi=135^{\circ}$ in Fig.~\ref{fig:3}. This fixes $A_{in}=2.7$~kT. The approximate size of $\Delta_1$ is fixed via the spacing of the breakdown orbits of the lowest set, giving $\Delta_1 \approx 0.08$~kT. $\Delta_2$ is fixed via the spacing between different sets, giving $\Delta_2 \approx 0.28$~kT and $\Delta_2^* \approx 0.275$~kT. These values represent a good match with experiment as shown in Fig.~\ref{fig:3} and Extended Data Fig.~\ref{EDI:Exp:Fig6}g.

However, the overall slope of those orbits that are maximal at $\varphi=90^{\circ}$ is too low. From this observation we infer that the lobes $\Delta_2^*$ are larger experimentally than calculated. Taking into account that the observed values for $\Delta_1$ are also larger than calculated, we conclude that sheet 4 exhibits lobes that are both more elongated and narrower than predicted in DFT. We note that we cannot mimmick this effect accurately by different \emph{a posteriori} rigid shifts of the individual bands, due to the avoided crossing induced by SOC. However, band shifts of $-9$~meV come closest to a satisfactory match. 

We note that the five sets of breakdown branches defined above exhibit a systematic hierarchy also in their cyclotron masses, which can be used to further corroborate the assignment. This is understood most easily by looking at Extended Data Fig.~\ref{EDI:Exp:Fig9}\,c2. When we neglect the SOC-induced avoided crossings at $j_1$ to $j_8$ for a moment, only two orbits would arise as shown in the inset: The dark red orbit stemming from the cuboid sheet 3 and the dark green orbit stemming from the octaeder-shaped sheet 4. These two have band masses of $1.5m_e$ and $3.2m_e$, respectively. Including SOC, avoided crossings at $j_1-j_8$ lead to the five sets of breakdown branches. Now, the \emph{highest} branch of set 1 corresponds to the non-SOC sheet 3, while the \emph{lowest} branch of set 5 corresponds to the non-SOC sheet 4. They have – of all branches -  the lowest and highest masses, respectively. The masses of the other branches are in between these two extremes, since the carriers travel partly on both non-SOC trajectories.

It follows, that the lowest branch in the first set has the highest mass within that set, since the carriers travel on the heavier sheet at all four corners, i.e., between $j_1-j_2$ and $j_3-j_4$ etc. The next higher branch of the same set is bit lighter, since it incorporates only three of the corners etc. The effect is, however, small in the calculated band masses. The same argument applies to all five sets. Thus, the masses are expected to increase overall from the lowest frequency set 1 to the highest frequency set 5, but decrease with increasing frequency \emph{within} each set.

We conclude that the vast majority of dHvA branches observed between the red and the green curve is due to the magnetic breakdown network of sheet 3 and sheet 4 (translucent yellow in Fig.~\ref{fig:3}). Branch $\rho$ is assigned to $3\Gamma$, $H$ is assigned to $4\Gamma$.  The total of 254 partly multiply degenerate orbits between these branches cannot all be resolved individually. Some of the stronger branches, due to both their probability and their degeneracy, are labelled $3\Gamma 4\Gamma (1-16)$ as tabulated in Extended Data Table~\ref{EDI:Exp:tab2}. The main orbits we associated with set 1 are $\rho, \sigma, \tau$ and $\upsilon$. Frequencies marked with a prime were observed in field windows from $11-16$\,T at $35$\,mK.

The assigned experimental branches match the effective mass evolution outlined above: Overall the masses increase from set one to set five from $\sim 10m_e$ to $\sim 16m_e$. However, inside the sets, a decrease of masses with increeasing frequency is observed within the experimental error bars, as can be seen for set 1 going from $10.9m_e$ ($\rho$) to $8.7m_e$ ($\upsilon$).

The overall fading of the observed signal strength with increasing frequency is also mainly due to the mass evolution from set 1 to set 5.

\newpage
\subsection{Sheets 5 \& 6}
The jungle-gym sheets 5 and 6 extend through the BZ surface and are thus degenerate on all nodal planes. In that sense, sheets 5 and 6 cannot be viewed as separate sheets, since the topological Fermi surface protectorates represent topological defects in the sheet structure. Rather, a charge carrier traveling on sheet 5 towards a nodal plane will smoothly cross over to sheet 6 and vice versa. The corresponding part of the FS is thus best viewed by doubling the BZ at the nodal plane and connecting "sheet 5" and "sheet 6" smoothly forming a joined sheet.

Sheet 5 and 6 give rise to two separate neck-type orbits, as discussed in the main text and shown in Extended Data Fig.~\ref{EDI:Exp:Fig9}\,d1. The neck orbit of sheet 5 (blue) $5\Gamma Y(1)$ can be readily assigned to $\xi_1$ based on a very good match in (i) frequency $f$, (ii) $f(\varphi)$ dispersion close to $f_0\cos \varphi $ expected for a neck with almost constant cross section, (iii) high signal strength due to the low curvature factor of the neck, and (iv) slight splitting into $\xi_1$ and $\xi_2$ due to a small warping of the neck that is imminent also in the DFT prediction.

Point (iv) is illustrated in more detail in Extended Data Fig.~\ref{EDI:Exp:Fig9}\,d2. 
The upper panel shows neck cross-sectional areas a (in kT) of sheet 5 (blue) and 6 (orange) vs $k\parallel B$ neglecting (dashed) and including (solid) SOC for $\varphi=90^{\circ}$ and $\varphi=180^{\circ}$. Neglecting SOC, two degenerate extremal orbits are present on the neck. Note that the crossing points between the orange and blue dashed lines do not correspond to extremal orbits. Including SOC, the degeneracy is lifted such that only one extremal cross section survives in our calculations.

The middle panel shows area $a$ vs $k_{\parallel}$ for sheet 5 for a series of field directions between $70^{\circ}-90^{\circ}$ and $160^{\circ}- 180^{\circ}$. The dashed line highlights the position of the single extremal area for fields around $90^{\circ}$. In contrast, around $180^{\circ}$, the neck is on the verge of developing a second minimum (highlighted by the gray shaded area). This can be seen more clearly in the lower panel, where the derivative $\frac{da}{dk_{\parallel}}$ is shown. Here, zero-crossings correspond to extremal orbits. The arrow highlights the region where $\frac{da}{dk_{\parallel}}$ is close to a saddle point around zero. This behavior highlights that a slightly stronger warping of the neck would lead to a splitting of the minimal neck orbit $5\Gamma Y(1)$ of sheet 5 into two minimal oribts $5\Gamma Y(1)$ and $5\Gamma Y(1)$ for $\varphi=180^{\circ}$ but not for $\varphi=90^{\circ}$. This corresponds to experiment. Note that, generically, such a splitting also implies the occurrence of a maximal orbit between the two. This maximal orbit would be however, quasi-degenerate with the second minimal orbit for low warping. The neck orbit $6\Gamma Y$ of sheet 6 is assigned to $\pi$ based on the excellent matching of criteria (i)-(iii).

The degenerate loop orbits $5U6U$ are always shared between sheets 5 and 6 for the rotation plane discussed here. They are assigned to $M_1$, excellently matching the absolute frequency and the $f(\varphi)$ dispersion. The sheets cut through four nodal planes for magnetic fields pointing along $[010]$, giving rise to the leaf-shaped orbits in Fig.~\ref{fig:4}\,b1 and b2 in the main text. When $\mathbf{B}$ is rotated away from $\left[ 010 \right]$ but remains in the $\left( 001 \right)$-plane, the former nodal-surface degeneracies at TP2 and TP4 turn into breakdown junctions, resulting in the two additional wedge-shaped orbits shown in Fig.~\ref{fig:4}\,b3 and b4. They are however - within the numerical uncertainty - all four quasi degenerate also for field directions in the plane that do not correspond to a high-symmetry direction. A connection of these four different orbits to a slight splitting of the experimental branch $M_1$ into $M_1$ and $M_2$ occurring around $\varphi=172.5^{\circ}$ but notably not around $\varphi=82.5^{\circ}$ (see Extended Data Fig.~\ref{EDI:Exp:Fig9}\,d3) can, however, not be ruled out completely. Another possible origin of $M_2$ from sheet 8 is discussed below. All orbits are matched simultaneously by a rigid band shift of $-4$~meV for both sheets 5 and 6.

The excellent agreement between the experimental data and the comprehensive theoretical assessment provides very strong evidence for topological nodal planes in MnSi. However, for the sake of completeness we wish to note that anticrossing and magnetic breakdown, in principle, could produce the same spectra under the strict precondition, that the tunneling amplitude is identical at all intersections and, more seriously, equal to 100\%, even under tilted magnetic field. As there is no theoretical justification we rule out such a very delicate situation: based on the DFT band structure we estimate that a FS splitting of $\sim 5$\,meV would already result in a reduction of the weight of the non-topological orbits by 50\%, which could easily be resolved in experiment. This corresponds to a $k$-space distance of about 1/500 of a reciprocal lattice vector, where the high resolution reflects the exponential dependence of the magnetic breakdown probability on the FS splitting.

\newpage
\subsection{Sheets 7 \& 8}
The properties of orbits on the pair of jungle-gym sheets 7 and 8 (Extended Data Fig.~\ref{EDI:Exp:Fig10}\,a1) are analogous to the orbits observed on sheets 5 and 6. The degenerate loop-type orbits are shared between the sheets in the same way as for sheets 5 and 6, with intersections at TP1 to TP4 as illustrated in Extended Data Fig.~\ref{EDI:Exp:Fig10}\,a2. For $\mathbf{B}$ along $[010]$, two exactly degenerate lentil-shaped orbits exist (upper two depictions in panel a3), while for field away from $[010]$ in the $\left( 001 \right)$-plane, two additional heart-shaped orbits are stabilized as illustrated in the lower two depictions in panel a3. These heart-shaped orbits evolve quasi-degenerately in frequency from the lentil-shaped orbits under rotation of the field direction. The frequency of the loop orbits on sheets 7 and 8 is much lower and the masses are much higher as compared to the orbits on sheets 5 and 6. 

The loop orbits are assigned to the experimental dHvA branch $\o$. This branch has a heavy effective quasiparticle mass and strong curvature which leads to a small signal amplitude. Moreover it has a very strong angular dispersion and exists only for a small angular range. Importantly, the existence of $\o$ and in particular the absence of topologically trivial branches well below and well above $\o$ prove the existence of the nodal planes in exactly the same way as discussed for the loop orbits of sheet 5 and sheet 6 discussed above. Notably, the hypothetical topologically trivial branch from sheet 8 would have a frequency of about 2~kT for field along $[010]$ and a band mass $m_b\sim 2m_e$ comparable to the prominent branch of sheet 5 around 2.4~kT (blue lines in Fig.~\ref{fig:3}a1). Its absence is thus unambiguous proof that the loop orbits of sheet 7 and 8 indeed also pierce through topological Fermi surface protectorates.

Sheet 7 is predicted to exhibit three distinct neck-orbits, two of which correspond to minimal cross sections. The third neck-orbit corresponds to a maximal cross section and is located between the other two along the $\Gamma-$Y$-\Gamma$-line (Extended Data Fig.~\ref{EDI:Exp:Fig10}\,a1). The degeneracy of the minimal cross sections is lifted by the distortion of the Fermi surface parallel to the field direction, which we call rubber canvas effect in the following. The maximal and the larger one of the minimal branches exhibit a characteristic eye-like shape between $5$ and $6$~kT in the $f(\varphi)$ dispersion shown in Fig.\ref{fig:3}. This feature is clearly observed experimentally in branches $K$ and $\Lambda$, which are consequently assigned to $7\Gamma Y(2)$ and $7Y$ respectively. We note that while one of the two branches can be brought into coincidence with experiment by a small rigid shift of $-4$~meV, this is not possible for both branches at the same time, as they originate from the same sheet. The observation of an eye-structure with smaller enclosing area in experiment thus means that either the warping of the neck of sheet 7 is smaller than predicted, or, the asymmetry induced by the rubber canvas effect is larger than predicted. The warping decreases and the rubber canvas effect increases with increasing SOC strength. It is thus likely that our calculations underestimate the effect of SOC for this sheet. Branch $7\Gamma Y(1)$ is assigned to $\Theta$.

Sheet 8 is predicted to exhibit one neck orbit only. The combination of a very heavy band mass of $4\,m_e$ and a very strong and narrow angular dispersion as depicted by the yellow line in Fig.~\ref{fig:3}\,a1 with the overall reduction of the strength of the magnetic torque near the $[010]$ high symmetry direction make the observation of this orbit extremely difficult. 
However, the slight splitting of $M_1$ into $M_2$ around $\varphi=172.5^{\circ}$ shown in Extended Data Fig.~\ref{EDI:Exp:Fig9}\,d3 (left panel) and the unexpected local minima in the torque amplitude of $M_1$ around  $\varphi=82.5^{\circ}$ and  $\varphi=100^{\circ}$ (right panel) coincide with the predicted crossing of the neck branch of sheet 8 and the loop branches of sheet 5 and 6 (Fig.~\ref{fig:3}). While the observed splitting could actually correspond to the observation of both branches, the local minima of the amplitude could arise from a long-period beating of the two nearly degenerate frequencies. Taking together the comprehensive theoretical assessment and the evidence for sheets 5, 6 and 7, this provides strong empirical evidence of the neck orbit of sheet 8 supporting our interpretation overall.

\newpage
\subsection{Sheet 9 \& 10}

Sheets 9 and 10 are located around the R point and may form TPs if they cross a BZ boundary representing a nodal plane.

The main branches in regime V with $f<0.35$~kT may be assigned to sheet 9 (except $\alpha$). The most important criterion for this assignment is the unique symmetry evident in the angular dispersion of the strength of the torque signal: For orbits $\beta$ and $\gamma$ and higher frequencies in regime V, $f(\varphi)$ is not extremal at $\varphi=90^{\circ}$, since here the torque $\tau \propto \frac{1}{f} \frac{df}{d\varphi}$ does not vanish but exhibits a large amplitude (Extended Data Fig.~\ref{EDI:Exp:Fig10}\,b4 and c4). By symmetry, only the pockets of sheet 9, located on the $\Gamma-$R lines, exhibits this property. All other sheets exhibit extrema in $f(\varphi)$ at $\varphi=90^{\circ}$ and thus a vanishing torque amplitude as may be seen for regimes I to IV in Fig.\,\ref{fig:3}. This can be corroborated by considering the data of the $\left( \bar 1 \bar 10\right)$ rotation plane in Extended Data Fig.~\ref{EDI:Exp:Fig6}\,g, showing that the corresponding branches exhibit their extrema at $\mathbf{B} \parallel \left[\bar 11 1 \right]$ as expected. In regime V in Extended Data Fig.~\ref{EDI:Exp:Fig6}\,g we have shown the calculated dHvA branches both including and neglecting SOC for this purpose.

Both, sheets (9,10) originate from flat bands that are very sensitive to the magnetic field direction, the effects of spin-orbit coupling and the the Fermi level. With increasing pocket size, the DFT band masses increase rapidly reaching $\sim 10m_e$ for orbits with $f\sim 1$~kT. Due to the extreme sensitivity to SOC and the direction of the spin quantization axis it is instructive to examine the calculated dHvA signatures of sheets 9 and 10 both in the absence and in the presence of SOC. 

We neglect SOC at first. Shifts on the scale of a few meV lead to significant changes in the properties and connectivity of sheet 9, i.e., from a singly-connected multi-spiked pocket at R to 8 separate pockets close to R but centered on the $\Gamma-$R lines, each of which resembles a bunch of three bananas. This situation is shown in Extended Data Fig.~\ref{EDI:Exp:Fig10}\,b1-b3 for three different viewpoints ($E_{\text{F}}$-shift of $-10$~meV). Including SOC the rubber canvas effect distorts sheet 9 strongly (Extended Data Fig.~\ref{EDI:Exp:Fig10}\,c1-3) and leads to a series of Lifshitz transitions as a function of field direction (see SI movie), giving rise to over $30$ different dHvA orbits. Notably, for a given field direction this results in general in small, intermediate and large banana bunches, depending on the orientation of the bunch with respect to the field.

Because sheet 9 can give rise to a variety of orbits with different numbers of copies and splittings with extreme sensitivity to SOC and the Fermi level, we do not discuss all possible orbits separately. Instead, we classify them into three different kinds:
\begin{compactitem}
	\item[(1)] Orbits that enclose one banana of one bunch.
	\item[(2)] Orbits enclosing two to three bananas of one bunch near the stem.
	\item[(3)] Orbits enclosing two or more bunches located in different octants of the BZ around the R-point. They only occur when banana bunches are connected for the given field direction. Also note that the degeneracy at the nodal plane needs to be lifted for this situation to occur, because otherwise sheet 9 would cross over smoothly to its partner sheet 10 at the nodal plane.
\end{compactitem}
We focus on rotations in the $\left(001\right)$-plane, where type (1) orbits can exhibit up to a $12$-fold splitting, type (2) orbits can exhibit up to a $4$-fold splitting and type (3) orbits exhibit two-fold or lower splitting.

Neglecting SOC and shifting $E_{\text{F}}$ by $-10$~meV would result in (Extended Data Fig.~\ref{EDI:Exp:Fig10}\,d):
\begin{compactitem}
	\item Orbits enclosing one banana of one bunch with $f\sim 40-70$~T
	\item Orbits enclosing 3 bananas of one bunch with $f\sim 190-240$~T
\end{compactitem}
Their existence would be, quite naturally, rather robust. The bunches in different octants of the BZ are never connected in this scenario and type (3) orbits as well as sheet 10 are absent.

Including SOC and shifting $E_{\text{F}}$ by $-11$~meV would result in (Extended Data Fig.~\ref{EDI:Exp:Fig10}c1-c4):
\begin{compactitem}
	\item Orbits enclosing one banana of one bunch exhibit frequencies around $f\sim 10$~T. Their extremal character is very fragile, as the bananas are close to a conical shape without extremal cross sections.
	\item Orbits enclosing 2 or 3 bananas of one (small) bunch exhibit frequencies around $f\sim 60-70$~T. Their existence is robust.
	\item Orbits enclosing 2 or 3 bananas of one (intermediate) bunch are in the regime $f\sim 120$~T. Their existence is only robust when they do not cross a nodal plane.
	\item Orbits enclosing 2 (large) bunches that are connected via sheet 10 exhibit frequencies around $f<500$~T. Further away from the cubic face directions, the bunches become disconnected due to the rubber canvas effect and the frequencies drop below $250$~T.
\end{compactitem}

The main branches $\beta,$ $\gamma$, $\delta$, $\epsilon$, $\mu$, are assigned to sheet 9 due to the symmetry of the frequency spectrum. Here, the overlapping frequency branches $\beta$ and $\gamma$ stem from either single bananas or coupled bananas within the same bunch.

The apparent splitting of these frequency branches into at least 3 distinct branches ($\gamma$, $\delta$, $\epsilon$) with similar $f(\varphi)$-dispersion may be attributed to the effects of SOC. For $\varphi = 90^{\circ}$, the 8 bunches of bananas split into 4 larger and four smaller bunches. For fields away from $\varphi = 90^{\circ}$, the 8 bunches of bananas split into  2 larger, 4 intermediate and 2 smaller bunches due to SOC. In the same way, $\mu$ is split into multiple branches. A small residual misalignment of the actual rotation plane with respect to the $\left(001 \right)$ plane would further split these branches as shown in by the theory lines in regime V. Following the reasoning above, we associate the branches $\theta, \iota, \tilde{\iota}$ around $\sim 250$~T with combination orbits of sheet 9 and sheet 10.

In conclusion, a systematic assignment of the sheet 9 and 10 dHvA branches to specific orbits in the theory is beyond the accuracy of our DFT calculations. However, the majority of the dHvA branches in regime V (except $\alpha$) exhibit a finite torque amplitude for fields along [010]. This is unique among all orbits of all FS sheets, allowing us to unambiguously assign all the branches with this property to the sheet pair 9 and 10 and determine its approximate size, even without knowing the exact shape and location of the orbits on the sheets.


\newpage
\section*{References}

\end{document}